\documentclass[12pt]{article}
\usepackage[english]{babel}
\usepackage[top=2.5cm, bottom=2.5cm, left=2.0cm, right=2.0cm]{geometry}
\usepackage{amscd,amsfonts,amsmath,amssymb,amsthm,bbm,bm,latexsym,mathrsfs}
\allowdisplaybreaks
\usepackage{bookmark}
\usepackage{enumitem}
\usepackage{adjustbox}
\usepackage{gensymb}
\usepackage{natbib}
\usepackage{float,rotating}
\hypersetup{hidelinks}

\usepackage[usenames]{color}
\usepackage{epsfig,epstopdf,graphics,graphicx,subfigure,longtable}
\usepackage[toc,page]{appendix}
\usepackage{chapterbib}
\usepackage[normalem]{ulem}% to highlight cancelled text
\usepackage{cancel}

\makeatother

\usepackage[section]{placeins}
\usepackage{setspace}

\theoremstyle{remark}

\theoremstyle{plain}

\def\I {\mathbb{I}}

\usepackage{nomencl}

\makenomenclature

\makeatletter
\def\@biblabel#1{\hspace*{-\labelsep}}
\makeatother
\begin{document}
\title{\vspace*{-5cm}
\hfill
\newline
\vspace*{2cm}
\renewcommand{\baselinestretch}{1}
\\
{ \textbf{ 
\\[2em]
COVID-19 spreading in financial networks: A semiparametric matrix regression model}\footnote{We thank the participants at the 14th International Conference Computational and Financial Econometrics (CFE 2020) for the provided comments. We also thank Maurizio La Mastra for the excellent research assistant.
This research used the SCSCF and the HPC multiprocessor cluster systems provided by the Venice Centre for Risk Analytics (VERA) at University Ca' Foscari of Venice.
Matteo Iacopini acknowledges financial support from the Marie Sk\l{}odowska-Curie Actions, European Union, Seventh Framework Program HORIZON 2020 under REA grant agreement n. 887220.}}
}

\author{
\Large Monica Billio\thanks{Ca' Foscari University of Venice, Email: billio@unive.it. (corresponding author)}\\
\Large Roberto Casarin\thanks{Ca' Foscari University of Venice, Email: r.casarin@unive.it.}\\
\Large Michele Costola\thanks{Ca' Foscari University of Venice, Email: michele.costola@unive.it.}\\
\Large Matteo Iacopini\thanks{Vrije Universiteit Amsterdam, Email: m.iacopini@vu.nl.}\\
}
\date{\today}
\maketitle

%\tableofcontents
\thispagestyle{empty} \centerline{\large \bf Abstract} \baselineskip 14pt %
\vskip 10pt
\noindent Network models represent a useful tool to describe the complex set of financial relationships among heterogeneous firms in the system.  
In this paper, we propose a new semiparametric model for temporal multilayer causal networks with both intra- and inter-layer connectivity.
A Bayesian model with a hierarchical mixture prior distribution is assumed to capture heterogeneity in the response of the network edges to a set of risk factors including the European COVID-19 cases.
We measure the financial connectedness arising from the interactions between two layers defined by stock returns and volatilities. 
In the empirical analysis, we study the topology of the network before and after the spreading of the COVID-19 disease.
\noindent

\vskip %
5pt\noindent \textbf{JEL Classification}: C11; C58; G10.

\newpage

\setcounter{page}{1}

\doublespacing

\section{Introduction}  \label{introduction}
The recent outbreak of the COVID-19 disease has severely affected the economy and the financial markets due to the consequences of the lockdowns and travel limitations. According to the International Monetary Fund \citep{imf2020}, the world growth in 2020 is projected to have a contraction of 4.4\%. During February-March of 2020, the global financial market has suffered multiple crashes having the largest drop of around 13\% on 16 March 2020.  To ensure the financial stability and avoid the breakdown of the markets, central banks supported the functioning of the system with asset purchase programs.
While the global financial crisis has been originated by the vulnerabilities of the US mortgage market, which in turn has been the root of the European sovereign debt crisis, the COVID-19 pandemic has represented an (unprecedented) exogenous shock to the financial system that could not be reasonably foreseen and hence, priced by the financial markets.
The analysis the financial connectedness through the network modeling can provide interesting insights to policy makers on the effect of the COVID-19 on the financial system.

The literature of financial connectedness and network modeling has rapidly increased after the recent financial crisis both in theoretical \citep[i.e.,][]{elliott2014financial,acemoglu2015systemic} and empirical analyses \citep[i.e.,][]{billio2012econometric,diebold2014network}.
Regarding the econometric methods, the methodologies proposed for extracting unobserved networks have been particularly flourishing, especially in the Bayesian framework. \cite{ahelegbey2016bayesian,Daniel2016} combine a Bayesian graphical approach with vector autoregressive (VAR) models where the contemporaneous and temporal causal structures of the model are represented by means of two distinct graphs.
In a graphical model framework, \cite{bianchi2019modeling} propose a Markov-switching graphical SUR model to investigate changes in systemic risk. The authors show that connectivity increases in during 1999-2003 and in 2008-2009. Using a time-varying parameter vector autoregressive model (TVP-VAR), \cite{geraci2018measuring} estimate a dynamic Granger Network in the S\&P 500 market and find a gradual decrease in network connectivity not detectable using a rolling window approach. Similarly, in the framework of forecast variance error decomposition \citep{diebold2009measuring,diebold2014network}.
%\cite{korobilis2018measuring} estimate a large Bayesian TVP-VAR model showing that the total spillover index better describes turning points with respect to the one obtained from rolling-window VAR estimates.
In high-dimensional multivariate time-series, \cite{billio2019bayesian} propose a Bayesian nonparametric Lasso prior for VAR models. The causal networks are extracted through clustering and shrinking effects, and well describe real-world networks features. \cite{bernardi2019high} propose a shrinkage and selection methodology for network inference through a regularised linear regression model with spike-and-slab prior on the parameters. The financial linkages are expressed in terms of inclusion probabilities resulting in a weighted directed network.
Referring to econometric models for networks, \cite{billio2018bayesianDT} propose a dynamic linear regression model for tensor-valued response variables and covariates with a parsimonious parametrization based on the parallel factor decomposition. \cite{billio2018bayesian} propose a Bayesian Markov-switching regression model for multidimensional arrays (tensors) of binary time series. The coefficient tensor can switch between multiple regimes in order to capture time-varying sparsity patterns of the network structure.
% Moreover, firm-level centrality is positively correlated to realized financial losses.

In this paper we propose a new semiparametric model for temporal multilayer causal networks. A Bayesian model with mixture prior distribution is assumed in order to model heterogeneity in the response of the network edges to a set of exogenous variables.

Recently, the literature has focused on multilayer networks where different channels of relationships, defined as layers, characterize a given set of nodes (system).
The study of connectedness involving the interdependence among these layers can improve the measurement of the network topology. For instance, \cite{wang2020multilayer} consider a multilayer Granger causality network composed by mean spillover, volatility spillover, and extreme risk spillover layers. The authors show that significant changes in connectivity on extreme risk and volatility spillover layers before a general financial turmoil. 
\cite{casarin2020multilayer} propose a Bayesian graphical vector autoregressive model to extract multilayer network in the international oil market and show that oil production network is a lagged driver for prices.

In this paper, we follow this stream of literature and consider multilayer networks with both intra- and inter-layer connectivity. We measure the financial connectedness arising from the interactions between two layers defined by assets returns and volatilities. These connectivity effects are represented at four levels: (i) return linkages (returns causes returns); (ii) volatility linkages (volatility causes volatility); (iii) risk premium linkages (volatility causes returns); and (iv) leverage linkages (return causes volatility). 

To investigate the role of selected risk factors on the dynamic evolution of the multilayer financial network, we propose a novel semiparametric model for panels of matrix-valued data.
The use of matrix-valued statistical models in time series econometrics has become increasingly popular over the last decades. In the seminal paper by \cite{HarrisonWest99BayesForecastDLM}, matrix-valued distributions were exploited for representing state space models. More recently, \cite{CarvWest07DynMatNormGraph,Wang09Bayes_matrixNormalGraph} used the matrix normal distribution in Bayesian dynamic linear models while \cite{Carvetal07HIW_Graph} applied the hyper-inverse Wishart distribution in a Gaussian graphical model.
Other applications of matrix-variate distributions include stochastic volatility \citep{Uhlig97Bayes_VAR_SV,Gourieroux09Wishart_AR,Golosnoy12conditional_Wishart_AR}, classification of longitudinal datasets \citep{Viroli11MatNorm}, models for network data \citep{Zhu17Network_VAR,Zhu19Network_Quantile_regression}, and factor models \citep{Chen19Matrix_DynamicFactor}.
With respect to the existing literature, our approach proposes two main contributions.
First, we extend the matrix-valued linear model to panels of matrix-valued data.
Second, we propose a hierarchical mixture prior to cope with overfitting and loss of efficiency in high-dimensional settings. This prior choice allows for a semiparametric model that grants higher flexibility in investigating the impact of covariates on matrix-valued response variables.
Our model and inference are well suited for the analysis of multilayer temporal networks, where the intra- and inter-layer connectivity at each point in time is encoded by a cross section of adjacency matrices.

Finally, an original application to a European financial network among 412 firms based in Germany, France, and Italy, shows that the proposed framework scales well in high-dimensions (i.e., hundreds of nodes) and can be successfully used to provide new insights on shock transmission in financial markets.
%of the network to define a novel econometric model for matrix-valued panel data, which allows us to study the temporal evolution of a cross section of adjacency matrices.
%
Inspired by the causality definition between return and volatility \cite[i.e.,][]{bekaert2000}, we label the intra-connectivity risk premium according to the time-varying risk premium hypothesis (volatility causes returns) and the leverage according to the leverage hypothesis (return shocks lead to changes in volatilities). In the proposed model, each adjacency matrix related to the four levels of connectivity is modelled as a function of a set of risk factors, including market returns, implied volatility, corporate credit risk, and the number of COVID-19 new confirmed European cases.
In the empirical analysis, we study the topology of the European financial network before and after the spreading of the COVID-19 disease.

Our findings highlight that COVID-19 is the most relevant factor in explaining the connectivity of the European financial network at firm and sector level, in particular industrial, real estate, and health care.
The probabilities of volatility and risk premium linkages are the most positively affected by the COVID-19, which at the same time has a negative effect on leverage linkages.
Moreover, we find evidence of a positive relationship between firm centrality and the number of  its linkages that are impacted by the COVID-19, across all layers except leverage.

The remaining of the paper is structured as follows.
Section~\ref{sec:netmodel} introduces a novel econometric framework for matrix-valued panel data, then Section~\ref{sec:inference} presents the Bayesian inference procedure.
Section~\ref{sec:empirical_analysis} illustrates the empirical analysis and our major results.
Finally, Section~\ref{sec:conclusion} concludes.

%\cite{bormetti2020stochastic} whose contribution is the introduction of a new family of discrete-time stochastic volatility option pricing models characterised by two measurement equations as well as by a transition equation for the latent conditional variance states described by an Heterogeneous Autoregressive Gamma process with leverage effects; the resulting SV-LHARG(p) model is applied on a large sample of the SP 500 index options and results to be more effective (with an high level of captured latent volatility persistence), with respect to other competitor models, in precisely capturing the dynamics of the short time-to-maturity ATM implied volatility with consequent better option pricing performance. The SV-LHARG(p) model is also strengthened by the application of a Bayesian inference procedure for both parameters and latent conditional variance as well as by the development of a Markov Chain Monte Carlo procedure for posterior approximation.

\section{Network model}    \label{sec:netmodel}
Let $\mathcal{G}_t=(G_{11t},G_{22t},G_{12t},G_{21t},E)$ a two-layer temporal network \citep{boccaletti2014structure}, where $G_{ijt}\subset E\times E$ is the connectivity graph between layers $i$ and $j$, $E=\{1,\ldots,N\}$ is the set of nodes (firms). 
In the proposed framework, each node represents a firm, the two layers are given by firm stock return (layer $1$) and volatility (layer $2$). Four graphs encode the connectivity between and within layers. 
Here, we focus on causal financial network where each adjacency matrix is directed and hence, asymmetric. This implies that each element of the matrix indicates a causal relationship between two nodes.
The connectivity is represented through the intra-layer adjacency matrices $Y_{11,t}$ and $Y_{22,t}$ and inter-layer adjacency matrices $Y_{12,t}$ and $Y_{21,t}$. 

For the intra-connectivity graphs, we label \textit{return linkages} the sub-network $Y_{11,t}$ (returns causes returns) and \textit{volatility linkages} the sub-network $Y_{22,t}$ (volatility causes volatility).
Regarding the inter-connectivity graphs, we label the two graphs inspired by the causality definition in asset pricing between return and volatility as discussed in \cite{bekaert2000}. 
The label \textit{risk premium linkages} for the sub-network $Y_{11,t}$ refers to the time-varying risk premium hypothesis (volatility causes return) while the label \textit{leverage linkages} for the sub-network $Y_{22,t}$ is based on the leverage hypothesis (return causes volatilities).

We propose the following matrix-variate linear model for studying the impact of a set of $R$ covariates $(f_{1,t},\ldots,f_{R,t})$ on the linkages
\begin{equation}
\begin{split}
Y_{11,t}=\sum_{r=1}^{R}B_{11,r}f_{r,t}+ E_{11,t},\quad E_{11,t}\sim \mathcal{MN}_{n,n}(O,\Sigma_{1,11},\Sigma_{2,11}) \\
Y_{22,t}=\sum_{r=1}^{R}B_{22,r}f_{r,t}+ E_{22,t},\quad E_{22,t}\sim \mathcal{MN}_{n,n}(O,\Sigma_{1,22},\Sigma_{22,2}) \\
Y_{12,t}=\sum_{r=1}^{R}B_{12,r}f_{r,t}+ E_{12,t},\quad E_{12,t}\sim \mathcal{MN}_{n,n}(O,\Sigma_{12,1},\Sigma_{12,2}) \\
Y_{21,t}=\sum_{r=1}^{R}B_{21,r}f_{r,t}+ E_{21,t},\quad E_{21,t}\sim \mathcal{MN}_{n,n}(O,\Sigma_{21,1},\Sigma_{21,2})
\end{split}
\label{eq:model_observables}
\end{equation}
where $B_{lk,r}$ are $(n\times n)$ matrices of coefficients and $\mathcal{MN}_{n,n}(O,\Sigma_1,\Sigma_2)$ denotes the zero-mean matrix normal distribution \citep[see][Ch.2, for further details]{gupta1999matrix} with two variance/covariance matrices $\Sigma_1$ and $\Sigma_2$.
If the $(n \times p)$ random matrix $X$ is distributed as a matrix normal with mean $M$ and covariance matrices $\Sigma_1,\Sigma_2$, with $\Sigma_1$ of size $(n\times n)$ and $\Sigma_1$ of size $(p\times p)$, written $X \sim \mathcal{MN}_{n,p}(M,\Sigma_1,\Sigma_2)$, then
\begin{equation}
P(X|M,\Sigma_1,\Sigma_2) = (2\pi)^{-np/2} |\Sigma_2|^{-p/2} |\Sigma_1|^{n/2} \exp\Big( -\frac{1}{2} \operatorname{tr}\big( \Sigma_2^{-1} (X-M)' \Sigma_1^{-1} (X-M) \big) \Big).
\label{eq:matrix_normal_pdf}
\end{equation}
For identification purposes, we assume $\Sigma_{lk,1}=I_n$ and $\Sigma_{lk,2}=\operatorname{diag}(\sigma^2_{1,lk},\ldots,\sigma^2_{n,lk})$, for each $l,k=1,2$.

\section{Bayesian Inference}   \label{sec:inference}
\subsection{Prior specification}
\begin{figure}[t!]
\centering
\includegraphics[trim= 0mm 0mm 0mm 0mm,clip,height= 5.0cm, width= 10.4cm]{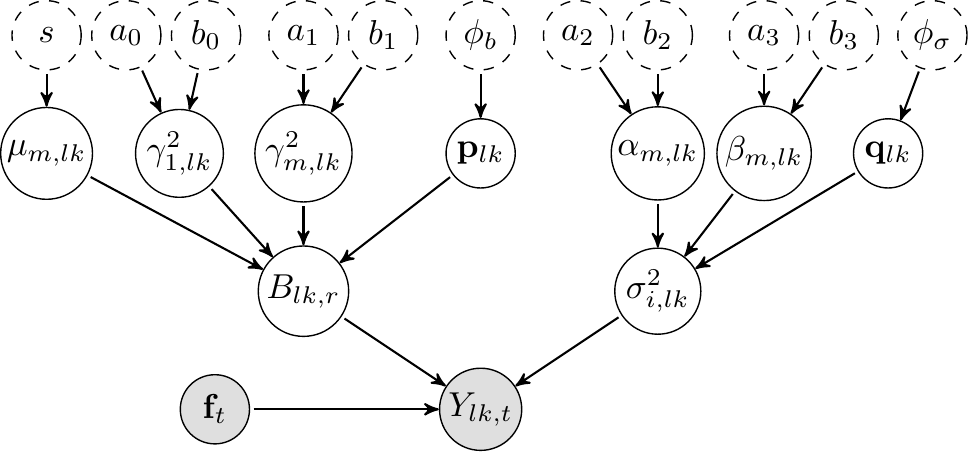}
\caption{Directed Acyclic Graph of the proposed Bayesian semiparametric model for multilayer networks. It exhibits the conditional independence structure of the observation model for $Y_{lk,t}$ with covariates $\mathbf{f}_{t}=(f_{1,t},\ldots,f_{R,t})'$ (grey circles), the parameters $\mathbf{p}_{lk}=(p_{1,lk},\ldots,p_{M_b,lk})$, $\mathbf{q}_{lk}=(q_{1,lk},\ldots,q_{M_\sigma,lk})$, $B_{lk,r},\sigma_{i,lk}^2$, the component-specific parameters $\mu_{m,lk},\gamma_{m,lk}^2,\alpha_{m,lk},\beta_{m,lk}$ (white solid circles) and the fixed hyperparameters $s,a_0,b_0,a_1,b_1,a_2,b_2,a_3,b_3$ (white dashed circles). The directed arrows show the causal dependence structure of the model.}
\label{fig:DAG}
\end{figure}

As regards the prior assumption on the model parameters, we choose the following independent mixture of Normal distributions for the coefficients $b_{ij,lk,r}$, for $i,j=1,\ldots,n$ (with $i  \neq j$), $l,k=1,2$, and $r=1,\ldots,R$
%\begin{align}
%b_{ij,lk,r}|\mathbf{p}_{lk},\boldsymbol{\mu}_{lk},\boldsymbol{\gamma}_{lk}^2 \sim p_{1,lk} \delta(b_{ij,lk}) + \sum_{m=2}^{M} p_{m,lk} \mathcal{N}(b_{ij,lk}|\mu_{m,lk},\gamma^2_{m,lk})
%\end{align}
%A more computationally efficient alternative is
\begin{align}
b_{ij,lk,r}|\mathbf{p}_{lk},\boldsymbol{\mu}_{lk},\boldsymbol{\gamma}_{lk}^2 \sim p_{1,lk} \mathcal{N}(b_{ij,lk,r}| 0,\gamma^2_{1,lk}) + \sum_{m=2}^{M_b} p_{m,lk} \mathcal{N}(b_{ij,lk}|\mu_{m,lk},\gamma^2_{m,lk}).
\label{eq:prior_Bijlkr}
\end{align}
To solve the label switching problem, we impose the identification constraint
$\mu_{2,lk} < \mu_{3,lk} < \ldots < \mu_{M_b,lk}$.
As regards the prior distribution for the variances $(\sigma^2_{1,lk},\ldots,\sigma^2_{n,lk})$ we assume the following mixture of Inverse Gamma distributions
\begin{equation}
\sigma^2_{i,lk,r}|\mathbf{q}_{lk},\boldsymbol{\alpha}_{lk},\boldsymbol{\beta}_{lk} \sim \sum_{m=1}^{M_\sigma} q_{m,lk} \mathcal{IG}(\sigma^2_{i,lk}|\alpha_{m,lk},\beta_{m,lk}),
\label{eq:prior_sigma2_ilk}
\end{equation}
and impose the identification constraint on the mean by assuming $\beta_{1,lk}/(\alpha_{1,lk}-1) < \beta_{2,lk}/(\alpha_{2,lk}-1) < \ldots < \beta_{M_\sigma,lk}/(\alpha_{M_\sigma,lk}-1)$.
Finally, for the hyper-parameter of the mixture prior, we assume the following Dirichlet prior and Normal-Inverse Gamma prior distributions:
\begin{eqnarray}
(p_{1,lk},p_{2,lk},\ldots,p_{M_b,lk}) & \sim & \mathcal{D}ir(\phi_b,\phi_b,\ldots,\phi_b) \\
(q_{1,lk},q_{2,lk},\ldots,q_{M_\sigma,lk}) & \sim & \mathcal{D}ir(\phi_\sigma,\phi_\sigma,\ldots,\phi_\sigma) \\
\mu_{1,lk}	    & =    & 0,  \\
\gamma_{1,lk}^2 & \sim & \mathcal{G}a(a_0,b_0),  \\
\mu_{m,lk}      & \sim & \mathcal{N}(0,s^2), \quad m=2,\ldots,M_b \\
\gamma_{m,lk}^2 & \sim & \mathcal{IG}(a_1,b_1), \quad m=2,\ldots,M_b \\
\alpha_{m,lk}   & \sim & \mathcal{G}a(a_2,b_2), \quad m=1,\ldots,M_\sigma \\
\beta_{m,lk}    & \sim & \mathcal{G}a(a_3,b_3), \quad m=1,\ldots,M_\sigma,
\end{eqnarray}
which is a standard choice in Bayesian mixture modeling \citep{fruhwirth2006finite}.
The first component of the mixture prior for coefficients $b_{ij,lk,r}$ is a Bayesian Lasso \citep{park2008bayesian} distribution, that corresponds to setting $a_0=1$.
This prior specification strategy overcomes overparametrization and overfitting issues by clustering coefficients into groups and by shrinking the coefficients in the first group toward zero, thus improving the estimation efficiency in high-dimensions.
The proposed hierarchical mixture prior distribution naturally induces a mixture model for the matrix-valued observations. For each matrix $Y_{lk,t}$, with $l,k=1,2$ and $t=1,\ldots,T$, by integrating out the parameters in the likelihood, one obtains the following marginal likelihood.
\begin{align}
P(Y_{lk,t} | \boldsymbol{\mu}_{lk},\boldsymbol{\gamma}_{lk}^2, \boldsymbol{\alpha}_{lk},\boldsymbol{\beta}_{lk}) & = \sum_{m'=1}^{M_\sigma^n} \sum_{m=1}^{M_b^{n^2}} \tilde{p}_{m,lk} \tilde{q}_{m',lk} P(Y_{lk,t} | \tilde{\boldsymbol{\theta}}_{m,lk}^b, \tilde{\boldsymbol{\theta}}_{m',lk}^\sigma),
\label{eq:marginal_likelihood}
\end{align}
where $\tilde{\boldsymbol{\theta}}_{m,lk}^b = (\tilde{\boldsymbol{\mu}}_{m,lk},\tilde{\boldsymbol{\gamma}}_{m,lk}^2)$, $\tilde{\boldsymbol{\theta}}_{m,lk}^\sigma = (\tilde{\boldsymbol{\alpha}}_{m',lk},\tilde{\boldsymbol{\beta}}_{m',lk})$, and
\begin{align*}
 & P(Y_{lk,t} | \tilde{\boldsymbol{\theta}}_{m,lk}^b, \tilde{\boldsymbol{\theta}}_{m',lk}^\sigma) = \\
 & = \int \!\!\!\int P(Y_{lk,t} | B_{lk,1},\ldots,B_{lk,R}, \boldsymbol{\sigma}_{lk}^2) P(\boldsymbol{\sigma}_{lk}^2 | \tilde{\boldsymbol{\theta}}_{m',lk}^\sigma) P(B_{lk,1},\ldots,B_{lk,R} | \tilde{\boldsymbol{\theta}}_{m,lk}^b) \: \mathrm{d}B_{lk,1} \cdots \mathrm{d}B_{lk,R} \mathrm{d}\boldsymbol{\sigma}_{lk}^2.
\end{align*}
See the Appendix for a proof.
We summarize our Bayesian semiparametric model in the Directed Acyclic Graph representation of Figure \ref{fig:DAG}.

\subsection{Posterior approximation}
Start by introducing two set of allocation variables $D_{ij,lk,r}^b$, for $i,j=1,\ldots,n$, $l,k=1,2$, $r=1,\ldots,R$, and $D_{i,lk}^\sigma$, for $i=1,\ldots,n$, $l,k=1,2$. Denote the collection of all parameters with $\boldsymbol{\theta} = (B_{lk,1},\ldots,B_{lk,R}, \sigma_{1,lk}^2,\ldots,\sigma_{n,lk}^2)'$, and let $\mathbf{Y},\mathbf{f}$ be the collection of all observed networks and risk factors, respectively.
Letting $\boldsymbol{\sigma}_{lk}^2 = (\sigma_{1,lk}^2,\ldots,\sigma_{n,lk}^2)'$, the likelihood of the model in Eq.~\eqref{eq:model_observables} is
\begin{align}
\notag
P(\mathbf{Y},\mathbf{f}|\boldsymbol{\theta}) & = \prod_{t=1}^T \prod_{l=1}^2 \prod_{k=1}^2 (2\pi)^{-n^2/2} |\operatorname{diag}(\boldsymbol{\sigma}_{lk}^2)|^{-n/2} |I_n|^{n/2} \\ \notag
 & \cdot \exp\Big( -\frac{1}{2} \operatorname{tr}\Big( \operatorname{diag}(\boldsymbol{\sigma}_{lk}^2)^{-1} \big(Y_{lk,t}-\sum_{r=1}^R B_{lk,r} f_{r,t} \big)' I_n^{-1} \big( Y_{lk,t}-\sum_{r=1}^R B_{lk,r} f_{r,t} \big) \Big) \Big) \\ \notag
 & =  (2\pi)^{-2n^2T} \prod_{l=1}^2 \prod_{k=1}^2 |\operatorname{diag}(\boldsymbol{\sigma}_{lk}^2)|^{-nT/2} \\ \notag
 & \cdot \exp\Big( -\frac{1}{2} \operatorname{tr}\Big( \sum_{l=1}^2 \sum_{k=1}^2 \operatorname{diag}(\boldsymbol{\sigma}_{lk}^2)^{-1} \sum_{t=1}^T \big(Y_{lk,t}-\sum_{r=1}^R B_{lk,r} f_{r,t} \big)' \big( Y_{lk,t}-\sum_{r=1}^R B_{lk,r} f_{r,t} \big) \Big) \Big).
\label{eq:likelihood}
\end{align}
Since the joint posterior distribution is not tractable, we implement an MCMC approach based on a Gibbs sampling algorithm to sample from the posterior distribution and to approximate all posterior quantities of interest.
The Gibbs sampler iterates over the following steps:
\begin{enumerate}
\item Draw $(b_{ij,lk,1},\ldots,b_{ij,lk,R})$ from the Normal distribution $P(b_{ij,lk,r} | -)$.
\item Draw $\sigma^2_{i,lk}$ from Inverse Gamma distribution $P(\sigma_{i,lk}^2 | -)$.
\item Draw the allocations $(d_{11,lk,r}^b,\ldots,d_{n1,lk,r}^b,d_{12,lk,r}^b,\ldots,d_{n2,lk,r}^b,d_{1n,lk,r}^b,\ldots,d_{nn,lk,r}^b)$ from the discrete distribution $P(d_{ij,lk,r}^b | -)$.
\item Draw the allocations $(d_{1,lk}^\sigma,\ldots,d_{n,lk}^\sigma)$ from the discrete distribution $P(d_{i,lk}^\sigma | -)$.
\item Draw $(p_{1,lk},\ldots,p_{M_b,lk})$ from the Dirichlet distribution $P(\mathbf{p}_{lk} | -)$.
\item Draw $(q_{1,lk},\ldots,q_{M_\sigma,lk})$ from the Dirichlet distribution $P(\mathbf{q}_{lk} | -)$.
\item Draw the hyperparameters:
\begin{enumerate}[label=\alph*)]
\item $\mu_{m,lk}$, for $m=2,\ldots,M_\sigma$, from the Normal distribution $P(\mu_{m,lk}|-)$.
\item $\gamma_{1,lk}^2$ from the Generalized Inverse Gaussian distribution $P(\gamma_{1,lk}^2|-)$.
\item $\gamma_{m,lk}^2$, for $m=2,\ldots,M_b$, from the Inverse Gamma distribution $P(\gamma_{m,lk}^2|-)$.
\item $\alpha_{m,lk}$, for $m=1,\ldots,M_\sigma$, from the distribution $P(\alpha_{m,lk}|-)$.
\item $\beta_{m,lk}$, for $m=1,\ldots,M_\sigma$, from the Gamma distribution $P(\beta_{m,lk}|-)$.
\end{enumerate}
\end{enumerate}

\section{Empirical Analysis} \label{sec:empirical_analysis}
In this section, we first describe the firms' dataset and the set of risk factors (source: Bloomberg and Eikon/Datastream), then we illustrate the network extraction procedure by means of Granger causality tests. Finally, we discuss the results of the proposed network model.

\subsection{Data description}

\paragraph{The European firms.}
The dataset includes 412 European firms (176 German, 162  French, 74 Italian) belonging to 11 GICS sectors: Financials (43 firms), Communication Services (38 firms), Consumer Discretionary (61 firms), Consumer Staples (17 firms), Health Care (48 firms), Energy (9 firms), Industrials (88 firms), Information Technology (45 firms), Materials (24 firms), Real Estate (22 firms), Utilities (12 firms), and not classified in a specific GICS sector (5 firms).\footnote{The list of the firms, the countries, and the information about their GICS sectors and industries are available upon request to the Authors.}

The data sample ranges from the 4th of January 2016 to the 30th of September 2020, at weekly frequency (Friday-Friday), thus including the period before and after the outbreak of the COVID-19. 
%For each firm, we have download the daily stock prices (opening, closing, high, low) and the daily stock total returns: \\ 
The weekly logarithmic return for firm $i$, $r_{i,t}$, is obtained from the total returns series, whereas the weekly volatility is computed using the estimator of the variance proposed by \cite{garman1980estimation}: 
\begin{equation}
\begin{split}
\hat{\sigma}^2_{i,t} & =0.511(H_{i,t} - L_{i,t})^2 - 0.383(C_{i,t} - O_{i,t})^2\\
 & \quad -0.019[(C_{i,t}-O_{i,t})(H_{i,t}+L_{i,t}-2O_{i,t}) - 2(H_{i,t} - O_{i,t})(L_{i,t} -O_{i,t})],
\end{split}
\end{equation}
where $H_{i,t}$ is the weekly logarithmic high price, $L_{i,t}$ is the weekly logarithmic low price, $O_{i,t}$ is the weekly logarithmic opening price, and $C_{i,t}$ is the logarithmic closing price. 
The weekly prices have been obtained by taking in a given week the maximum among the daily high prices (weekly High Price), the minimum among the daily low prices (weekly Low Price), the opening price of the first available day in a week (weekly Opening Price), and the closing prices of the last available day in a week (weekly Closing Price).
%Figure \ref{fig:cross_ret_vol} shows the inter-quantile range at the 95\% (gray area) and the median (blue solid line) of the cross-sectional distribution of the returns and the estimated volatilities over time. The lowest (highest) peak in returns (volatilities) is on the 13th (20th) of March 2020 in the correspondence of the COVID-19 worldwide outbreak.

%\begin{figure}[htbp!]
%\center
%\begin{tabular}{c}
%\vspace{7pt}\includegraphics[width=400pt, height=150pt]{figures/median_ret-eps-converted-to.pdf}\\
%
%\vspace{7pt}\includegraphics[width=400pt, height=150pt]{figures/median_vol-eps-converted-to.pdf}
%\end{tabular}
%\caption{The cross-section median (solid blue line) and the inter-quantile range at 95\% (gray area) of the returns (top panel) and the estimated volatilities (bottom panel) for the considered European firms over time.}
%\label{fig:cross_ret_vol}
%\end{figure}

\begin{figure}[t!h]
\centering
\begin{tabular}{ccccc}
\begin{rotate}{90} \hspace*{30pt} {\footnotesize SX5E} \end{rotate} \hspace*{-12pt} &
\includegraphics[trim= 0mm 0mm 0mm 0mm,clip,height= 3.0cm, width= 7.0cm]{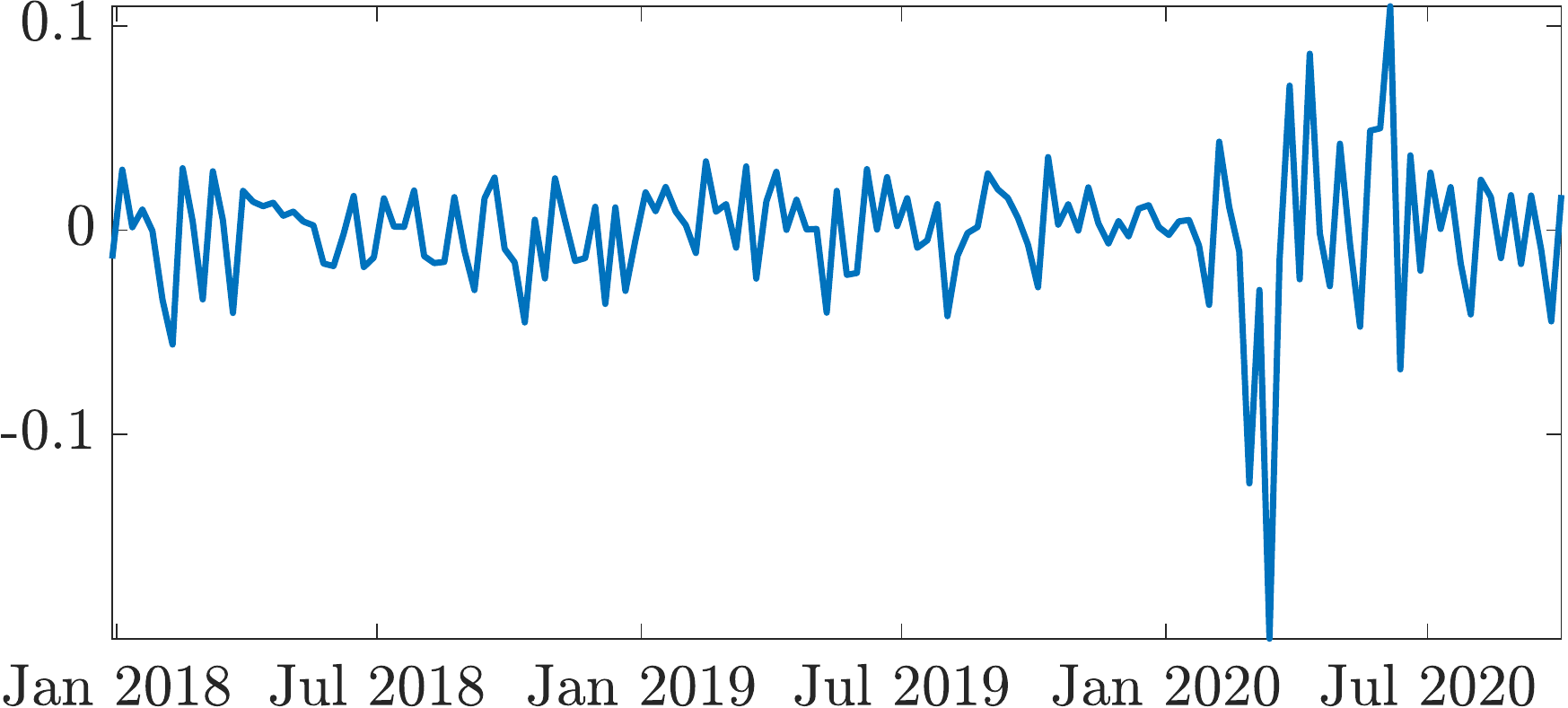} & &
\begin{rotate}{90} \hspace*{29pt} {\footnotesize V2X} \end{rotate} \hspace*{-12pt} &
\includegraphics[trim= 0mm 0mm 0mm 0mm,clip,height= 3.0cm, width= 7.0cm]{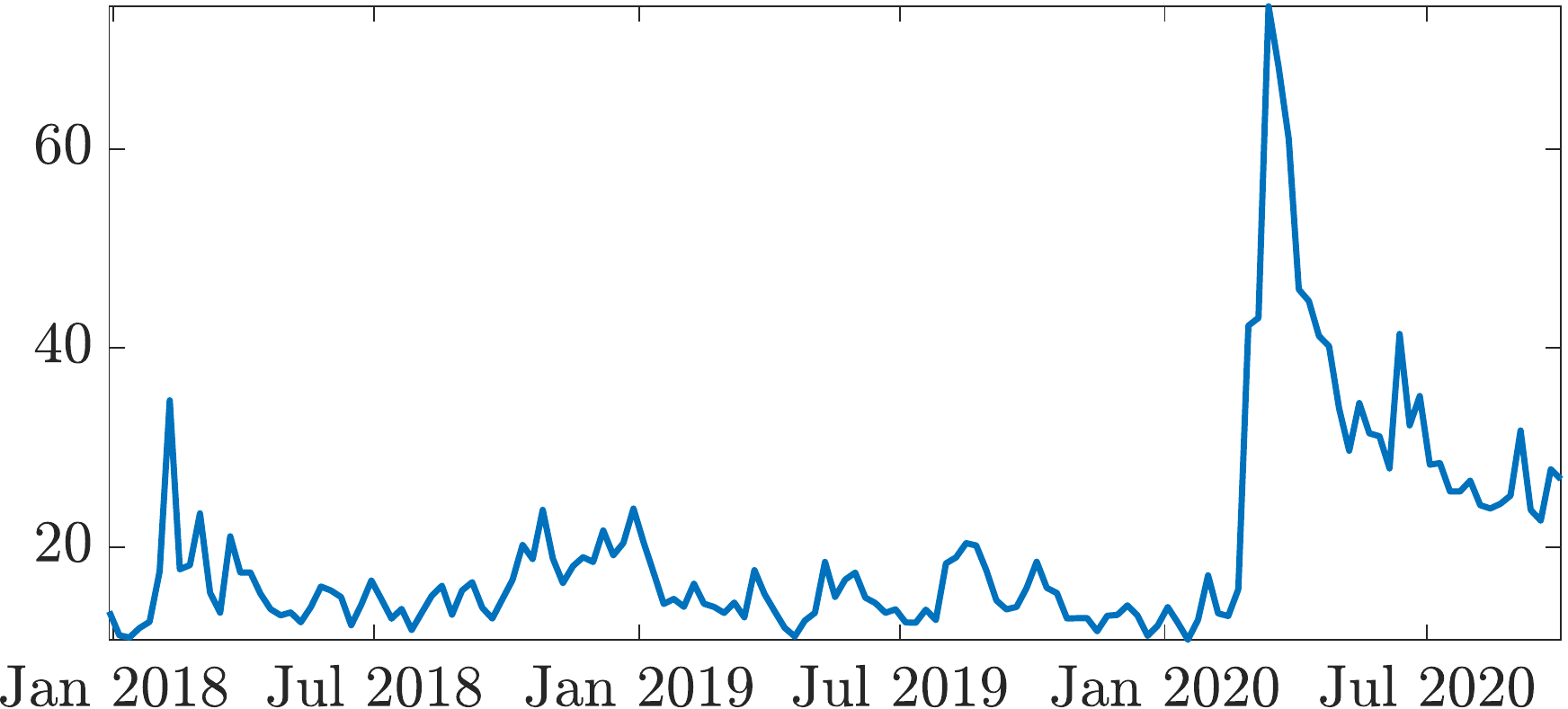} \\[10pt]
\begin{rotate}{90} \hspace*{15pt} {\footnotesize LECPOAS} \end{rotate} \hspace*{-12pt} &
\includegraphics[trim= 0mm 0mm 0mm 0mm,clip,height= 3.0cm, width= 7.0cm]{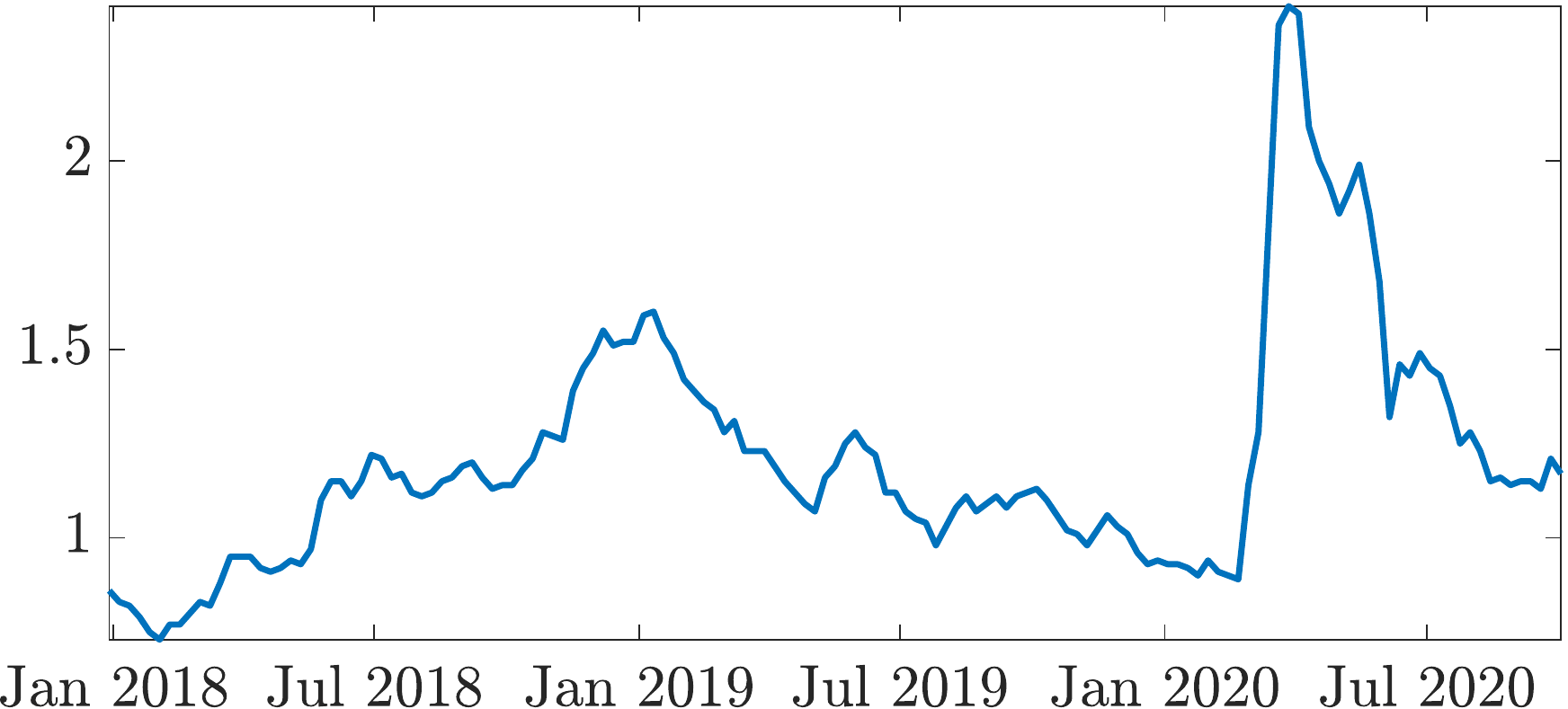} & &
\begin{rotate}{90} \hspace*{14pt} {\footnotesize NCOVEUR} \end{rotate} \hspace*{-12pt} &
\includegraphics[trim= 0mm 0mm 0mm 0mm,clip,height= 3.0cm, width= 7.0cm]{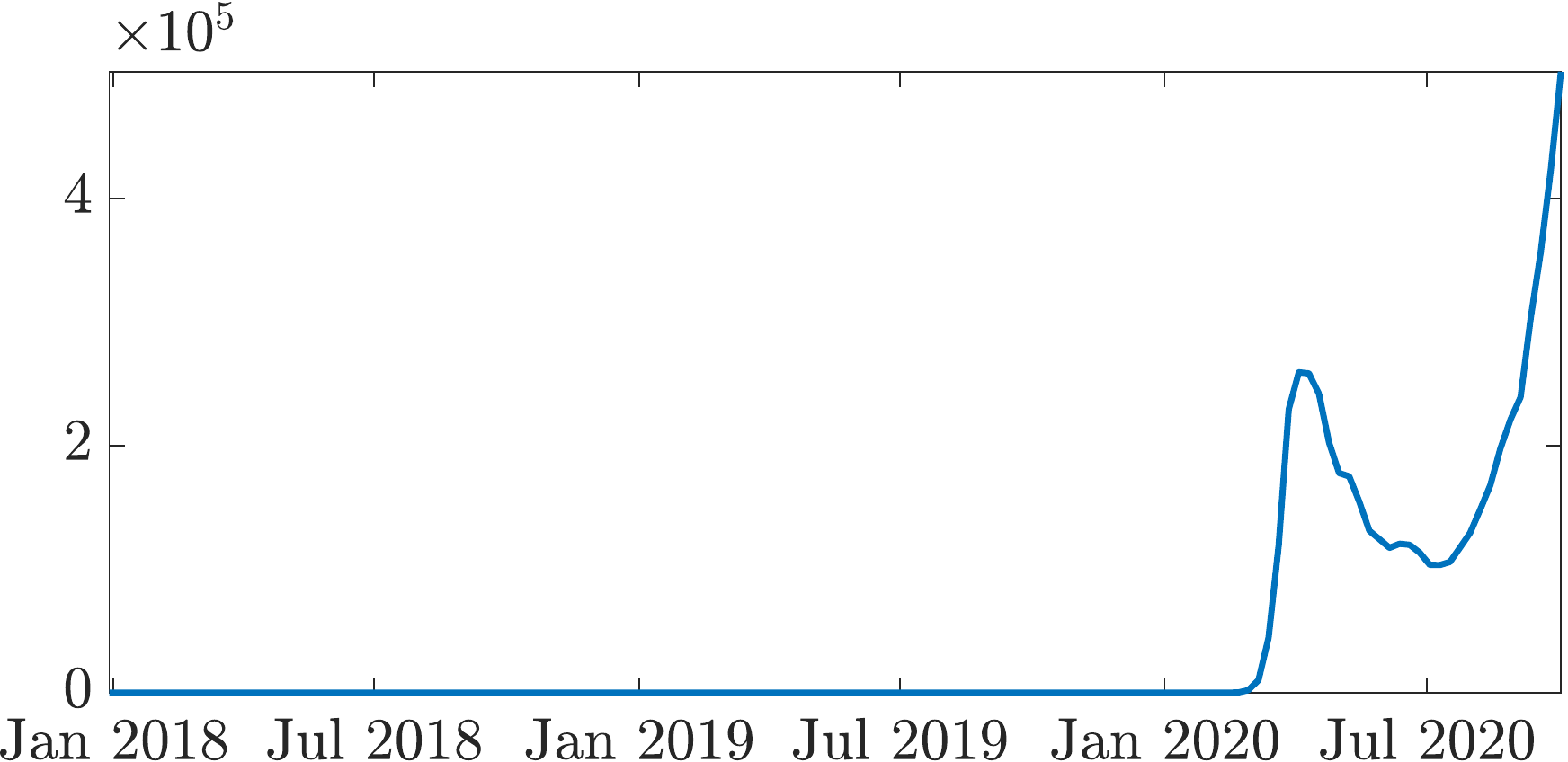}
\end{tabular}
\caption{The considered risk factors in the analysis: the returns on the Euro STOXX 50 index (top-left), the implied volatility on the Euro STOXX 50 index (top-right), the Bloomberg Barclays EuroAgg Corporate Average OAS (bottom-left) and the new European COVID-19 cases (bottom-right) for the considered European firms over time.}
\label{fig:riskfactors}
\end{figure}

\paragraph{The risk factors.}
We consider the following risk factors: (i) the log-returns on the Euro STOXX 50 index (SX5E), (ii) the implied volatility on the Euro STOXX 50 index (V2X), (iii) the Bloomberg Barclays EuroAgg Corporate Average OAS (LECPOAS) as a proxy for corporate credit risk, and (iv) the new European COVID-19 cases (NCOVEUR).
Figure~\ref{fig:riskfactors} reports the plots of the risk factors time series for the period under investigation and shows an abrupt change on the financial factors during the COVID-19 outbreak (bottom-right panel).

\subsection{Network extraction}  \label{sec:granger}
We estimate the dynamic networks of European financial institutions using pairwise Granger-causality \citep[e.g., see][]{billio2012econometric}. 
%We consider daily returns $\mathbf{r}_t$ and use a rolling window approach with a length of 252 observations (approximately 1 year). 
In this respect, we make use of a rolling window approach (104 observations, that is 2 years) and consider the intra- and inter-connectivity, namely, the return linkages, the volatility linkages, the risk premium linkages, and the leverage linkages:
\begin{equation}
%\Phi(L)\mathbf{r}_t=\mathbf{V}_t,
\begin{split}
 & x_{i,t} = \sum_{l=1}^{m}b_{11l} x_{i,t-l}+\sum_{l=1}^{m}b_{12l} x_{j,t-l}+\varepsilon_{it}\\
 & x_{j,t} = \sum_{l=1}^{m}b_{21l} x_{i,t-l}+\sum_{l=1}^{m}b_{22l} x_{j,t-l}+\varepsilon_{jt}
\end{split}
\end{equation}
where $i,j=1,\dots,k$ and $x = \{r,\hat{\sigma}\}$. Each entry $(i,j)$ of the adjacency matrix $Y_{lk,t}$ associated to layer $lk$, with $l,k=\,2$, is defined as $Y_{ij,lk,t} = pval(b_{ij})$ for $i\neq j$. Therefore, the element $Y_{ij,lk,t}$ represents the observed probability that the relationship between $x_{i,t}$ and $x_{j,t}$ is statistically different from zero. 
We estimate a total of $145 \times 4$ adjacency matrices, $145$ matrices for each layer, covering the period from the 29th of December 2017 to the 20th of October 2020.\footnote{The estimation algorithm has been parallelized and implemented in MATLAB on two nodes at the High Performance Computing (HPC) cluster (VERA - Ca' Foscari University). Each node has 2 CPUs Intel Xeon with 20 cores 2.4 Ghz and 768 GB of RAM.}

\begin{figure}[t!]
\centering
\setlength{\abovecaptionskip}{-1pt}
\begin{tabular}{ccccc}
\begin{rotate}{90} \hspace*{34pt} {$\substack{Y_{11,t}\\\text{return}}$} \end{rotate} \hspace*{-4pt} &
\includegraphics[trim= 0mm 0mm 0mm 0mm,clip,height= 3.1cm, width= 7.4cm]{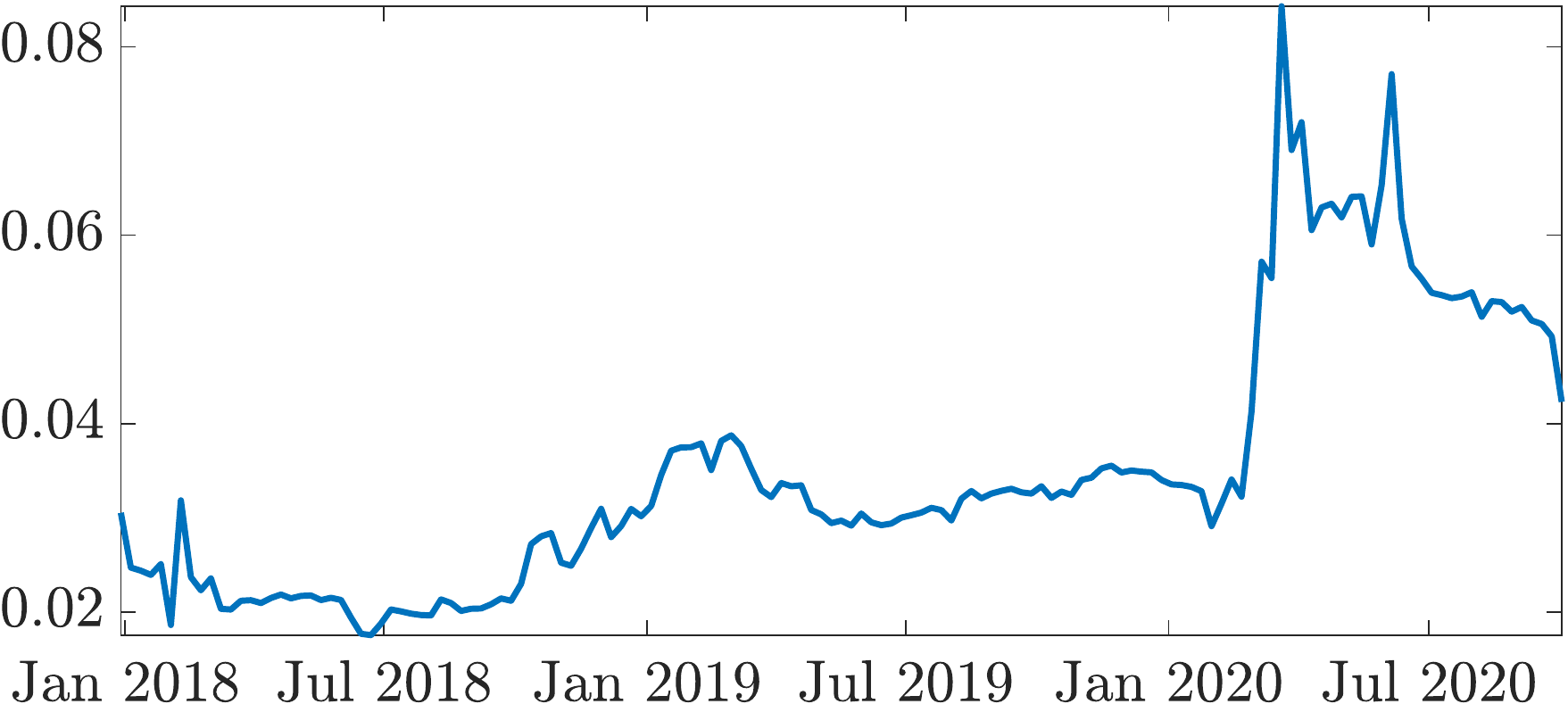} & &
\begin{rotate}{90} \hspace*{28pt} {$\substack{Y_{22,t}\\\text{leverage}}$} \end{rotate} \hspace*{-4pt} &
\includegraphics[trim= 0mm 0mm 0mm 0mm,clip,height= 3.1cm, width= 7.4cm]{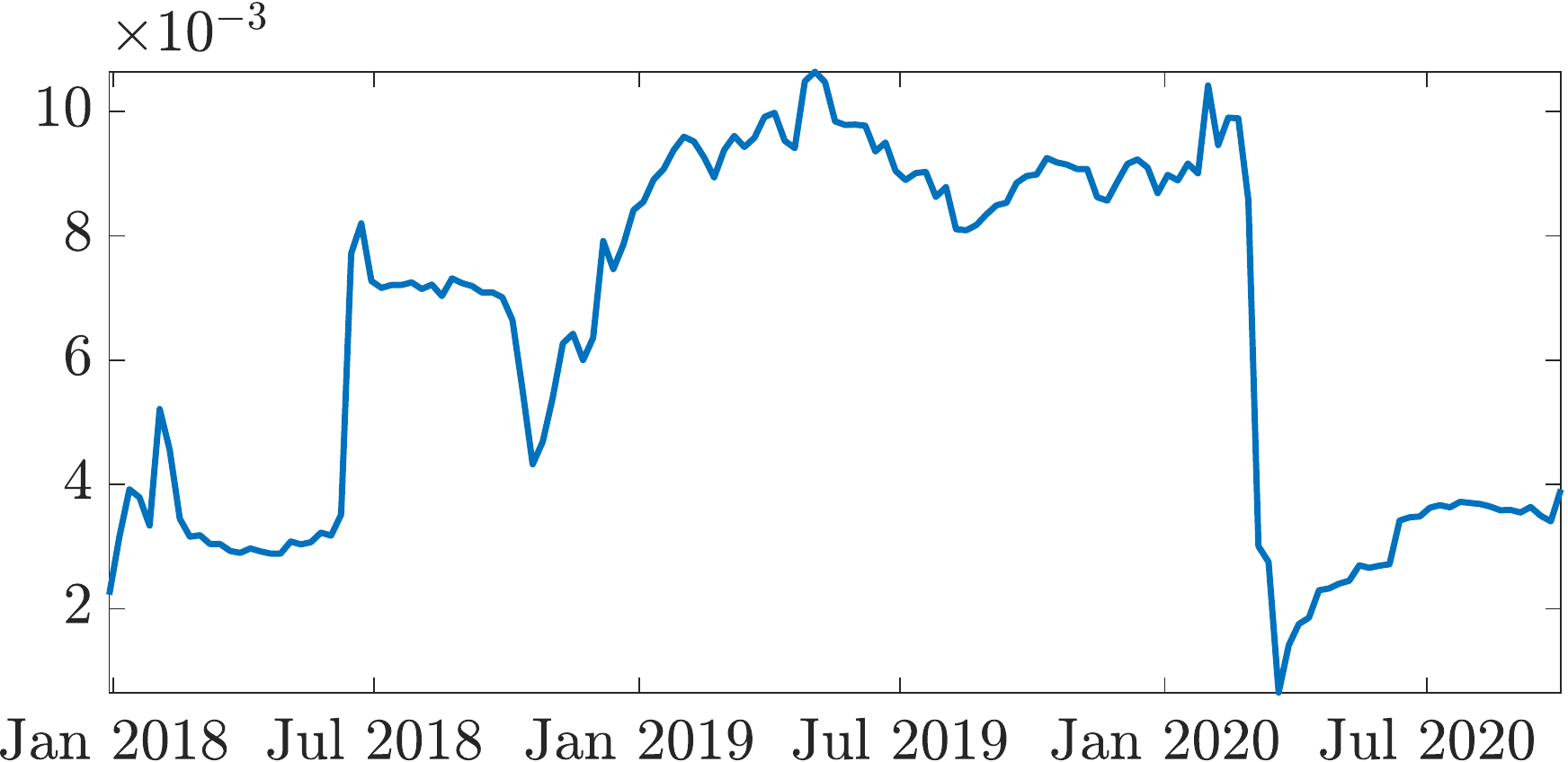} \\[10pt]
\begin{rotate}{90} \hspace*{31pt} {$\substack{Y_{21,t}\\\text{volatility}}$} \end{rotate} \hspace*{-4pt} &
\includegraphics[trim= 0mm 0mm 0mm 0mm,clip,height= 3.1cm, width= 7.4cm]{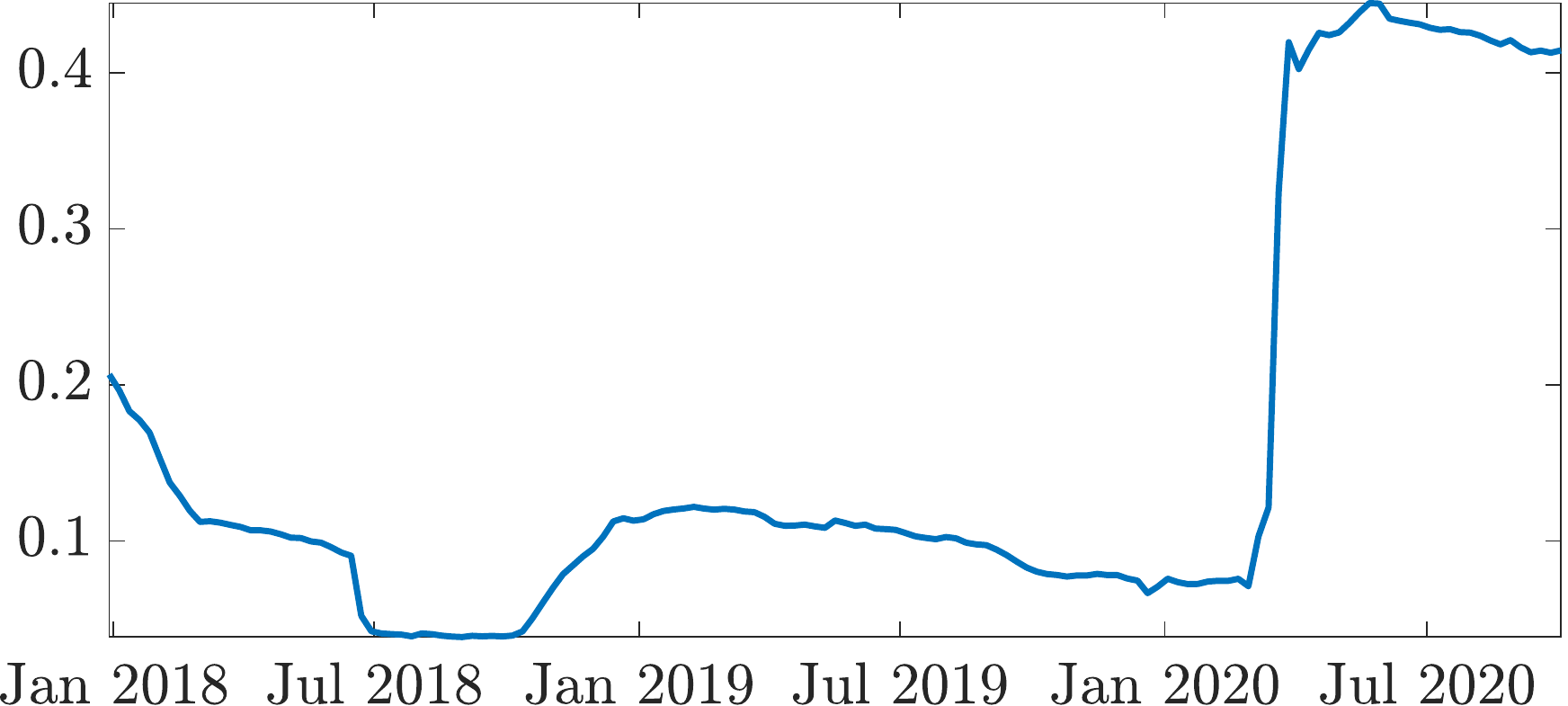} & &
\begin{rotate}{90} \hspace*{20pt} {$\substack{Y_{12,t}\\\text{risk premium}}$} \end{rotate} \hspace*{-4pt} &
\includegraphics[trim= 0mm 0mm 0mm 0mm,clip,height= 3.1cm, width= 7.4cm]{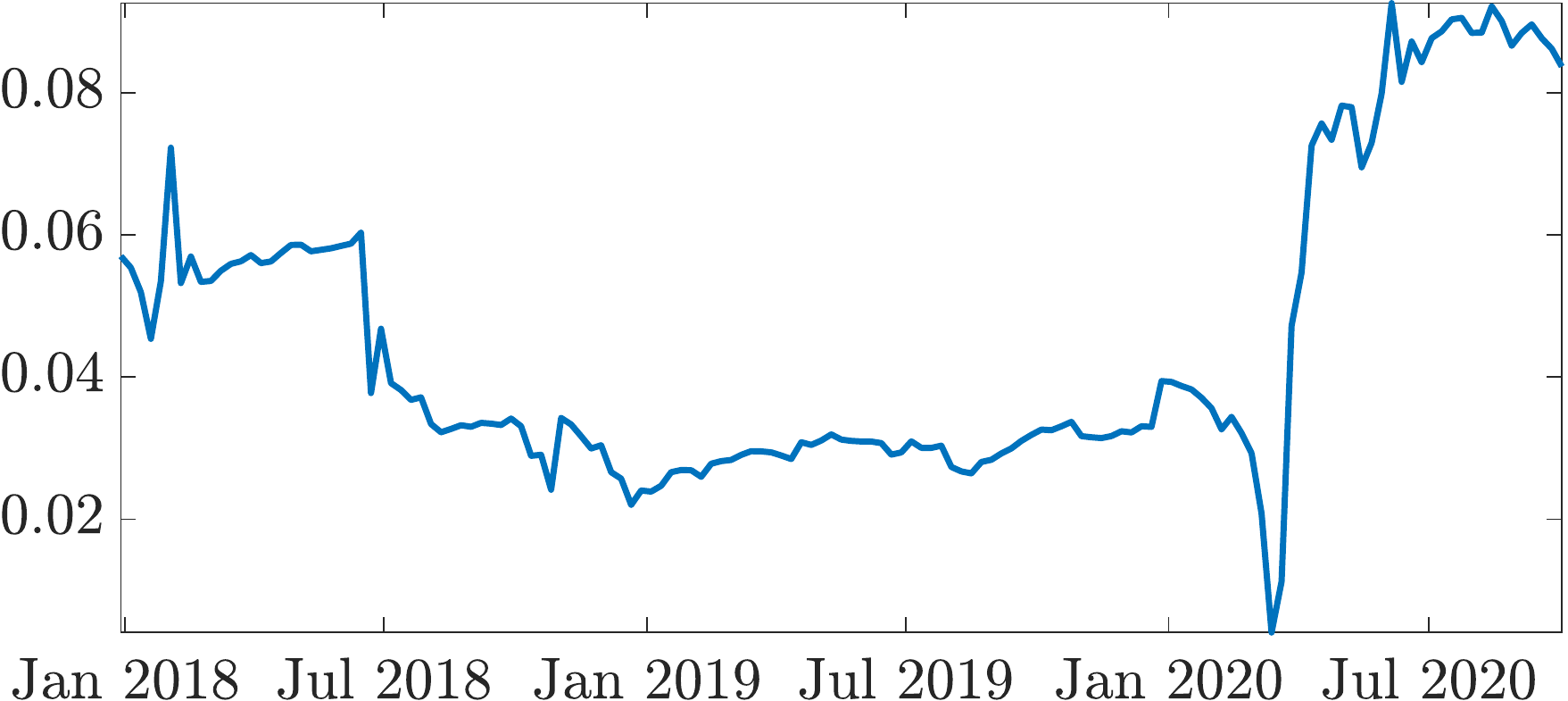}
\end{tabular}
\caption{The network densities (solid blue line) at 1\% level of statistical significance for the return linkages (top-left), the volatility linkages (bottom-left), the leverage linkages (top-right), and the risk premium (bottom-right) for the considered European firms over time.}
\label{fig:density}
\end{figure}

\begin{figure}[t!]
\centering
\setlength{\abovecaptionskip}{-2pt}
\begin{tabular}{cccc}
\begin{rotate}{90} \hspace*{28pt} {$\substack{\text{return,}\\\text{volatility}}$} \end{rotate} \hspace*{-10pt} & 
\includegraphics[trim= 0mm 4mm 0mm 0mm,clip,height=3.2cm,width=7.7cm]{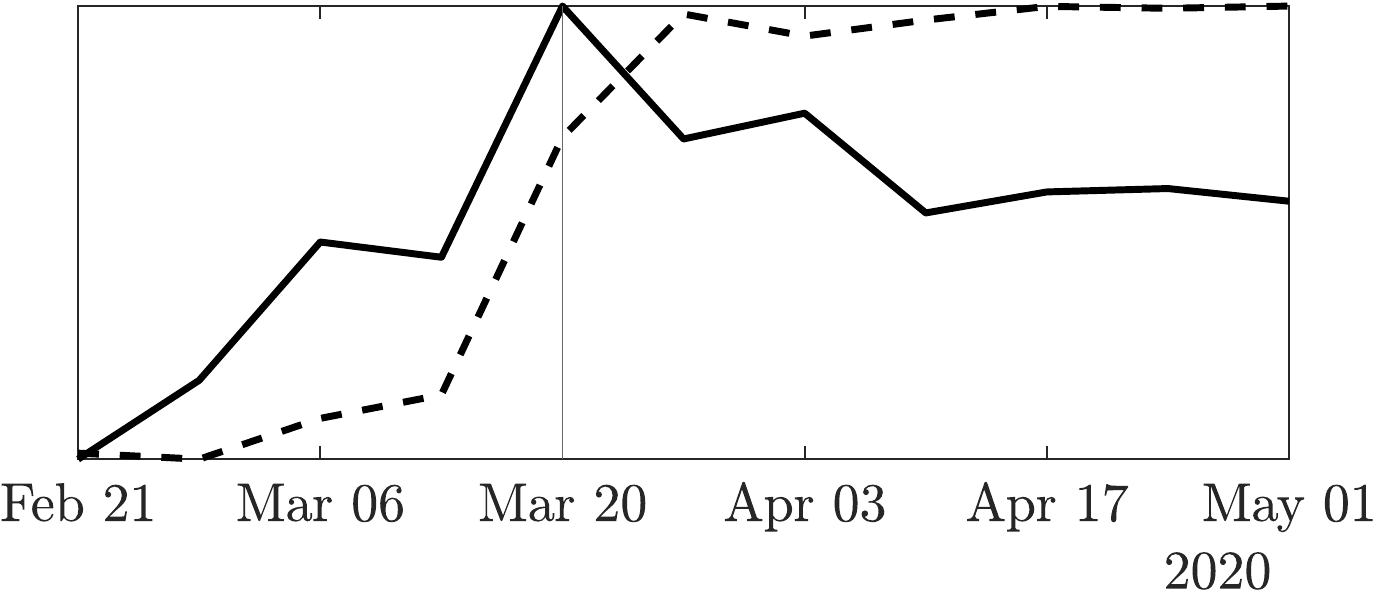} &
\begin{rotate}{90} \hspace*{25pt} {$\substack{\text{leverage,}\\\text{risk premium}}$} \end{rotate} \hspace*{-10pt} & 
\includegraphics[trim= 0mm 4mm 0mm 0mm,clip,height=3.2cm,width=7.7cm]{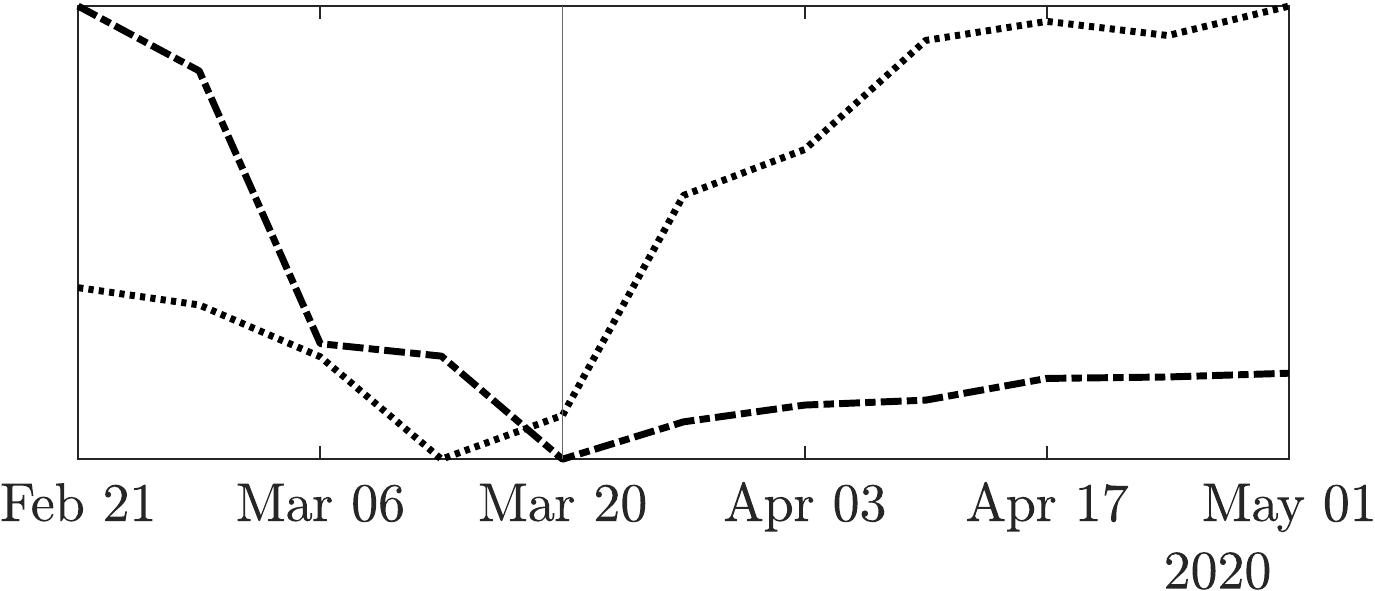}
\end{tabular}
\caption{The network densities at 1\% level of statistical significance from the 21st of February 2020 to the 1st of May 2021 (same scale). The return linkages (solid line), the volatility linkages (dashed line), the leverage linkages (dashed-dotted line) and the risk premium (dotted line) for the considered European firms over time.}
\label{fig:zoomed_density}
\end{figure}

Figure~\ref{fig:density} provides the density of the four sub-networks over time at 1\% level of statistical significance\footnote{The density of the network is defined as the total number of observed linkages over the total number of possible linkages. If the density is $0$, no connections exist, while if density is $1$, the network is fully connected.}.
Despite sharing some similarities, the dynamics of the intra- and inter-connectivity linkages can provide different signals on shock propagation in the financial markets.
This calls for a joint modeling of the four connectivity layers.
On average, the density is higher for the volatility linkages ($0.16$) followed by the risk premium linkages ($0.045$), the return linkages ($0.035$), and the leverage linkages ($0.006$). 
The density on the return linkages (top-left plot of Figure~\ref{fig:density}) shows two peaks on the 20th of March 2020 and the 5th of June 2020, respectively, with a reversion towards the mean after the first peak. Conversely, the density on the volatility linkages (bottom-left) shows a jump on the 27th of March 2020 with a new persistent higher level.
Interestingly, the leverage linkages (returns causes volatility) have an opposite behaviour with an abrupt drop of the density in March 2020. Finally, the risk premium linkages (volatility causes returns) exhibit a drop in the same month followed by a peak in April 2020.
The four sub-networks (i.e., layers) highlight that shocks on return and volatility have played different roles with heterogeneous timing during the outbreak of COVID-19 and its aftermath, thus affecting in different manner the intra- and inter-connectivity. 

The latter stylized fact can be better visualized by zooming in the period from the 21st of February 2020 to the 1st of May 2021.
Accordingly, Figure~\ref{fig:zoomed_density} reports the plot of the four densities on the same scale for this sub-period.
The density of the returns linkages (solid line) increases until the 20th of March 2020, whereas that of inter-linkages decreases (dashed-dotted and dotted lines). The volatility linkages (dashed line) slowly decrease until the 13th of March 2020, then suddenly jumps reaching the peak on the 27th of March 2020. 
At the same time, the density on the risk-premium linkages increases following a similar, but lagged, pattern as for the volatility linkages, whereas the leverage linkages start to slowly increase. This indicates that shocks on returns have been followed by shocks on volatility, where the latter have brought a persistent change in the level of connectivity.

\begin{figure}[t!h]
\centering
\hspace*{-20pt}
\setlength{\abovecaptionskip}{0pt}
\begin{tabular}{ccccc}
\begin{rotate}{90} \hspace*{90pt} {\large $\substack{Y_{11}\\\text{return}}$} \end{rotate} \hspace*{-8pt} &
\includegraphics[trim= 10mm 70mm 10mm 70mm,clip,height= 8.0cm, width= 8.0cm]{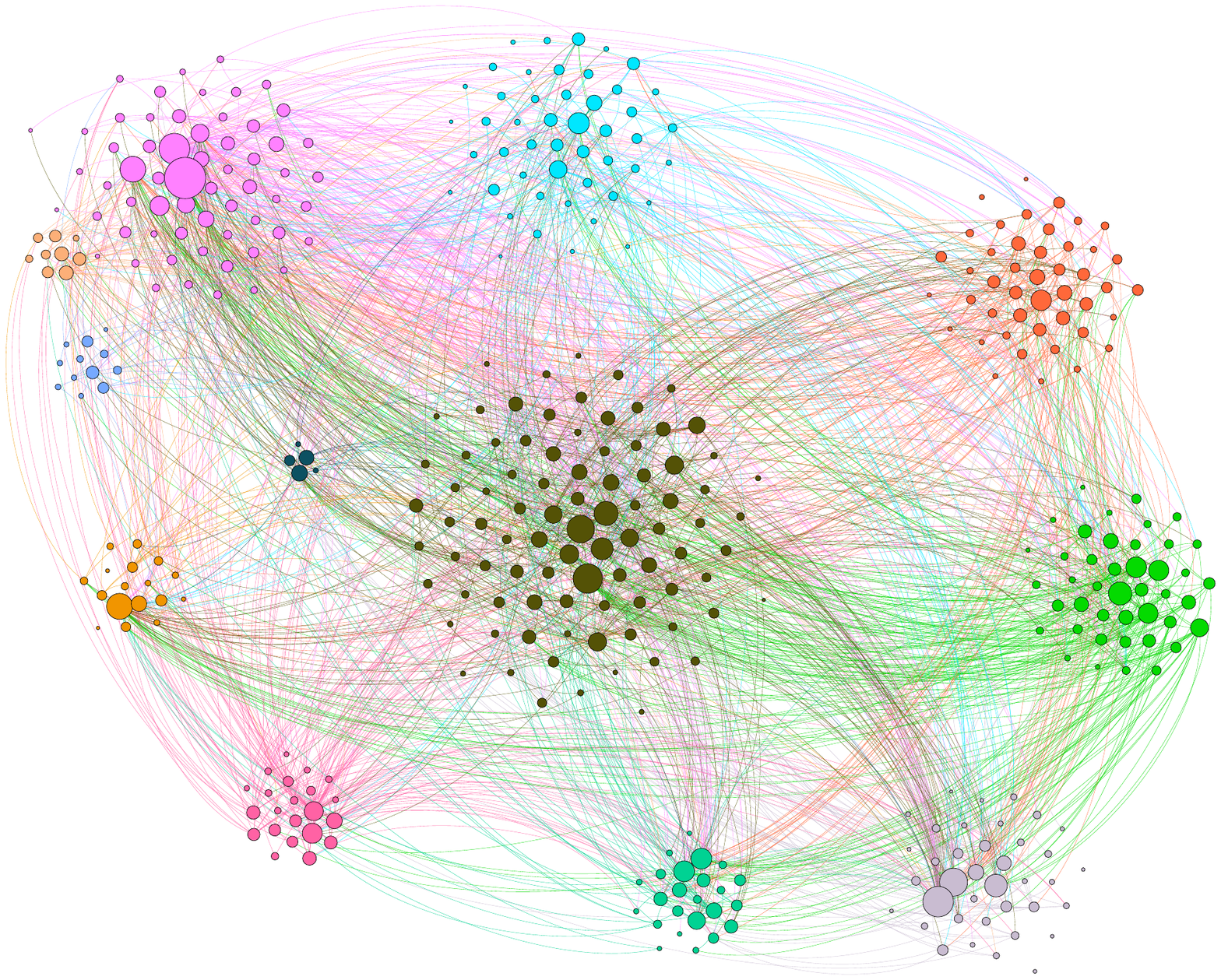} & \quad &
\begin{rotate}{90} \hspace*{90pt} {\large $\substack{Y_{11}\\\text{return}}$} \end{rotate} \hspace*{-8pt} &
\includegraphics[trim= 10mm 70mm 10mm 70mm,clip,height= 8.0cm, width= 8.0cm]{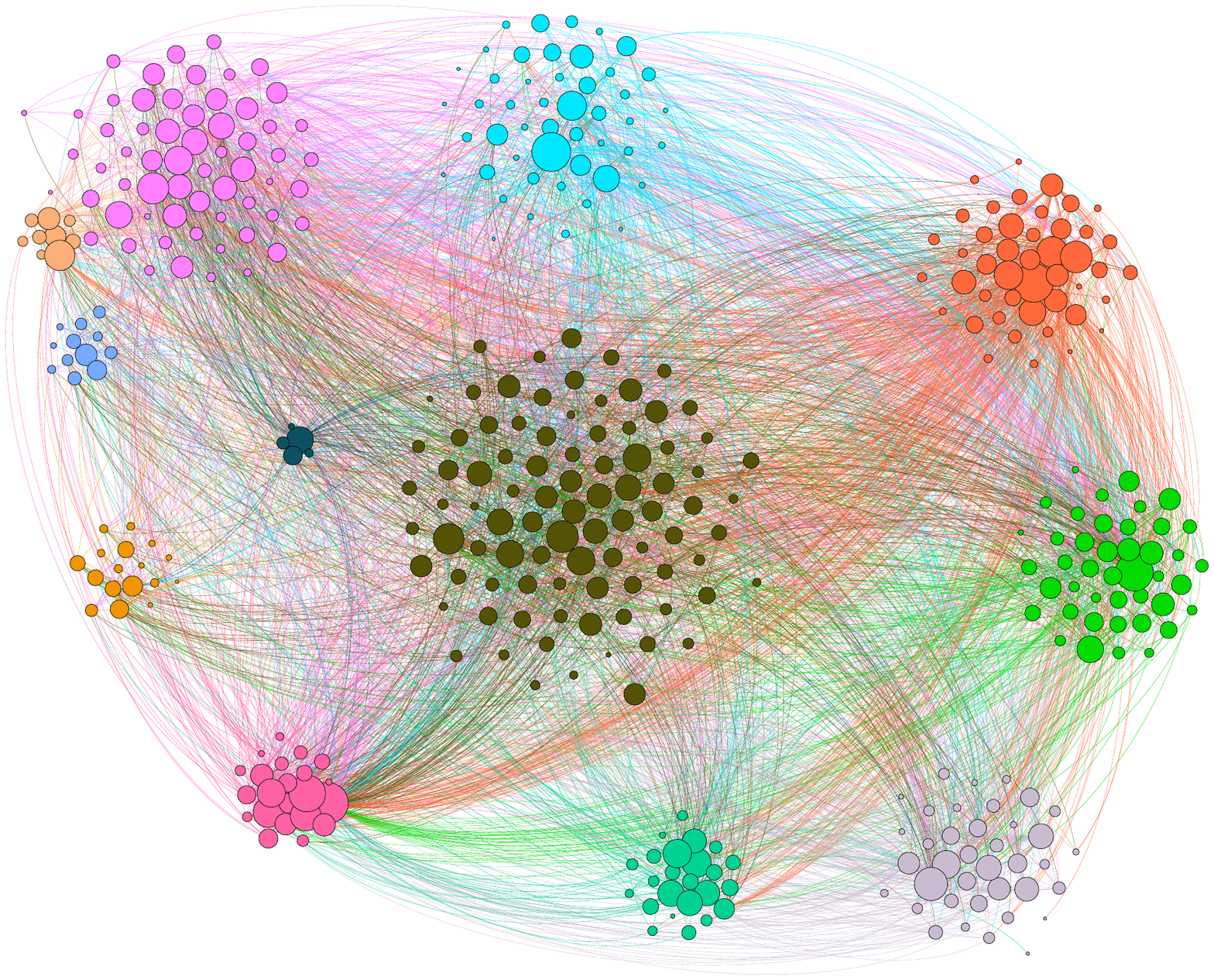} \\[10pt]
\begin{rotate}{90} \hspace*{90pt} {\large $\substack{Y_{22}\\\text{volatility}}$} \end{rotate} \hspace*{-8pt} & 
\includegraphics[trim= 10mm 65mm 10mm 65mm,clip,height= 8.0cm, width= 8.0cm]{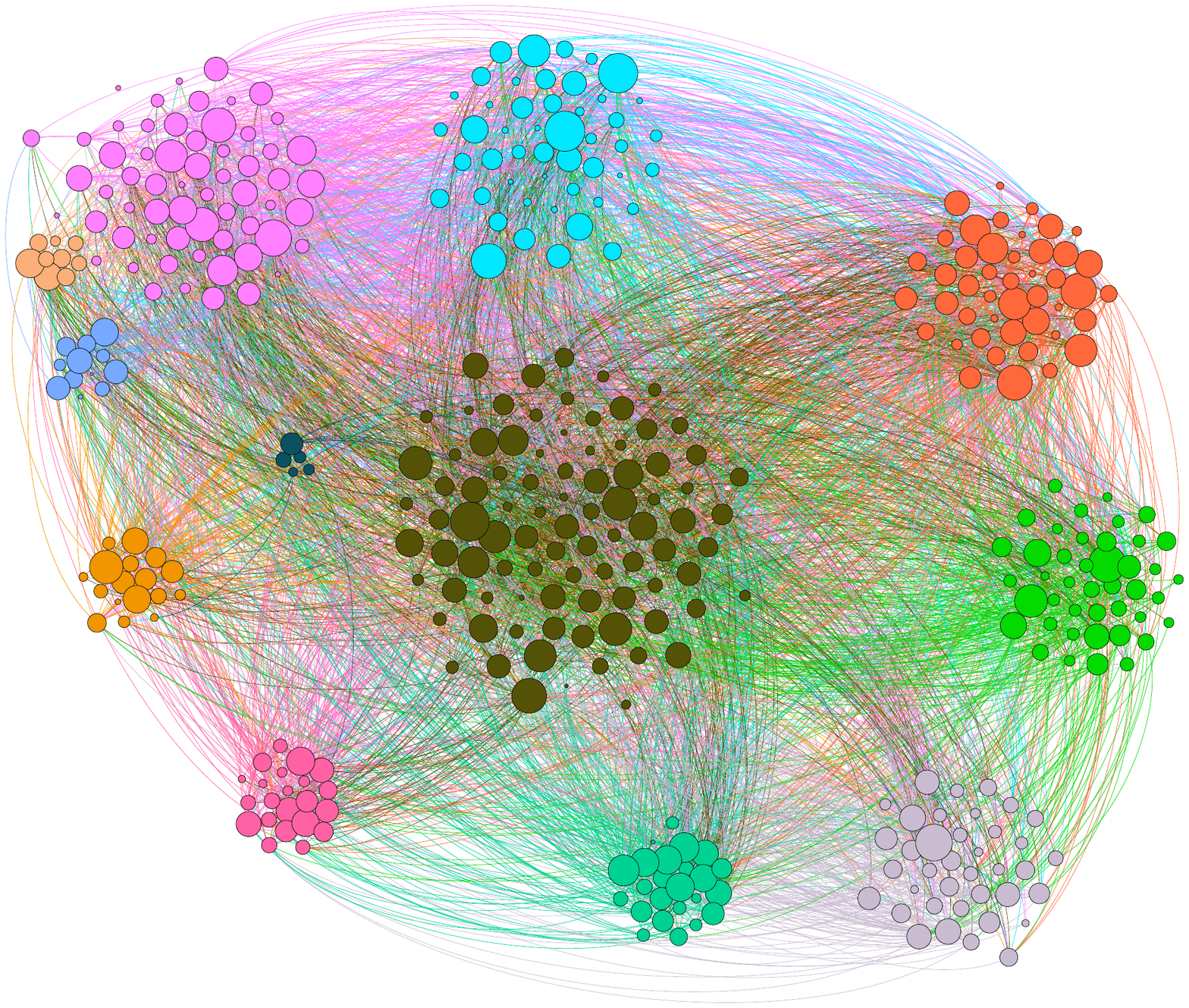} & \quad &
\begin{rotate}{90} \hspace*{90pt} {\large $\substack{Y_{22}\\\text{volatility}}$} \end{rotate} \hspace*{-8pt} &
\includegraphics[trim= 10mm 65mm 10mm 65mm,clip,height= 8.0cm, width= 8.0cm]{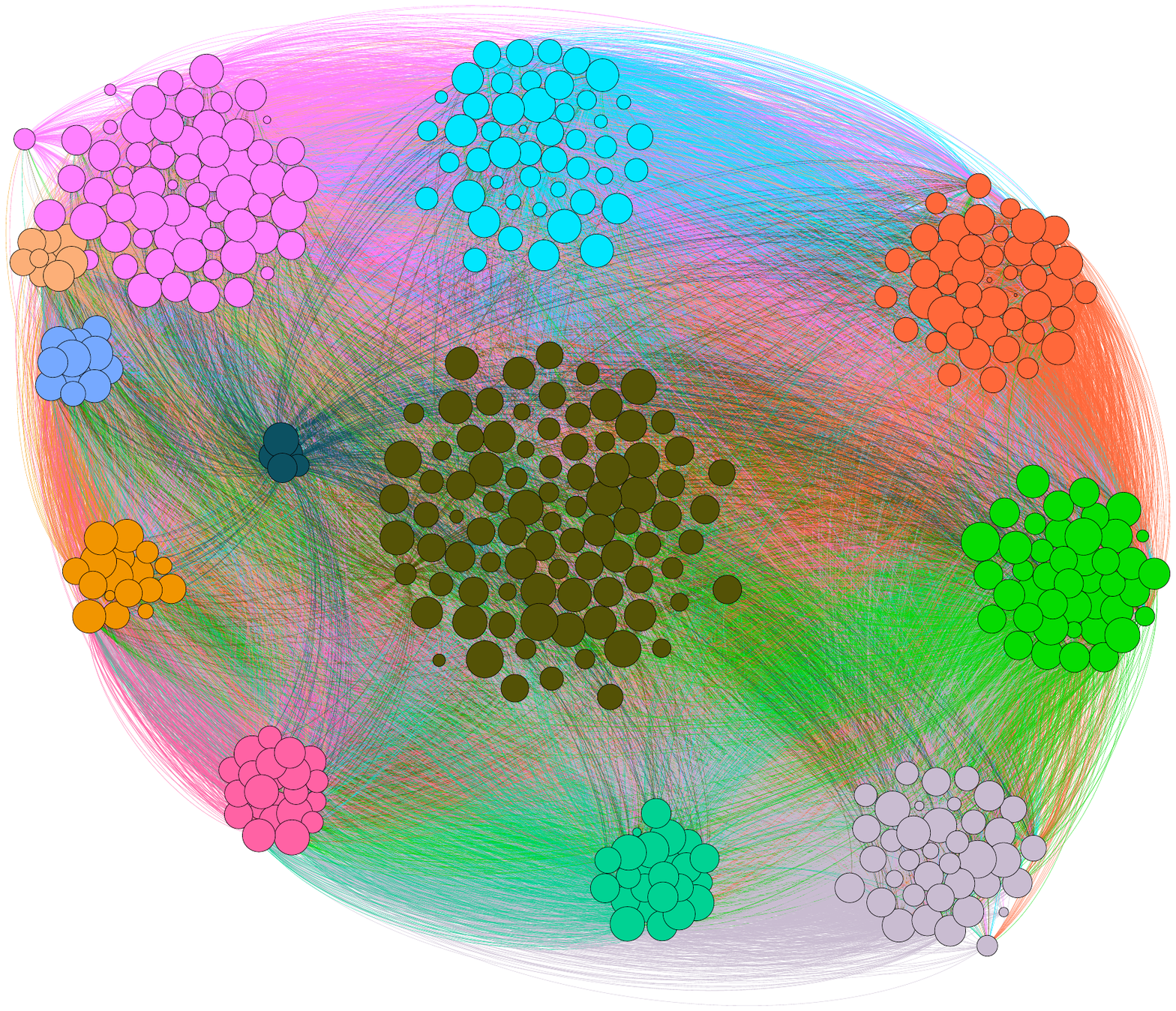}
\end{tabular}
\caption{The intra-layer directed networks: returns linkages (top) and volatility linkages (bottom) on the 17th of January 2020 (left) and the 27th of March 2020 (right). Edges are clockwise directed.
Node size: proportional to the total degree averaged over time within each regime. Edge color according to the Industry of the source node: Financials (light green), Communication Services (violet), Consumer Discretionary (pink), Consumer Staples (orange), Health Care (light blue), Energy (light orange), Industrials (olive green), Information Technology (dark orange), Materials (green), Real Estate (dark pink), Utilities (blue), and not classified in a specific GICS sector (dark green).
For exposition purposes, we drop edges with p-value larger than 0.1\%.}
\label{fig:intra_graphs}
\end{figure}

\begin{figure}[t!h]
\centering
\hspace*{-20pt}
\setlength{\abovecaptionskip}{0pt}
\begin{tabular}{ccccc}
\begin{rotate}{90} \hspace*{90pt} {\large $\substack{Y_{12}\\\text{risk premium}}$} \end{rotate} \hspace*{-8pt} &
\includegraphics[trim= 5mm 65mm 10mm 65mm,clip,height= 8.0cm, width= 8.0cm]{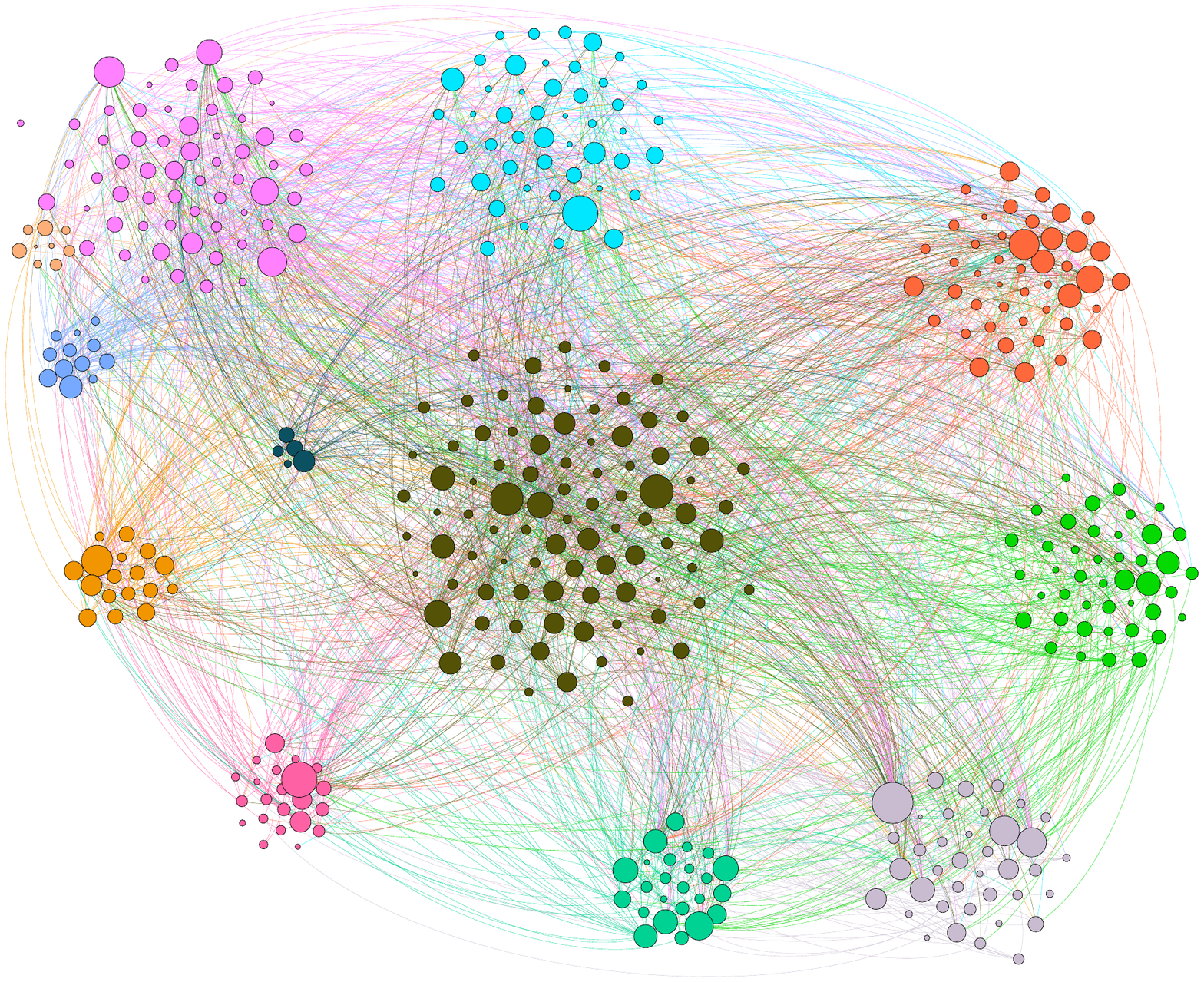} & \quad &
\begin{rotate}{90} \hspace*{90pt} {\large $\substack{Y_{12}\\\text{risk premium}}$} \end{rotate} \hspace*{-8pt} &
\includegraphics[trim= 5mm 65mm 10mm 65mm,clip,height= 8.0cm, width= 8.0cm]{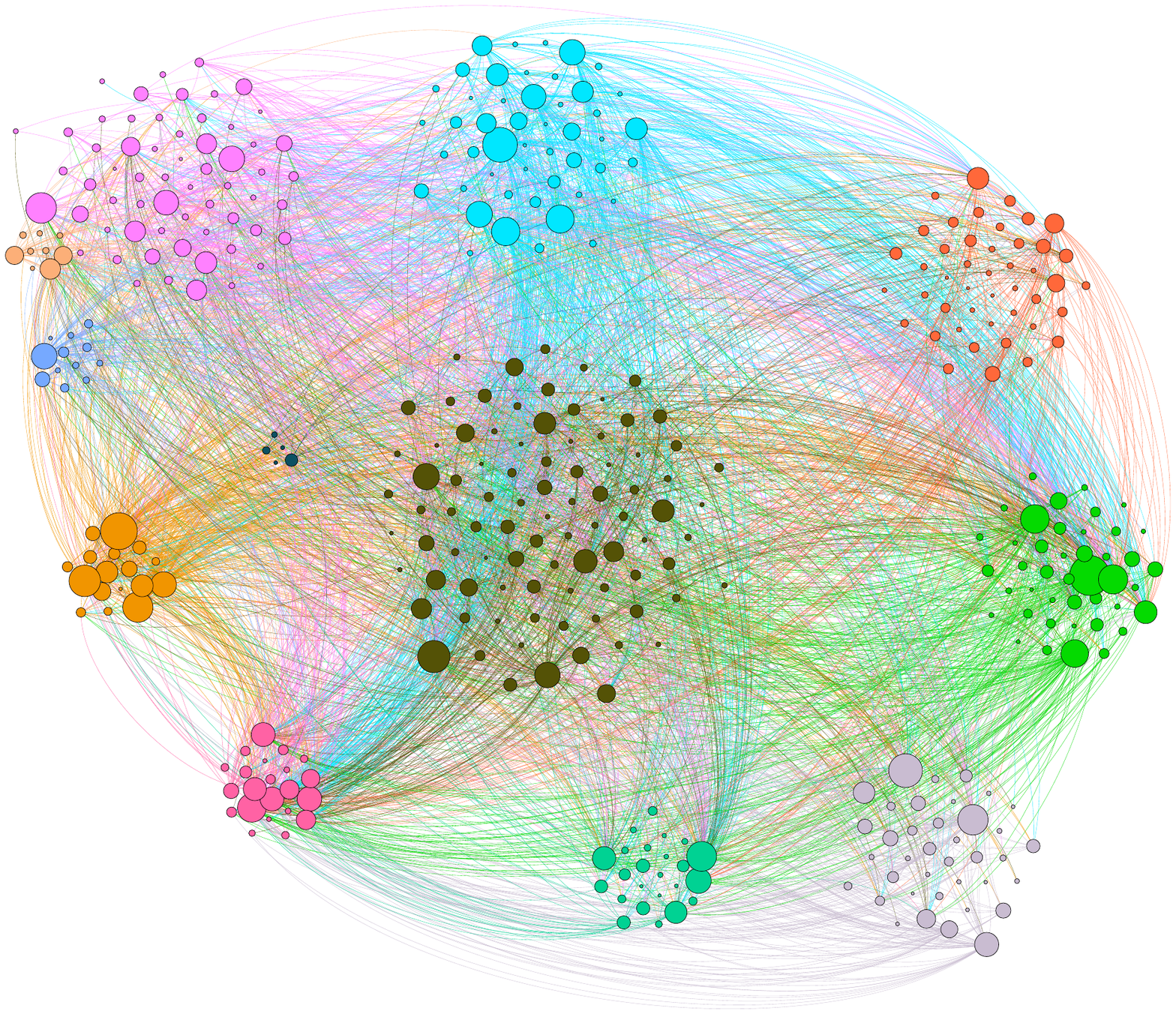} \\[10pt]
\begin{rotate}{90} \hspace*{90pt} {\large $\substack{Y_{21}\\\text{leverage}}$} \end{rotate} \hspace*{-8pt} & 
\includegraphics[trim= 5mm 65mm 10mm 65mm,clip,height= 8.0cm, width= 8.0cm]{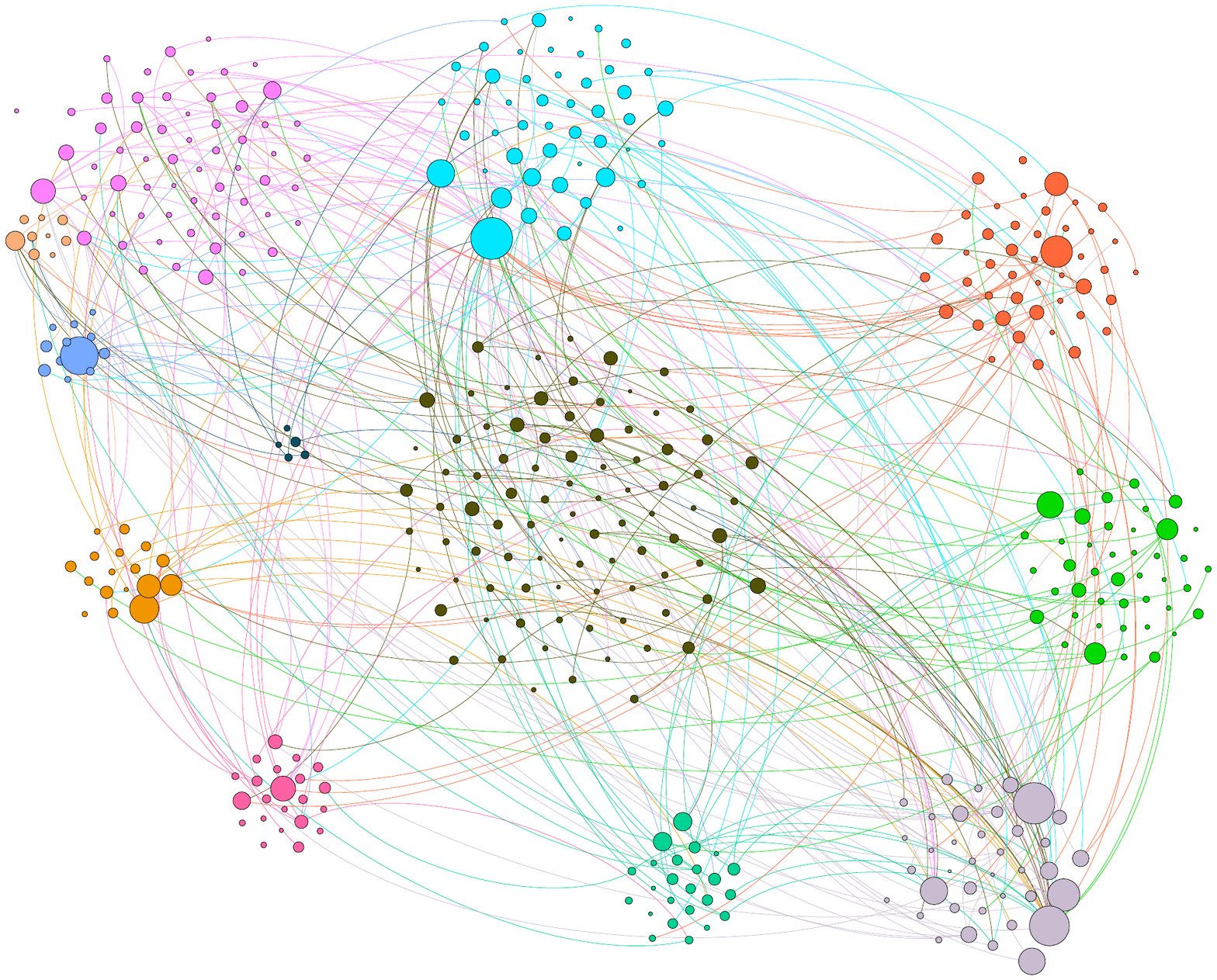} & \quad &
\begin{rotate}{90} \hspace*{90pt} {\large $\substack{Y_{21}\\\text{leverage}}$} \end{rotate} \hspace*{-8pt} &
\includegraphics[trim= 4mm 65mm 10mm 65mm,clip,height= 8.0cm, width= 8.0cm]{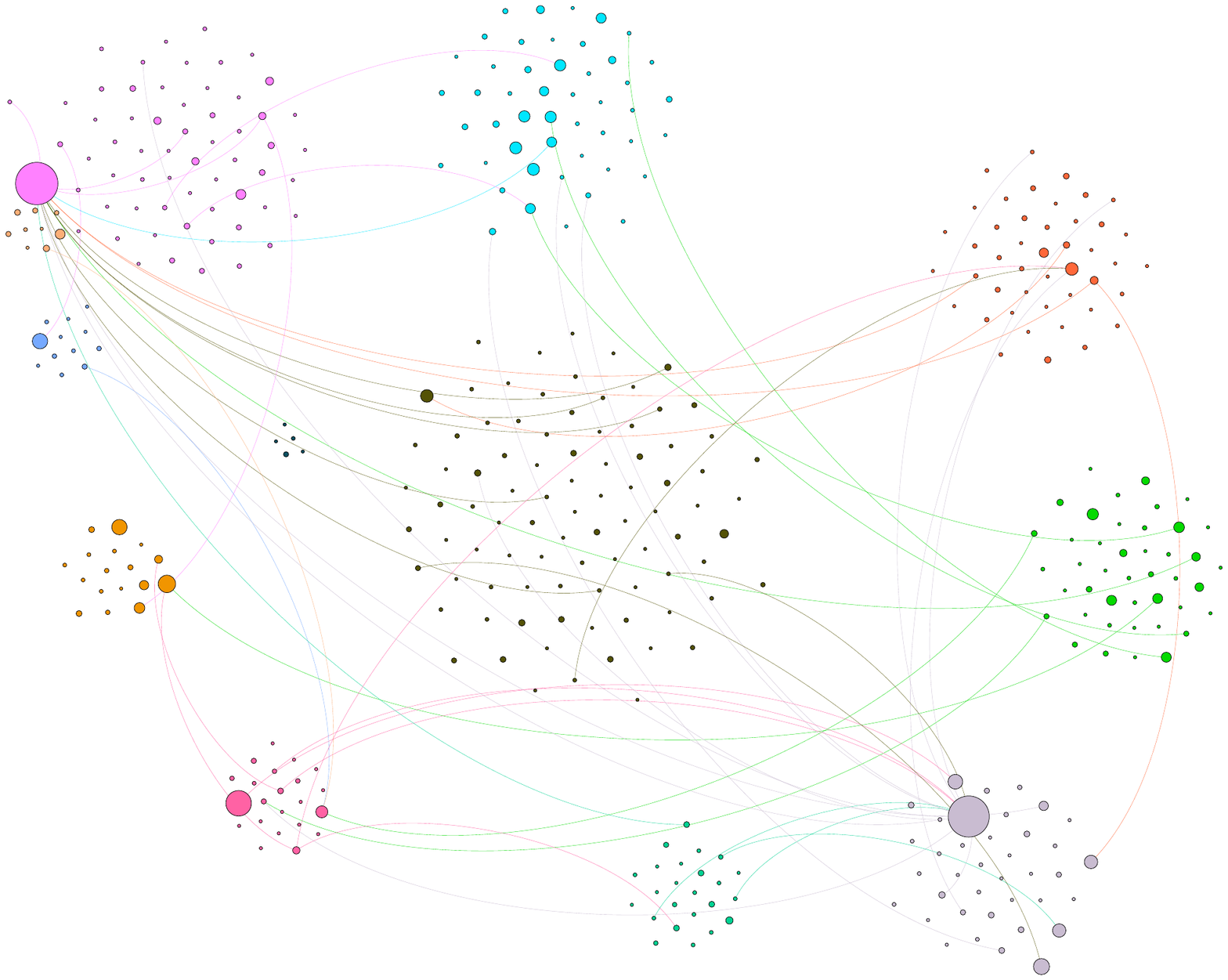}
\end{tabular}
\caption{The inter-layer directed networks: risk premium linkages (top) and leverage linkages (bottom) on the 17th of January 2020 (left) and the 27th of March 2020 (right). Edges are clockwise directed.
Node size: proportional to the total degree averaged over time within each regime. Edge color according to the Industry of the source node: Financials (light green), Communication Services (violet), Consumer Discretionary (pink), Consumer Staples (orange), Health Care (light blue), Energy (light orange), Industrials (olive green), Information Technology (dark orange), Materials (green), Real Estate (dark pink), Utilities (blue), and not classified in a specific GICS sector (dark green).
For exposition purposes, we drop edges with p-value larger than 0.1\%.}
\label{fig:inter_graphs}
\end{figure}

For the sake of completeness, we include in Figures~\ref{fig:intra_graphs}-\ref{fig:inter_graphs} the intra- and inter-layer directed networks on the 17th of January 2020 (one week before the first European COVID-19 case) and the 27th of March 2020.
As shown in the Figures, the connectivity increases in the intra-layer networks for the return linkages and especially, on the volatility linkages. Regarding the inter-layer directed networks, the connectivity of the risk premium linkages increase while it decreases significantly in the leverage linkages. 
It is therefore interesting to apply the proposed network model to measure the impact of the COVID-19 and the other risk factors, on the intra- and inter-linkages of the considered European firms.

%Even if the shocks occurred on returns on February 2020 have affected the return linkages and the inter-connectivity in the risk premium linkages (the initial drop on the density) and , 

\subsection{Results}
In this section, we apply the model and inference proposed in Sections~\ref{sec:netmodel}-\ref{sec:inference} to estimate the impact of the risk factors on the multilayer European financial network.
Since our analysis is focused on investigating the role of COVID-19, we describe the effects of the other risk factors (returns on the Euro STOXX 50 index, the implied volatility on the Euro STOXX 50 index, and the Bloomberg Barclays EuroAgg Corporate Average OAS) on financial linkages in the Appendix.
Figure~\ref{fig:posterior_mean} reports the impact of COVID-19 on each linkage $(i,j)$, from firm $j$ to firm $i$, across all layers. In particular, we report only non-null effects, that is coefficients such that zero is not included in the corresponding 95\% posterior high probability density interval (HPDI). The blue color indicates a positive impact on the probability of an edge from firm $j$ to firm $i$, while the red color indicates a negative impact.
Since our response variable is an increasing transformation of the p-value of Granger causality test performed on a pair $(i,j)$ of time-series, a negative (positive) COVID-19 coefficient implies an increase (decrease) in the p-value, hence in the probability to observe a linkage from $j$ to $i$.
As shown in the Figure, the COVID-19 has a weaker effect on return linkages in terms of magnitude and number of impacted linkages. 
Conversely, it increases the probability of volatility and risk premium linkages, whereas, in most of the cases, it reduces the probability of leverage linkages.
Overall, among the selected risk factors, the COVID-19 has the greatest impact on the European financial networks.
The other risk factors, such as the market returns, the implied volatility, and the corporate credit risk, have some effect on the volatility and leverage linkages and very weak effect on return and risk premium linkages (see Figure~\ref{fig:posterior_mean_apdx} in the Appendix).

\begin{figure}[t!h]
\centering
\begin{tabular}{ccccc}
\begin{rotate}{90} \hspace*{63pt} {\small return} \end{rotate} &
\includegraphics[trim= 0mm 0mm 0mm 0mm,clip,height= 6.2cm, width= 6.7cm]{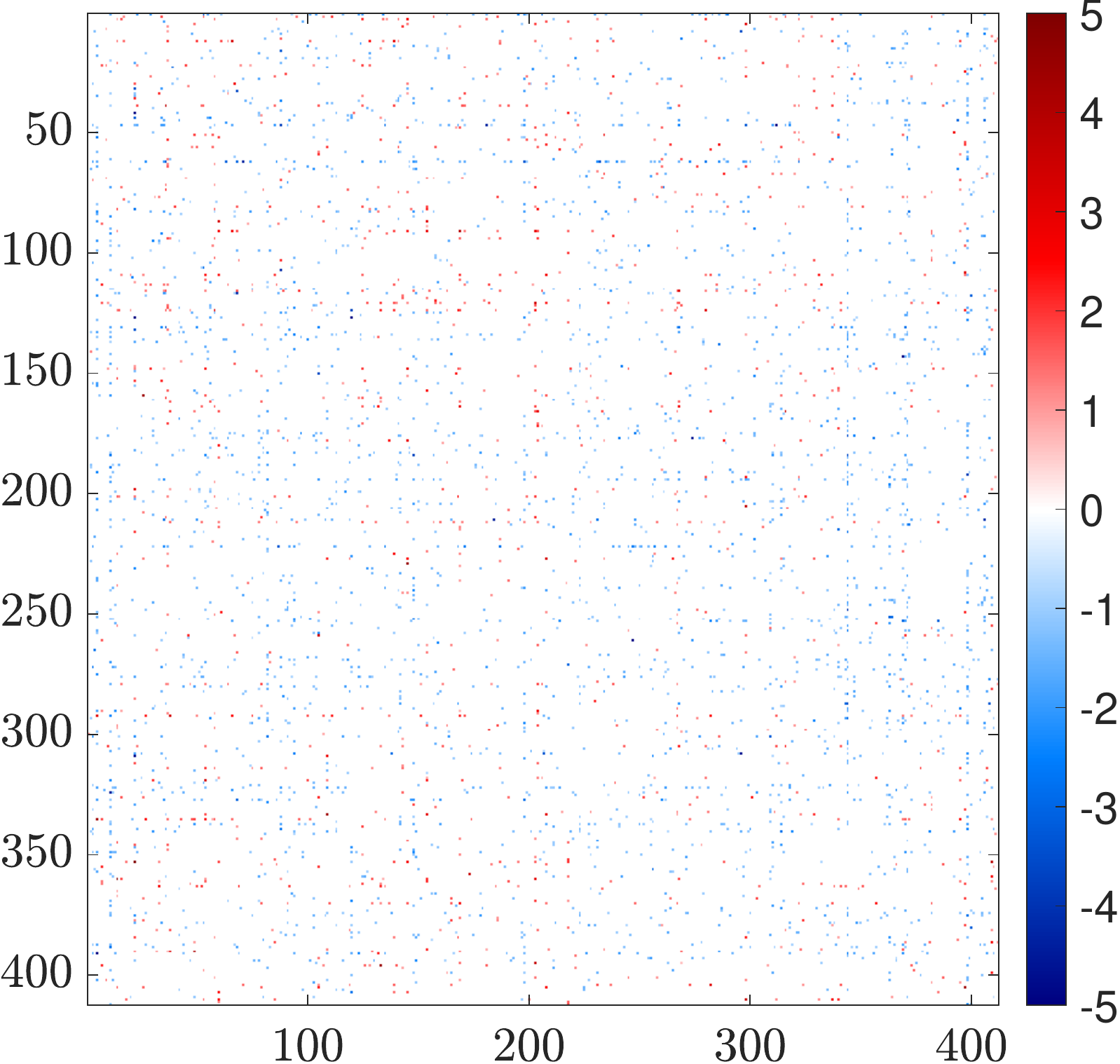} & \qquad &
\begin{rotate}{90} \hspace*{50pt} {\small risk premium} \end{rotate} &
\includegraphics[trim= 0mm 0mm 0mm 0mm,clip,height= 6.2cm, width= 6.7cm]{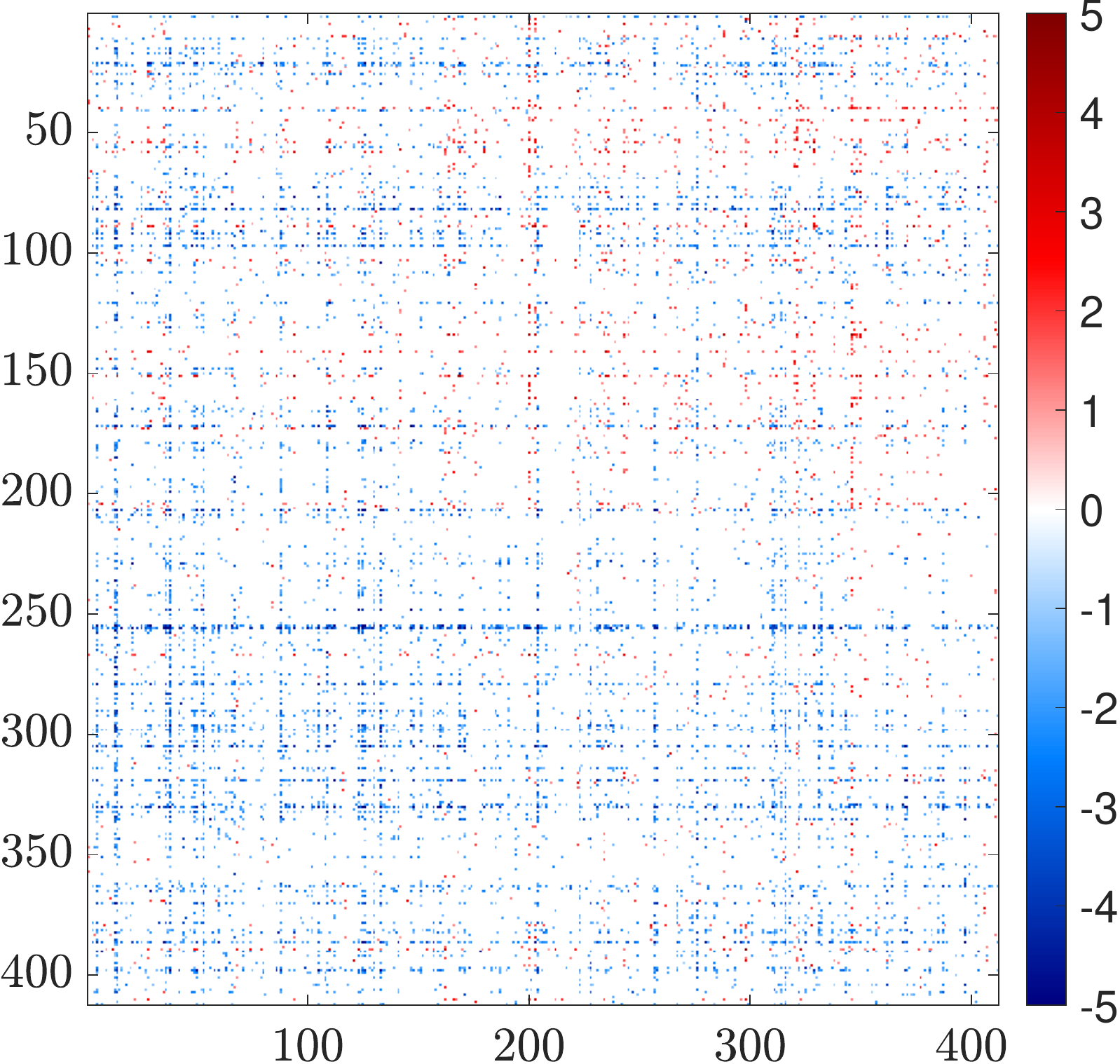} \\
\begin{rotate}{90} \hspace*{62pt} {\small volatility} \end{rotate} &
\includegraphics[trim= 0mm 0mm 0mm 0mm,clip,height= 6.2cm, width= 6.7cm]{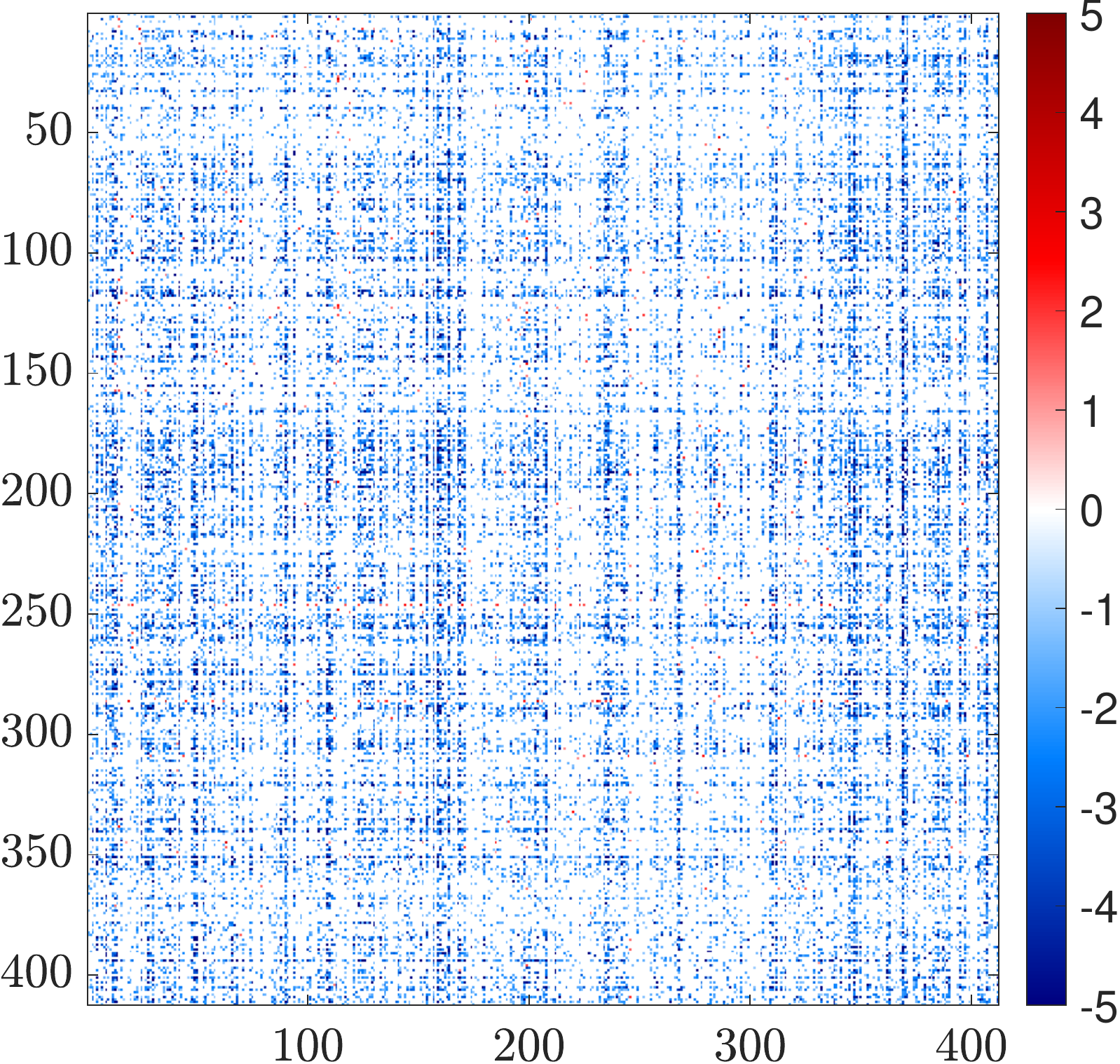} & \qquad &
\begin{rotate}{90} \hspace*{62pt} {\small leverage} \end{rotate} &
\includegraphics[trim= 0mm 0mm 0mm 0mm,clip,height= 6.2cm, width= 6.7cm]{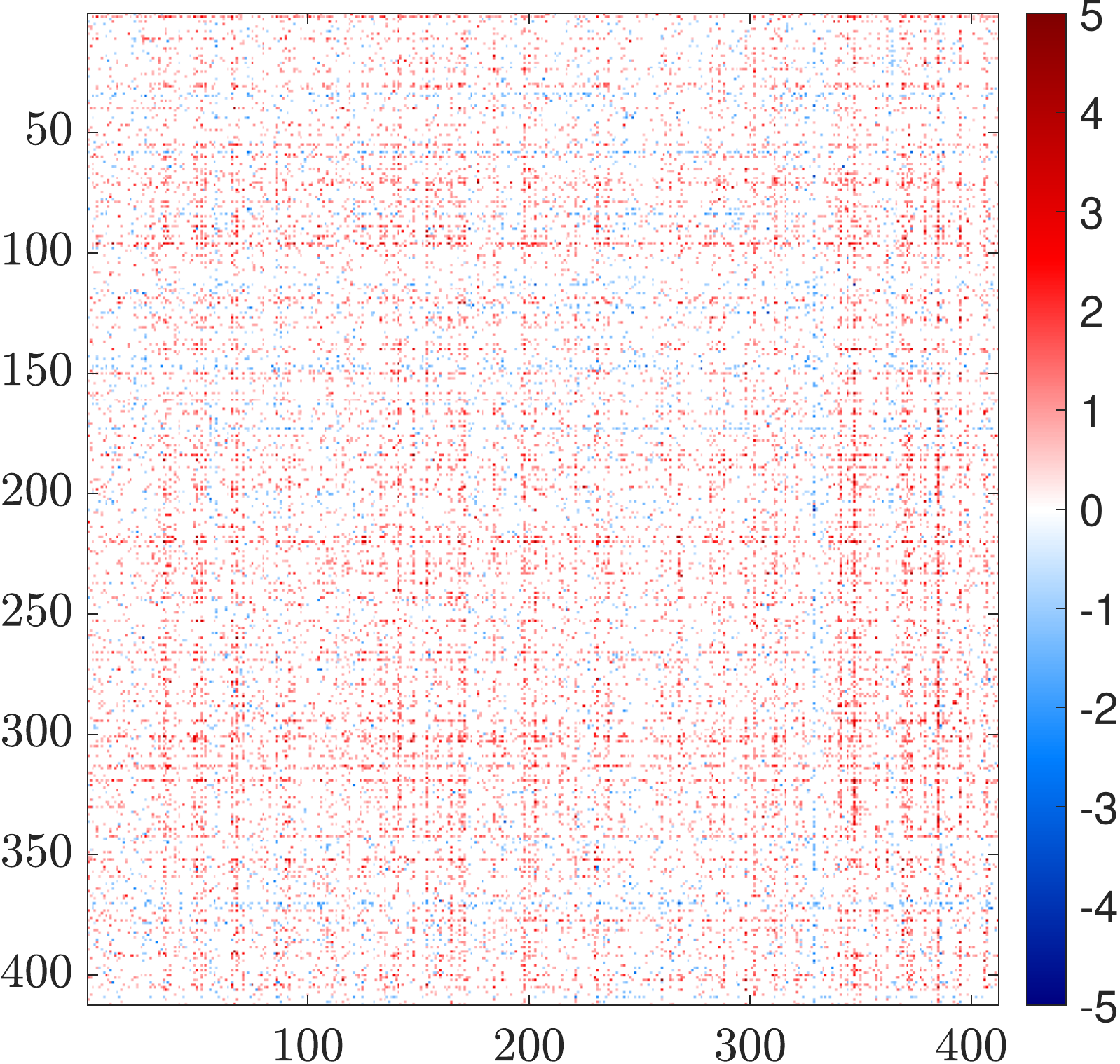}
\end{tabular}
\caption{Impact of COVID-19 on financial linkages for intra-layer (left column) and inter-layer (right column) networks.
In each plot, the coefficient in position $(i,j)$ refers to the impact of COVID-19 on the edge from firm $j$ to firm $i$.
Only non-null coefficients are reported: blue indicates positive impact on edge existence, red indicates negative impact on edge existence.
A coefficient is considered null if its posterior HPDI contains zero.
\label{fig:posterior_mean}
}
\end{figure}

\begin{figure}[t!h]
\centering
\setlength{\abovecaptionskip}{-2pt}
\begin{tabular}{ccccc}
\begin{rotate}{90} \hspace*{83pt} {\small return} \end{rotate} &
\includegraphics[trim= 0mm 0mm 0mm 0mm,clip,height= 6.2cm, width= 7.5cm]{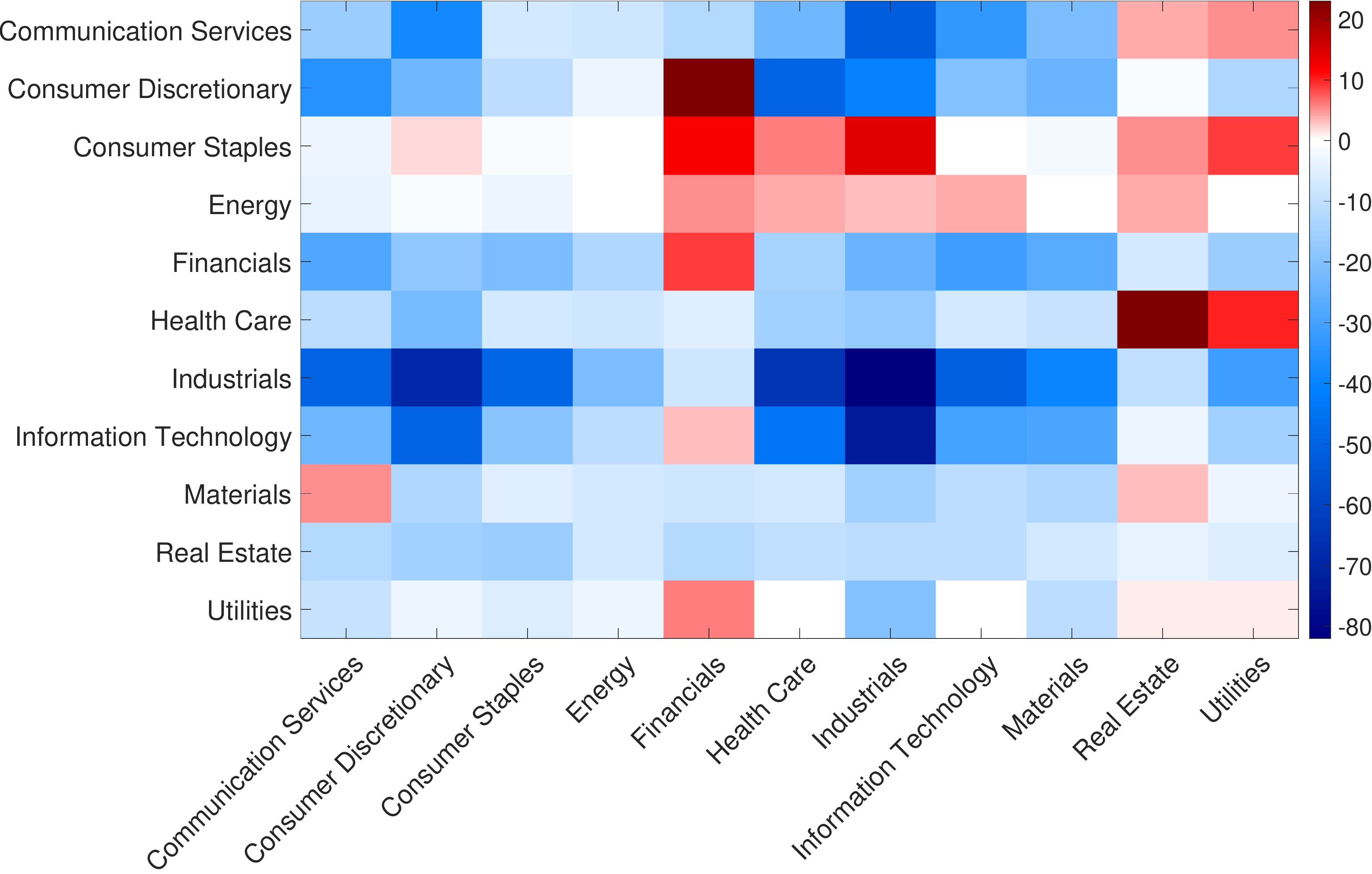} & \qquad &
\begin{rotate}{90} \hspace*{75pt} {\small risk premium} \end{rotate} &
\includegraphics[trim= 0mm 0mm 0mm 0mm,clip,height= 6.2cm, width= 7.5cm]{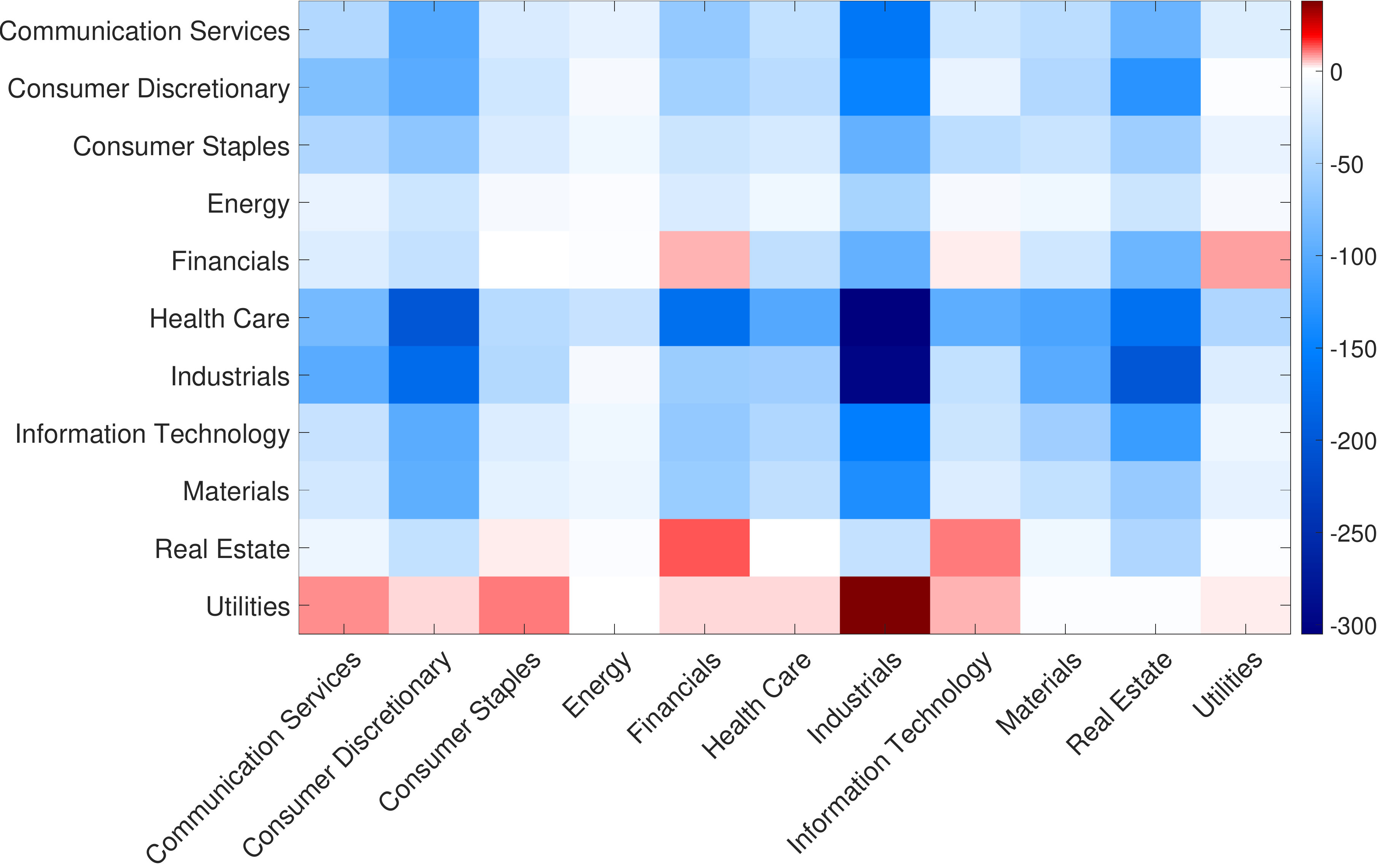} \\
\begin{rotate}{90} \hspace*{80pt} {\small volatility} \end{rotate} &
\includegraphics[trim= 0mm 0mm 0mm 0mm,clip,height= 6.2cm, width= 7.5cm]{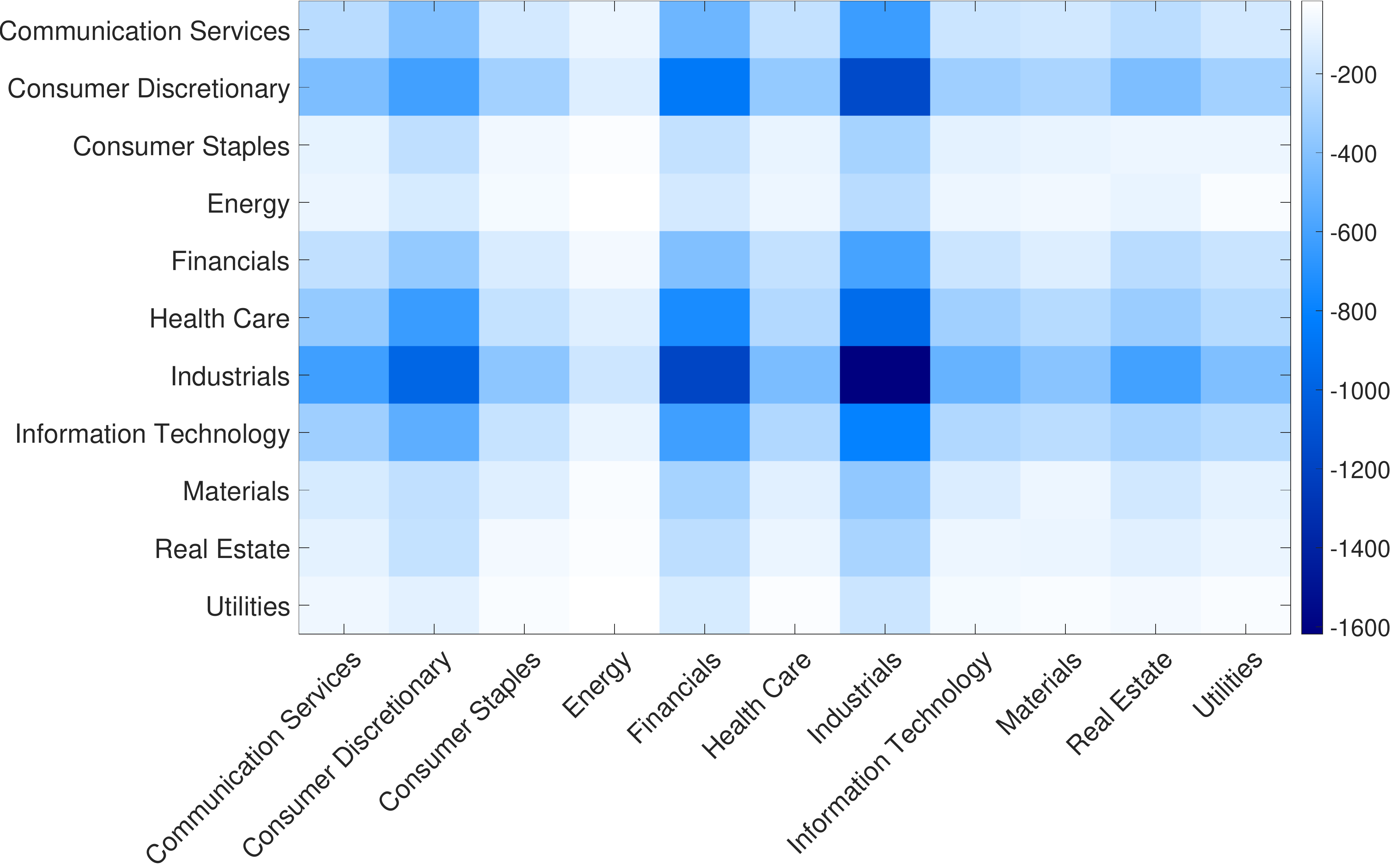} & \qquad &
\begin{rotate}{90} \hspace*{80pt} {\small leverage} \end{rotate} &
\includegraphics[trim= 0mm 0mm 0mm 0mm,clip,height= 6.2cm, width= 7.5cm]{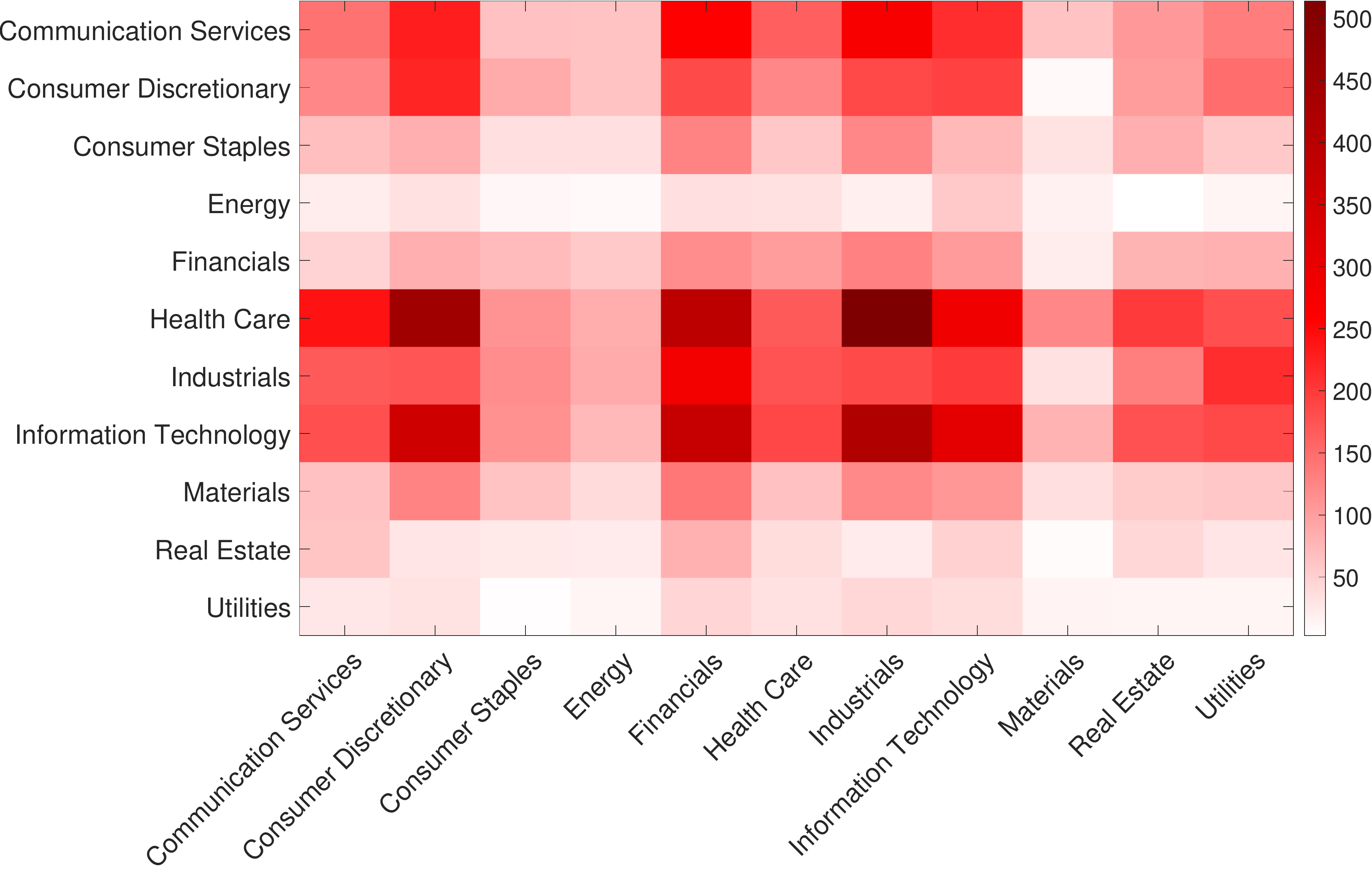}
\end{tabular}
\caption{Impact of COVID-19 on sector linkages for intra-layer (left column) and inter-layer (right column) networks.
In each plot, the block in position $(i,j)$ refers to the number of linkages (net effect) from sector $j$ to sector $i$ impacted by COVID-19.
Blue (red) indicates an increase (decrease) of the number of linkages.
\label{fig:posterior_mean_sector_net}
}
\end{figure}

Figure~\ref{fig:posterior_mean_sector_net} shows the net effect of COVID-19 on the linkages between sectors. 
In each panel, the block in position $(i,j)$ refers to the number of linkages (net effect) from sector $j$ to sector $i$ impacted by COVID-19. The main empirical findings are the following:
\begin{itemize}
\item there is evidence of a heterogeneous net impact on return and risk premium linkages, of a substantial increase in volatility linkages, and of a decrease in the leverage linkages;
\item the Industrial sector plays a pivotal role in the connectivity structure of the multilayer network. In return linkages, risk premium, and volatility linkages, there is an increase in the connectivity from and to the other sectors (except for Consumer Staples, Energy, and Utilities). The Industrial sector exhibits the largest increase in the connectivity level within sector (except in the leverage layer);
\item in the return linkages, the Financial sectors shows the largest decrease in the connectivity to the other sectors (red squares in the column). Conversely, in the risk premium and volatility linkages, there is an increase in the connectivity to the other sectors (except for Utilities);
\item utilities are the unique sector that exhibits a decrease in the connectivity from other sectors in the risk premium linkages, especially from the Industrial sector. A similar behavior can be found also in the return linkages for the Financial, Real Estate, and Utilities sectors;
\item energy is the only sector that does not affect and is not unaffected by COVID-19 in all the multilayer network (except for the incoming connectivity in the return linkages);
\item the Health Care and the information technology are the sectors that show the largest decrease in the connectivity from other sectors in the leverage linkages.  
\end{itemize}
In conclusion, the results discussed above show that the COVID-19 has affected the connectivity of the European financial network between the sectors.
We further investigate the relationship between the impact of COVID-19 on financial linkages and the centrality of each firm in the network.

\begin{figure}[t!]
\centering
\setlength{\abovecaptionskip}{-1pt}
\begin{tabular}{cccc}
 {\small return} & {\small risk premium} & {\small leverage} & {\small volatility} \\
%\begin{rotate}{90} \hspace*{10pt} {\small TOT degree} \end{rotate} &
\includegraphics[trim= 0mm 0mm 0mm 0mm,clip,height= 3.2cm, width= 3.7cm]{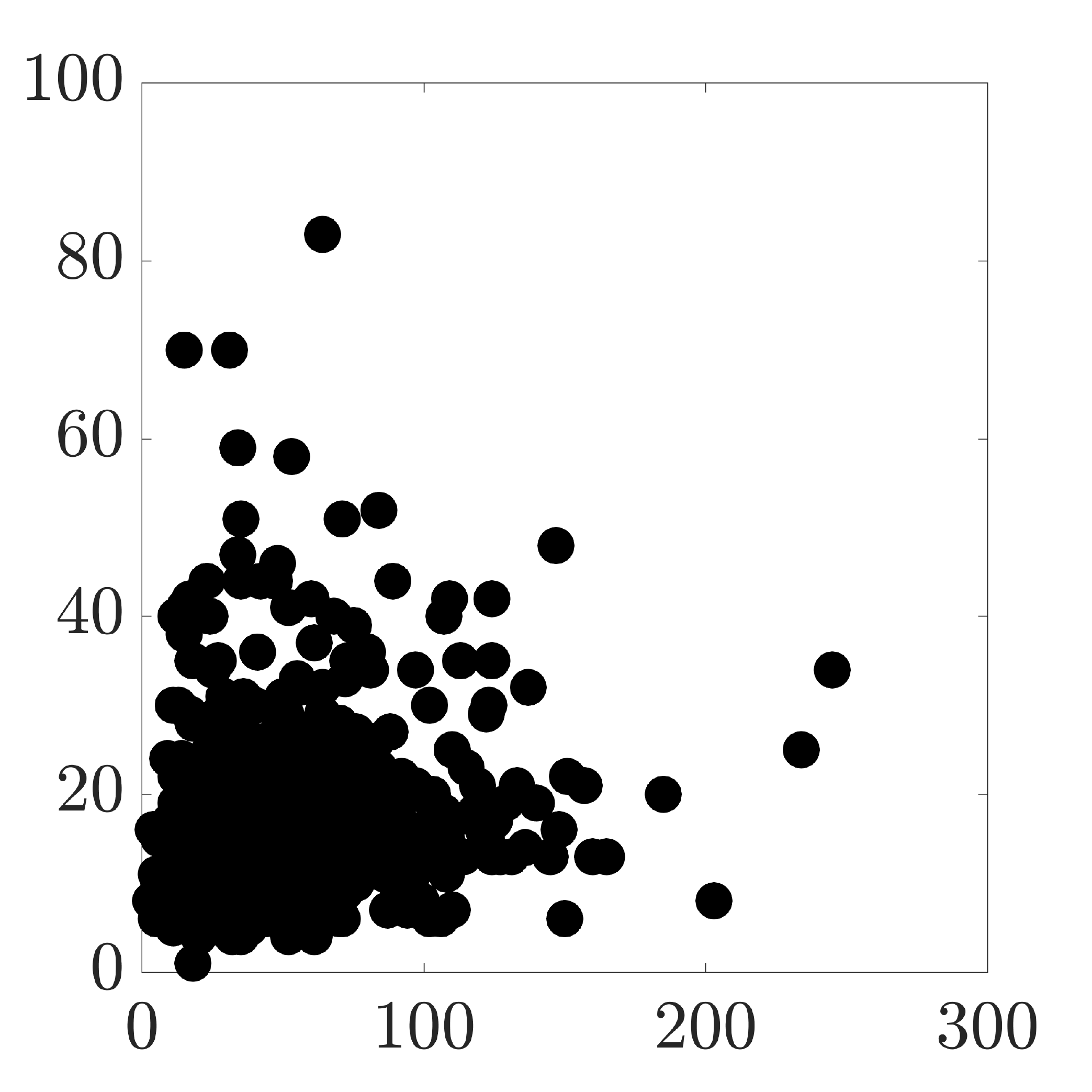} & 
\includegraphics[trim= 0mm 0mm 0mm 0mm,clip,height= 3.2cm, width= 3.7cm]{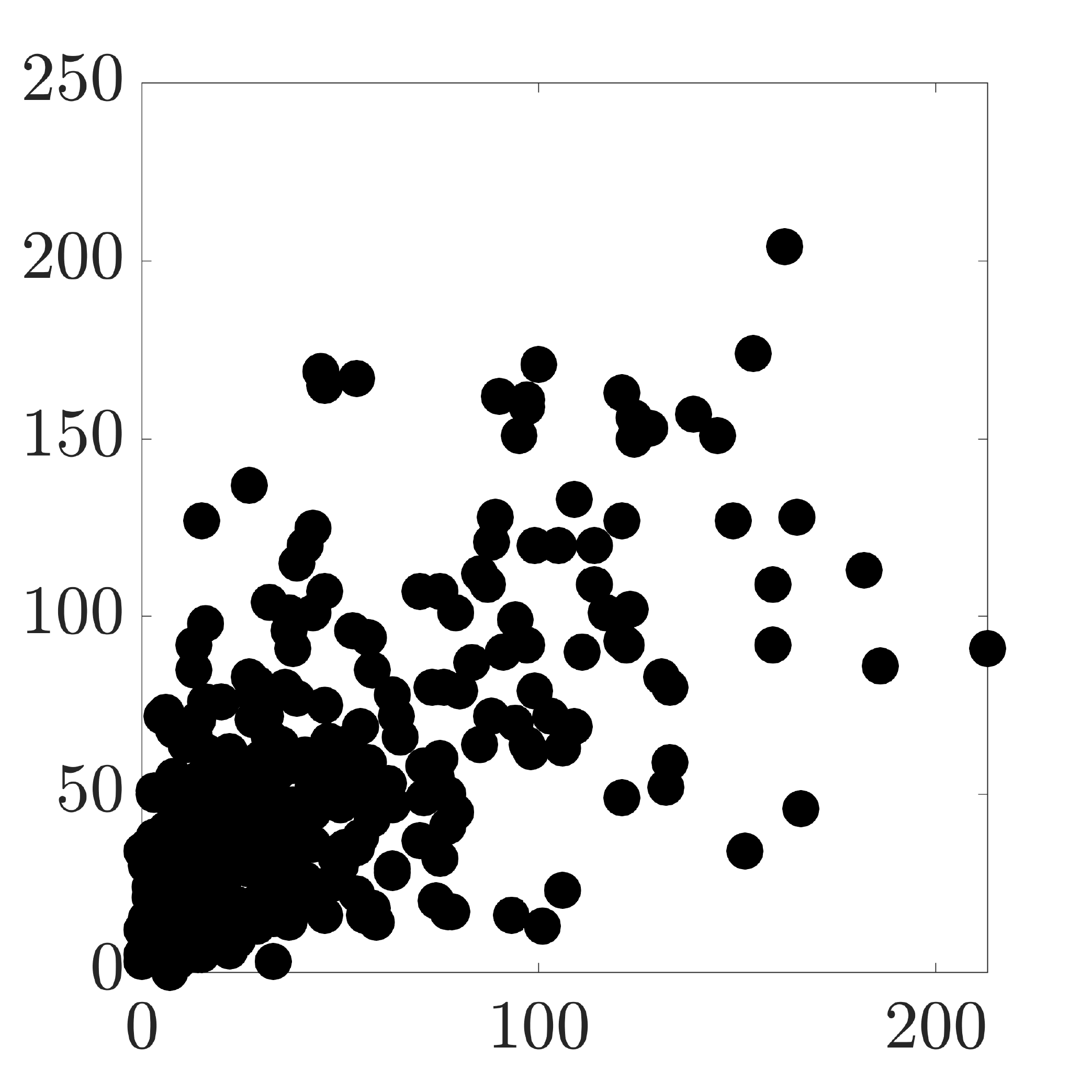} & 
\includegraphics[trim= 0mm 0mm 0mm 0mm,clip,height= 3.2cm, width= 3.7cm]{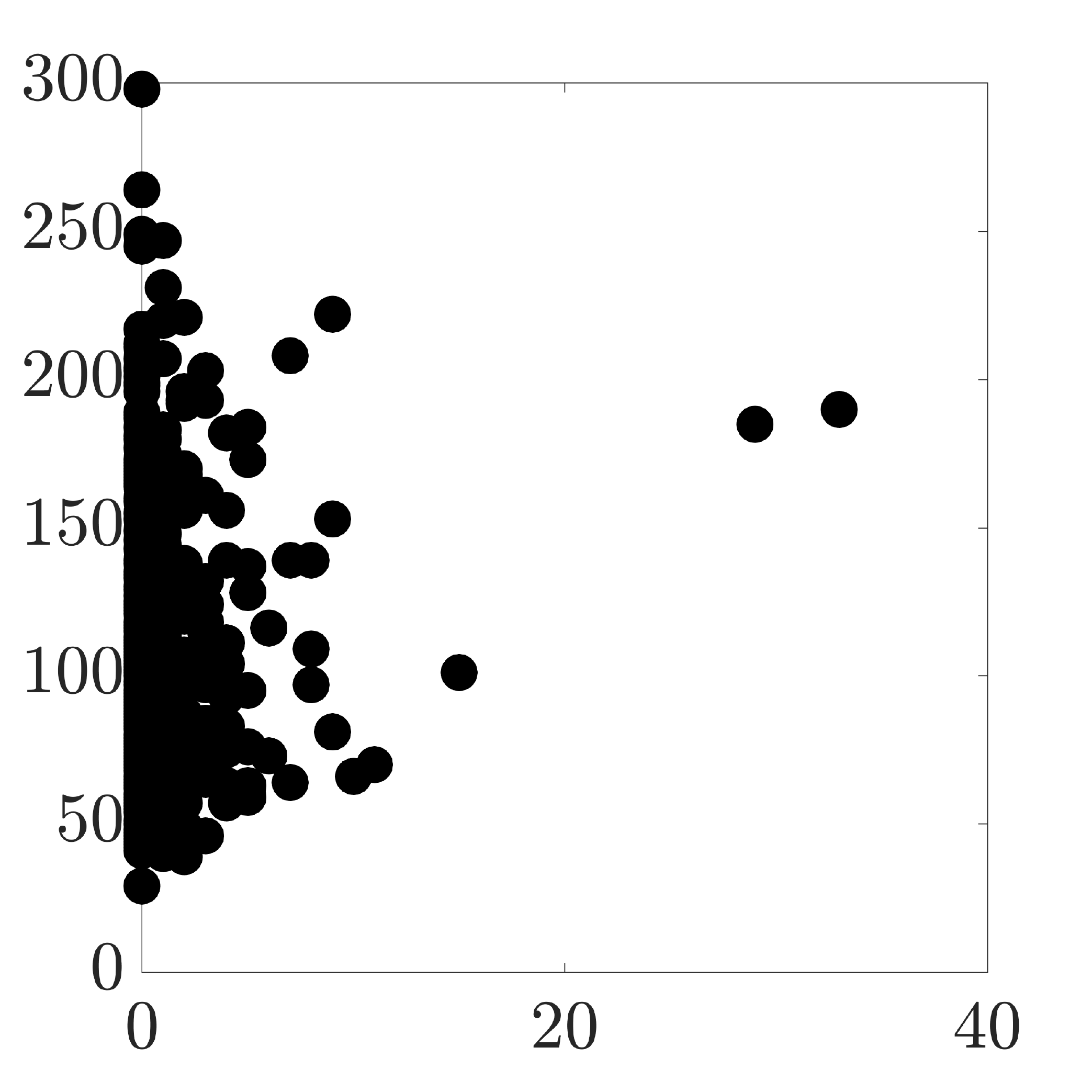} & 
\includegraphics[trim= 0mm 0mm 0mm 0mm,clip,height= 3.2cm, width= 3.7cm]{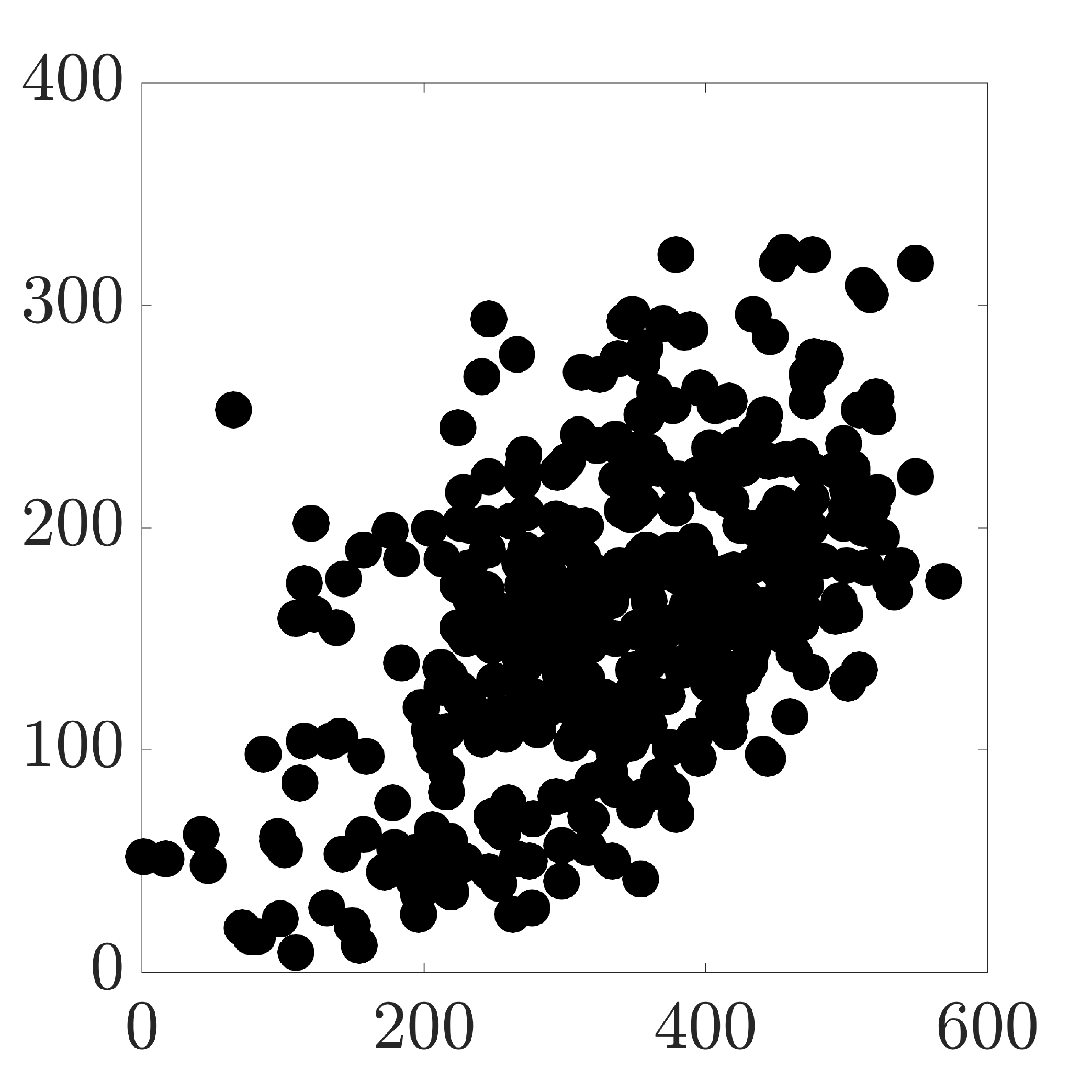}
\end{tabular}
\caption{Number of edges impacted by COVID-19, across layers (columns), for the nodes' total degree.
In each scatterplot: node total degree on the 27th March 2020 (horizontal axis) versus the number of non-null coefficients.
Only non-null coefficients are reported: blue indicates positive impact on edge existence, red indicates negative impact on edge existence.
A coefficient is considered null if its posterior HPDI contains zero.
}
\label{fig:numberCOVcoeff}
\end{figure}

Centrality is measured either by in-/out-degree or betweenness centrality.
Betweenness centrality quantifies the number of times a node acts as a bridge along the shortest path between two other nodes. Firms with large betweenness contribute to spreading contagion in the networks, thus requiring to be monitored for the stability of the financial system.
We measure the impact by considering the sum of the negative (blue) and positive (red) node coefficients of a risk factor, that is:
\begin{equation}
\begin{split}
\tilde{b}_{i,lk,r}^{IN,+} & = \sum_{j=1}^n \hat{b}_{ij,lk,r} \I(\hat{b}_{ij,lk,r} >0), \qquad
\tilde{b}_{i,lk,r}^{IN,-} = \sum_{j=1}^n \hat{b}_{ij,lk,r} \I(\hat{b}_{ij,lk,r} <0), \\
\tilde{b}_{i,lk,r}^{OUT,+} & = \sum_{j=1}^n \hat{b}_{ji,lk,r} \I(\hat{b}_{ji,lk,r} >0), \qquad
\tilde{b}_{i,lk,r}^{OUT,-} = \sum_{j=1}^n \hat{b}_{ji,lk,r} \I(\hat{b}_{ji,lk,r} <0), \\
\tilde{b}_{i,lk,r}^{BTW,+} & = \sum_{j=1}^n \hat{b}_{ij,lk,r} \I(\hat{b}_{ij,lk,r} >0) + \sum_{j=1}^n \hat{b}_{ji,lk,r} \I(\hat{b}_{ji,lk,r} >0), \\
\tilde{b}_{i,lk,r}^{BTW,-} & = \sum_{j=1}^n \hat{b}_{ij,lk,r} \I(\hat{b}_{ij,lk,r} <0) + \sum_{j=1}^n \hat{b}_{ji,lk,r} \I(\hat{b}_{ji,lk,r} <0).
\end{split}
\end{equation}
Figure~\ref{fig:numberCOVcoeff} shows the number of linkages of each firm which are impacted by COVID-19 versus the firms' total degree.
We find evidence, across the different inter- and inter-layer networks, of a positive relationship between firm centrality and the effect of COVID-19, except for leverage linkages.
In particular, in the risk premium and volatility layer, almost half of the linkage of each node have been impacted by the COVID-19.

\begin{figure}[t!]
\centering
\setlength{\abovecaptionskip}{0pt}
\begin{tabular}{ccccc}
 & return & risk premium & leverage & volatility \\
\begin{rotate}{90} \hspace*{15pt} {\small IN degree} \end{rotate} &
\includegraphics[trim= 0mm 0mm 0mm 0mm,clip,height= 3.2cm, width= 3.7cm]{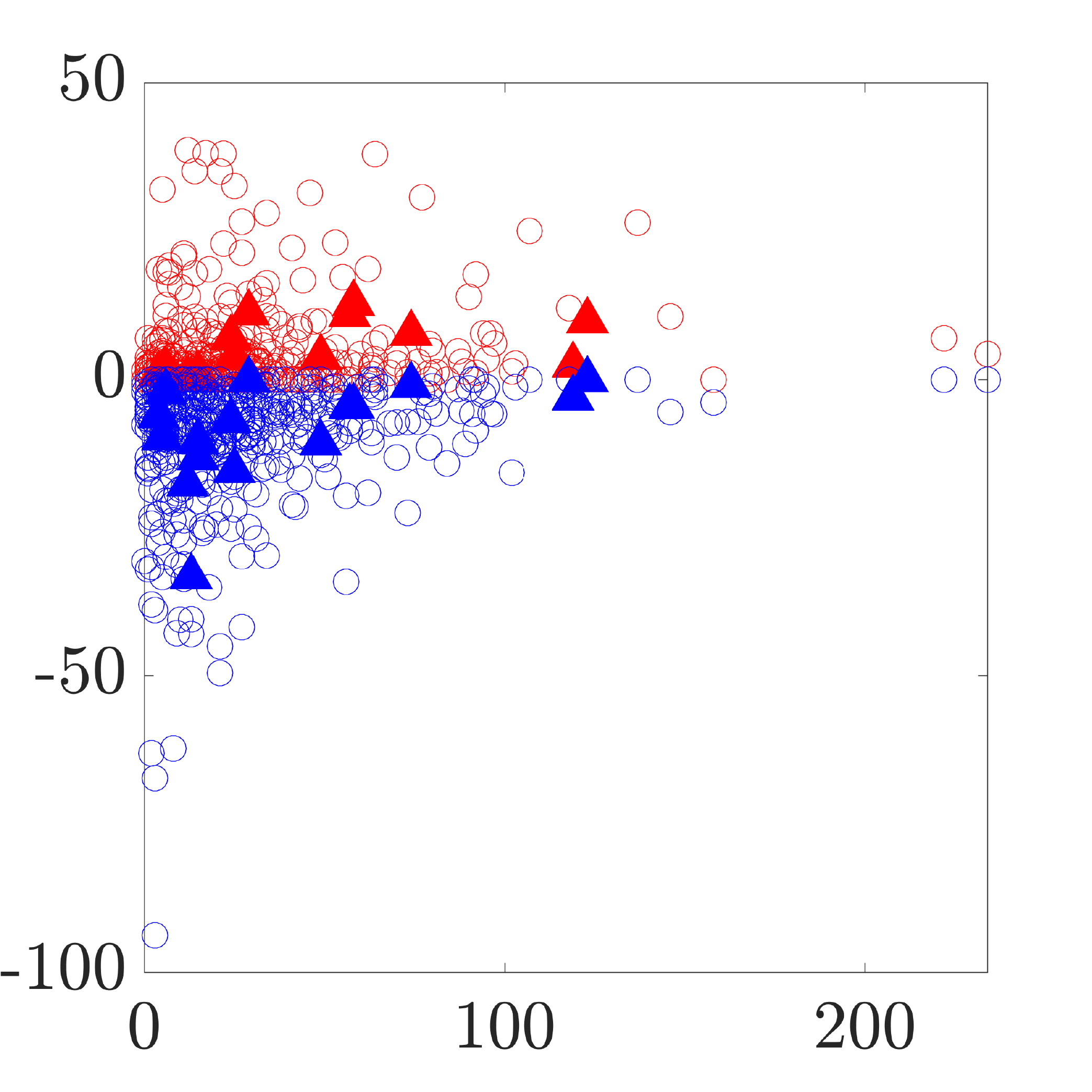} & 
\includegraphics[trim= 0mm 0mm 0mm 0mm,clip,height= 3.2cm, width= 3.7cm]{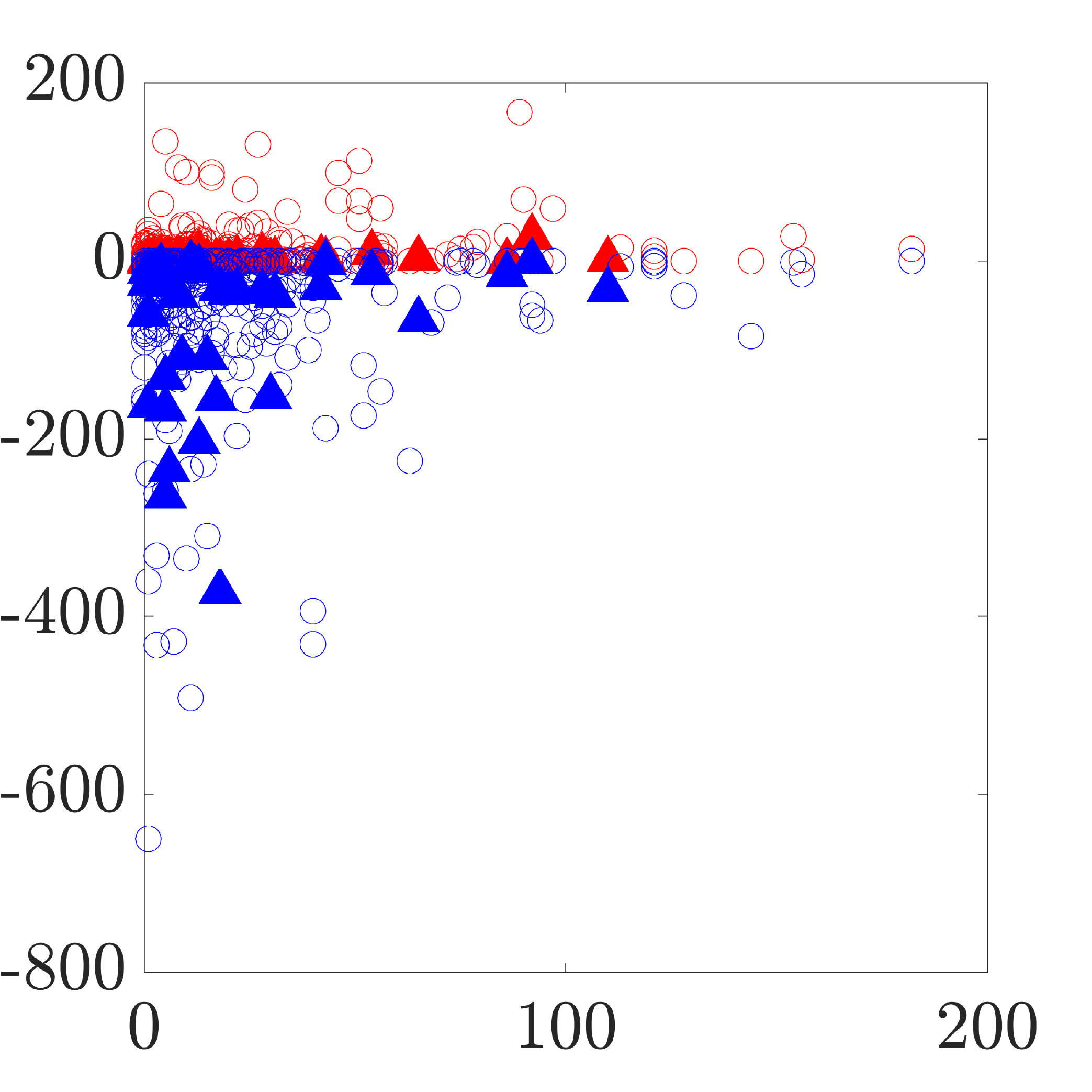} & 
\includegraphics[trim= 0mm 0mm 0mm 0mm,clip,height= 3.2cm, width= 3.7cm]{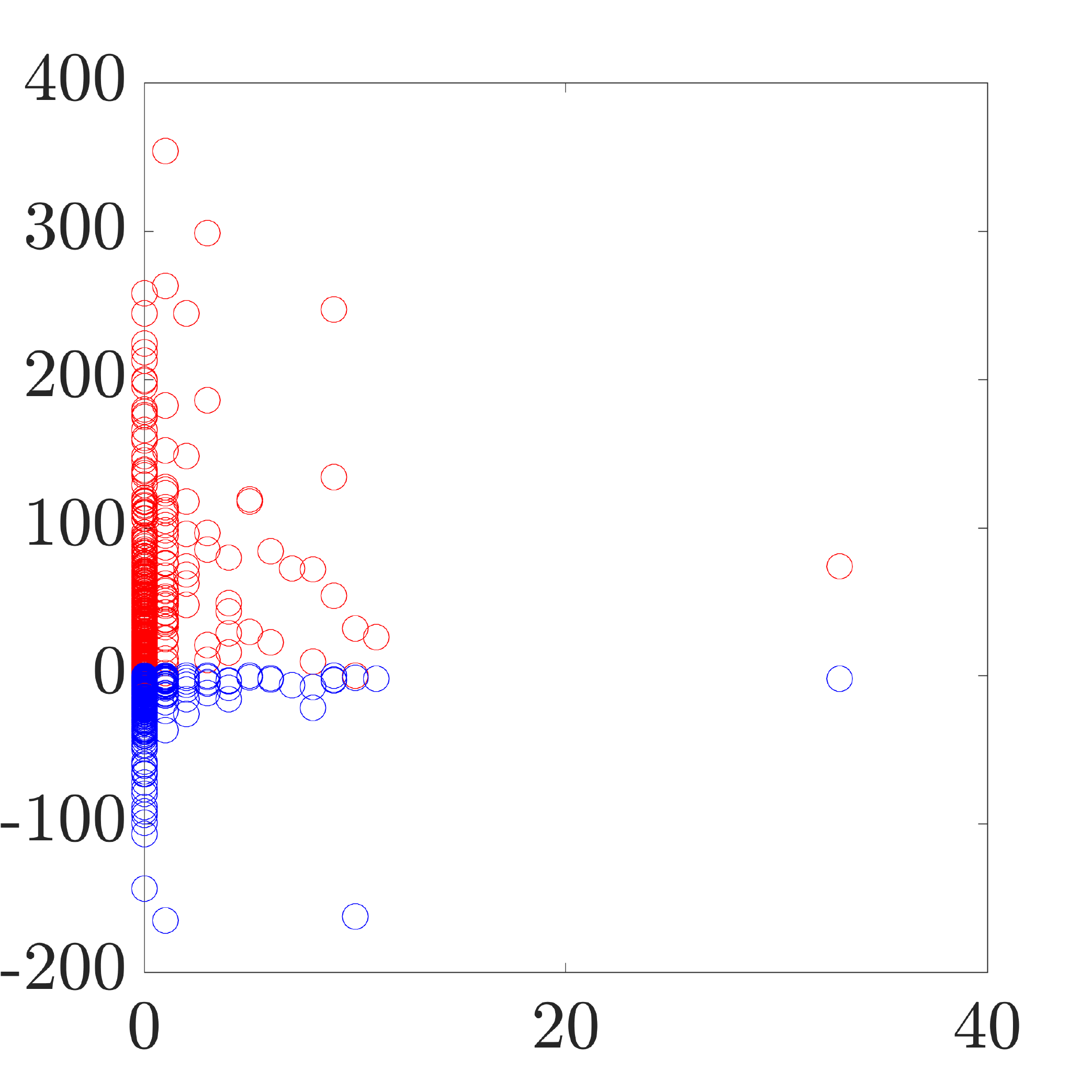} & 
\includegraphics[trim= 0mm 0mm 0mm 0mm,clip,height= 3.2cm, width= 3.7cm]{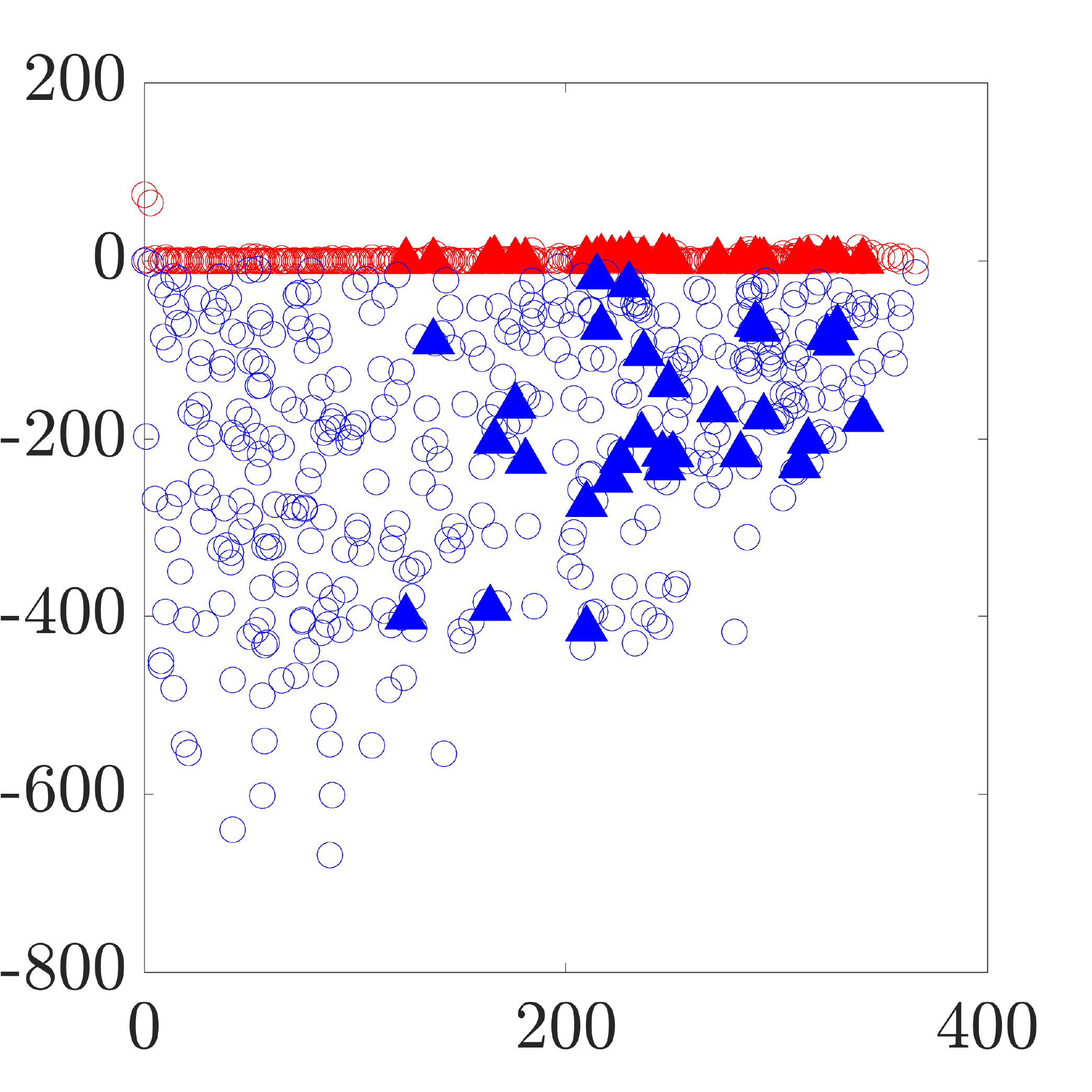} \\
\begin{rotate}{90} \hspace*{10pt} {\small OUT degree} \end{rotate} &
\includegraphics[trim= 0mm 0mm 0mm 0mm,clip,height= 3.2cm, width= 3.7cm]{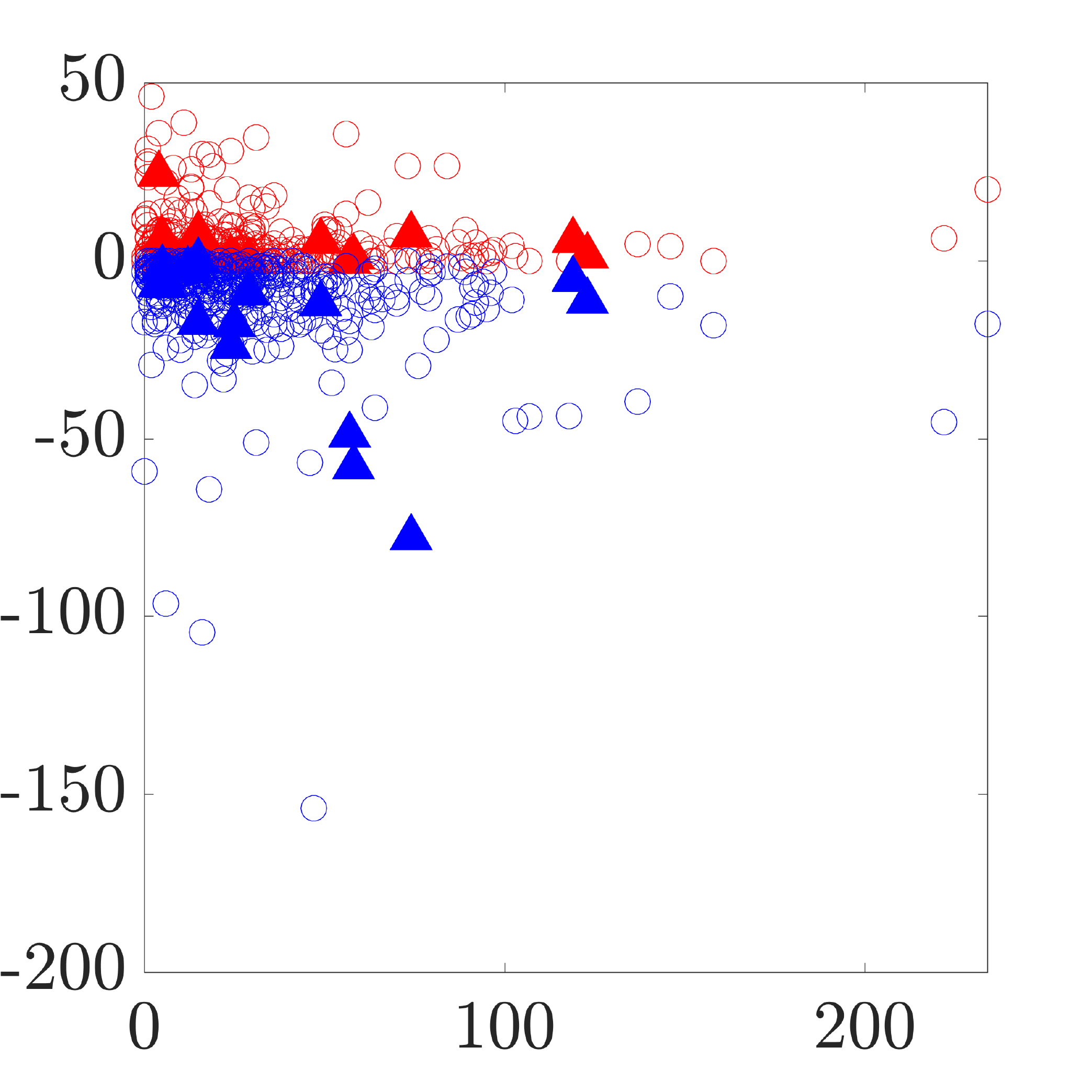} & 
\includegraphics[trim= 0mm 0mm 0mm 0mm,clip,height= 3.2cm, width= 3.7cm]{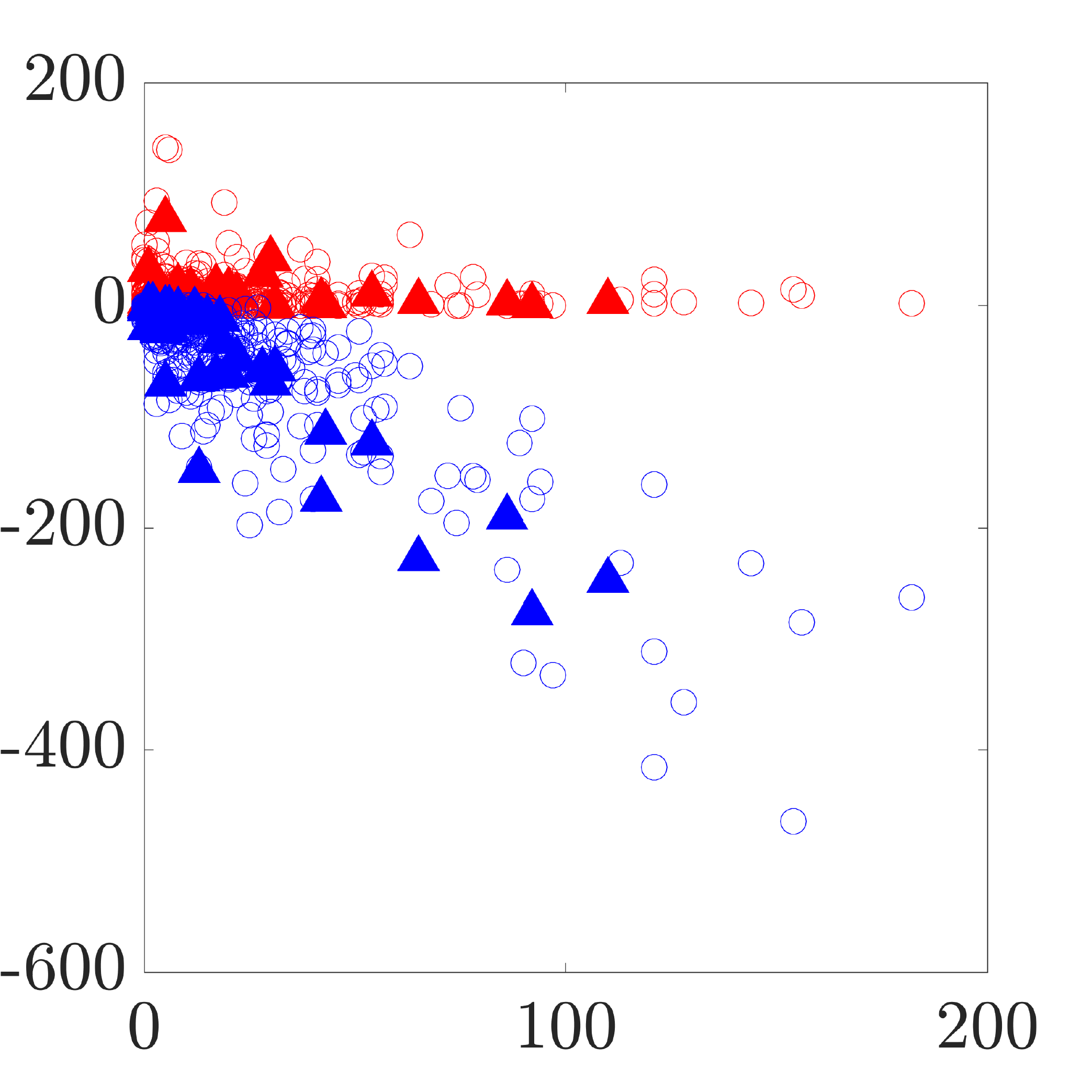} & 
\includegraphics[trim= 0mm 0mm 0mm 0mm,clip,height= 3.2cm, width= 3.7cm]{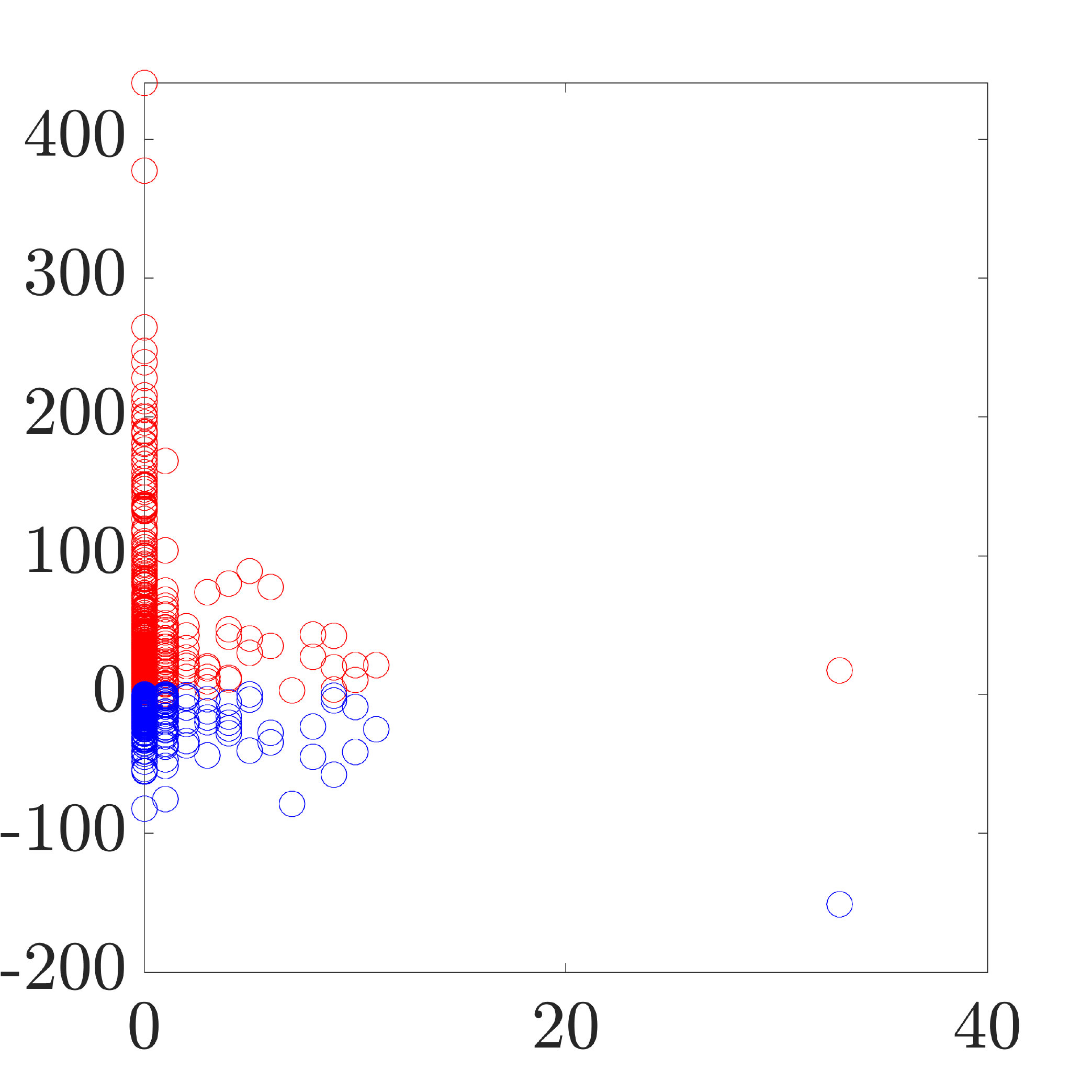} & 
\includegraphics[trim= 0mm 0mm 0mm 0mm,clip,height= 3.2cm, width= 3.7cm]{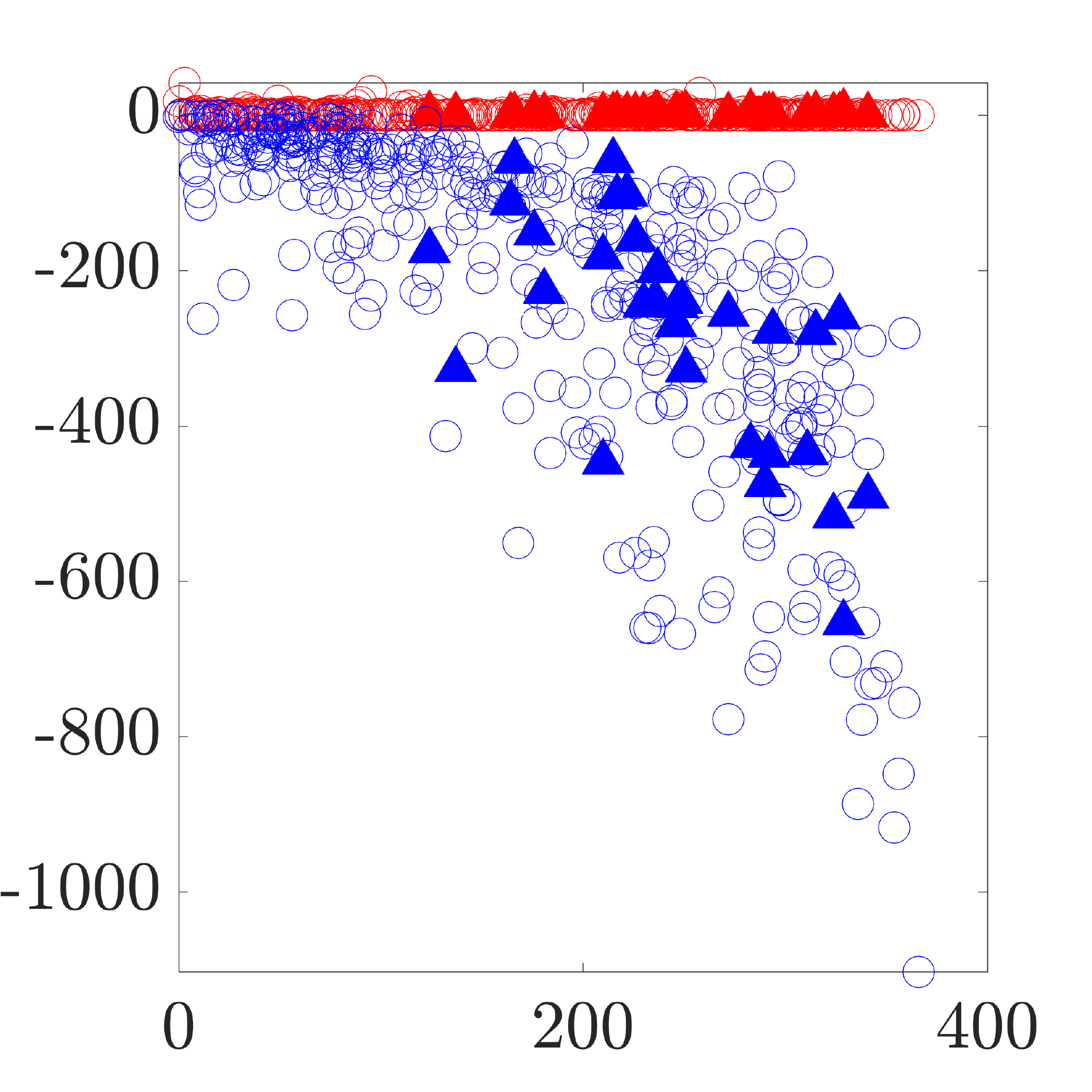} \\
\begin{rotate}{90} \hspace*{8pt} {\small betweenness} \end{rotate} &
\includegraphics[trim= 0mm 0mm 0mm 0mm,clip,height= 3.2cm, width= 3.7cm]{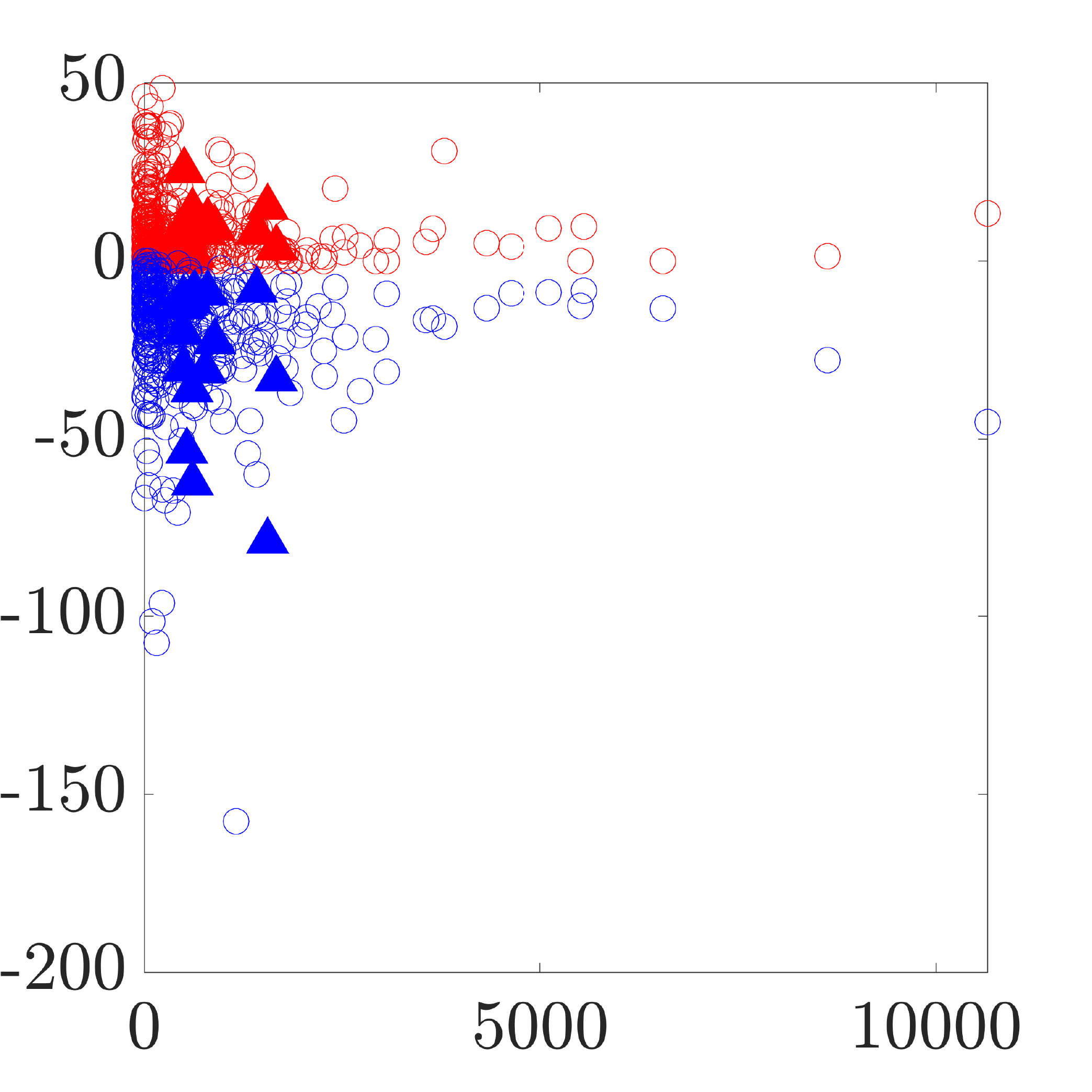} & 
\includegraphics[trim= 0mm 0mm 0mm 0mm,clip,height= 3.2cm, width= 3.7cm]{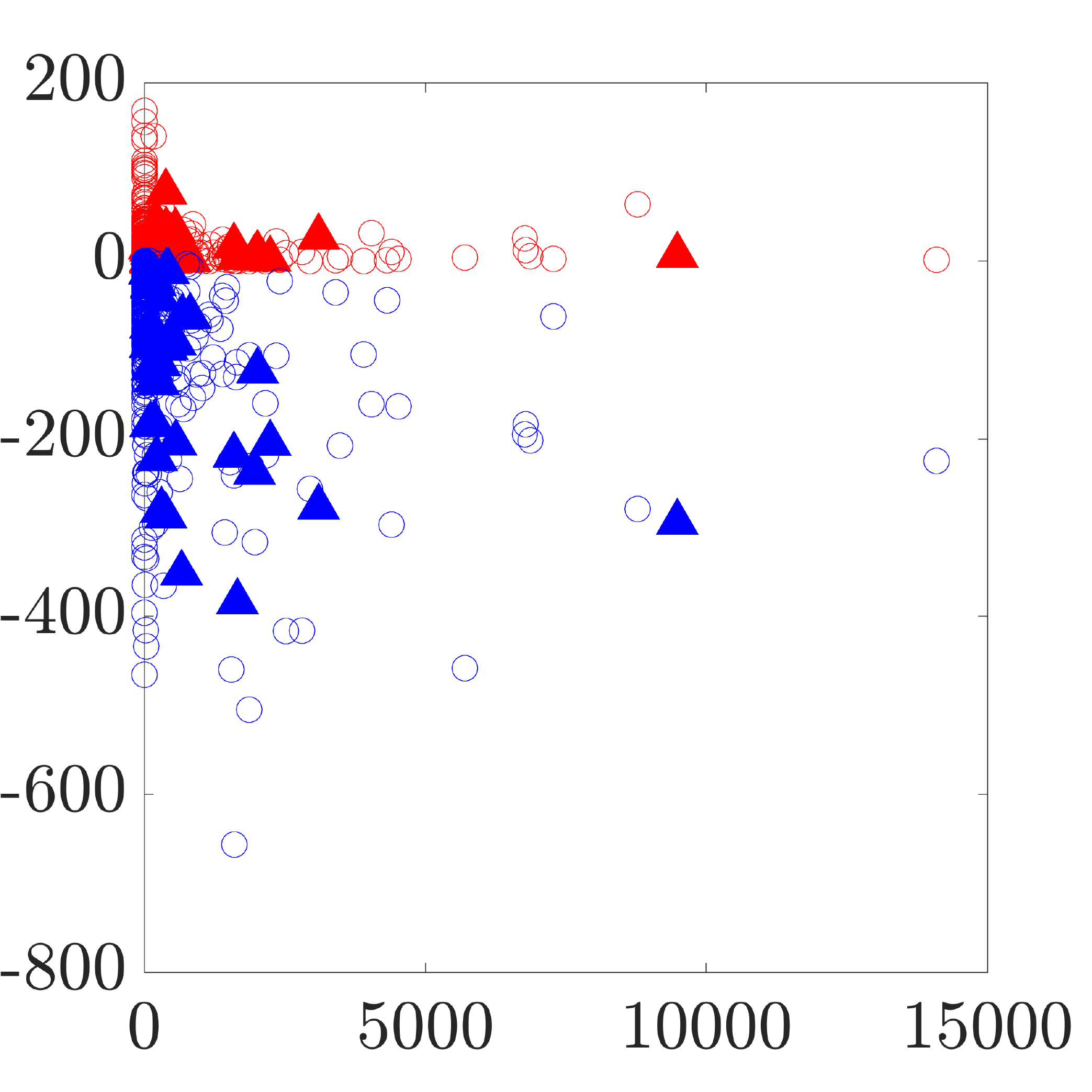} & 
\includegraphics[trim= 0mm 0mm 0mm 0mm,clip,height= 3.2cm, width= 3.7cm]{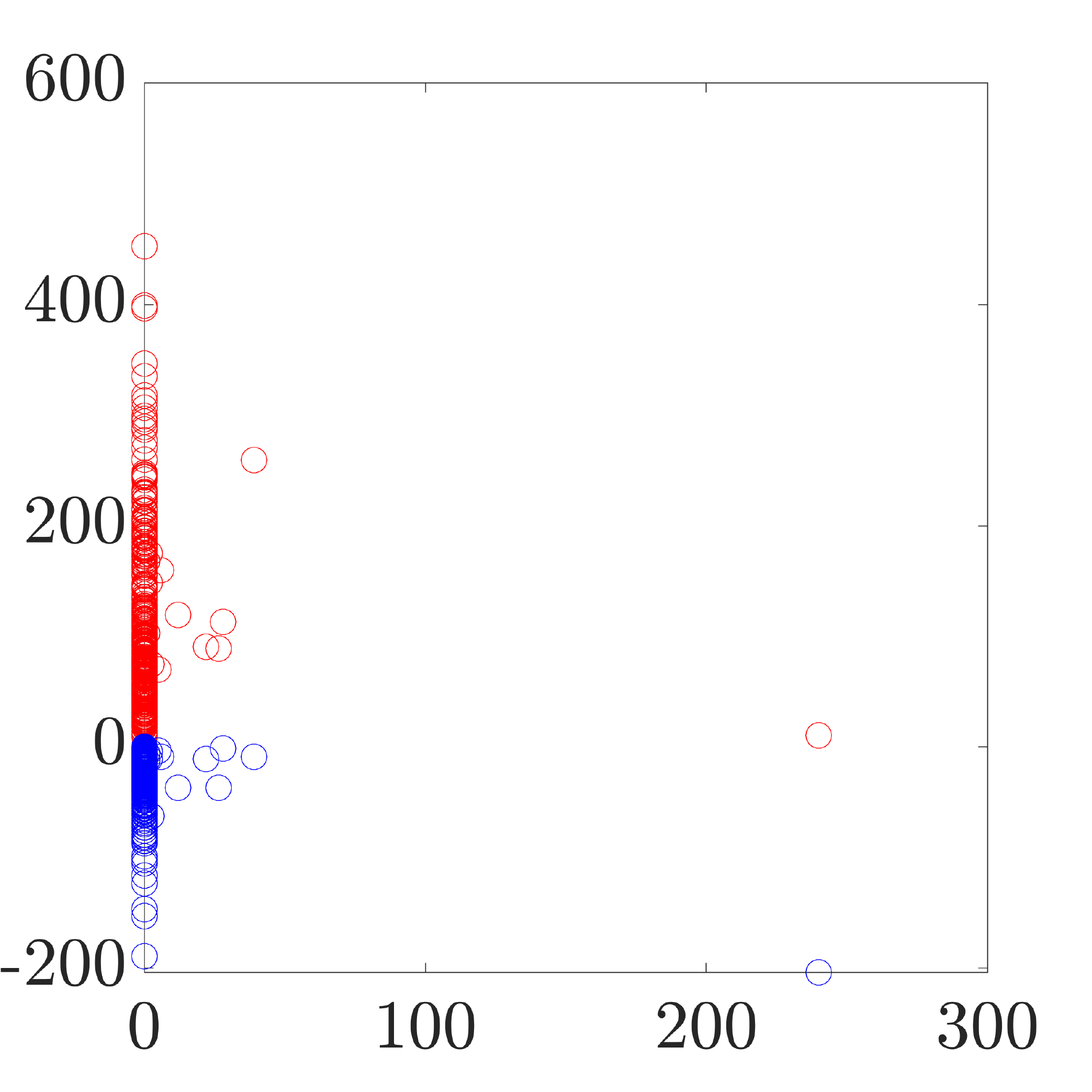} & 
\includegraphics[trim= 0mm 0mm 0mm 0mm,clip,height= 3.2cm, width= 3.7cm]{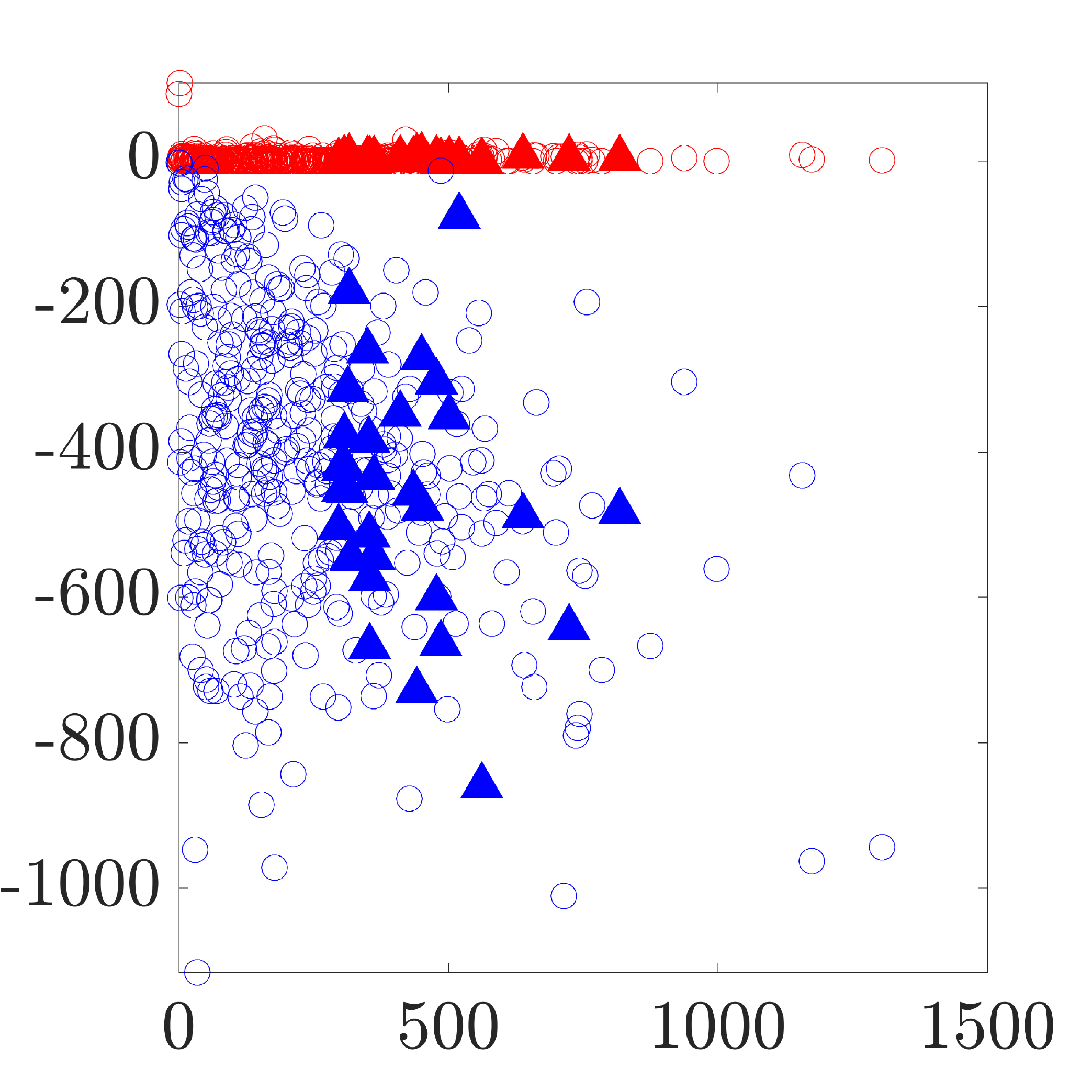}
\end{tabular}
\caption{Impact of the COVID-19 factors on financial linkages versus firm centrality, across layers (columns) and centrality measures (rows).
In each scatterplot: node centrality on the 27th March 2020 (horizontal axis) versus the sum of the negative (blue) and positive (red) node coefficients of a given variable (vertical axis).
Filled triangles indicate firms that moved from the 1st tercile of the betweenness centrality distribution on the 17th January 2020 to the 3rd tercile on the 27th March 2020.
Only non-null coefficients are reported: blue indicates positive impact on edge existence, red indicates negative impact on edge existence.
A coefficient is considered null if its posterior HPDI contains zero.
\label{fig:centrality}
}
\end{figure}

Figure~\ref{fig:centrality} shows the node centrality on the 27th March 2020 versus the sum of the negative (blue) and positive (red) node coefficients.
In each plot, the triangles indicate the firms with an increased betweenness centrality after the outbreak of the COVID-19. More specifically, we identify the firms which moved from the 1st tercile of the betweenness centrality distribution on the 17th January 2020 to the 3rd tercile on the 27th March 2020.

The negative impact of COVID-19 on edge existence (red color) uniformly affects the firms with low and high centrality in the multilayer networks. This is also true for the positive impact of COVID-19 (blue color) in the leverage linkages.
%There are heterogeneous findings for the return linkages where the positive impact XXXXXXXXXX
%
The most interesting findings concern the leverage and risk premium linkages.
In the volatility linkages, there is a positive relationship between the COVID-19 coefficients and the IN degree (last column in the first row) which indicates that the connectivity of firms with higher IN degree is less affected by COVID-19.
%That is, the negative coefficients implies a reduction on the p-value of the performed Granger causality tests and hence, an increase on the probability of the incoming connections.
Conversely, there is a negative relationship between the COVID-19 coefficients and the OUT degree (last column in the second row) which indicates that the connectivity of firms with higher OUT degree is more affected by the COVID-19.
Therefore, firms with higher OUT degree become more prone in transmitting volatility shocks to the system (last column, second row), with impact on other firms' volatility and returns (second column).
Similar conclusions can be drawn by considering the effect of COVID-19 on firms with large betweenness centrality (last column, last row).
% shows a negative relationship in the OUT degree and in the betweenness centrality indicating that firms with higher centrality are more affected by the COVID-19.

\section{Conclusion} \label{sec:conclusion}
In this paper, we have proposed a novel semiparametric framework for matrix-valued panel data.
The model has been applied to study multilayer temporal networks among European financial firms  (France, Germany, Italy).
We measure the financial connectedness arising from the interactions between two layers defined by assets returns and volatilities. The connectivity effects are represented at four levels: (i) return linkages; (ii) volatility linkages; (iii) risk premium linkages; and (iv) leverage linkages.

We have investigated the impact of COVID-19 on the structure of the network, which represents an unprecedented case since no previous disease outbreak had affected the real economy and the financial markets as the COVID-19 pandemic.
There is evidence of the explanatory power of COVID-19 for the connectivity of the European financial network at firm and sector level (e.g., Industrial, Real Estate, and Health Care).
The COVID-19 has a heterogeneous effect across layers, increasing the probabilities of volatility and risk premium linkages while decreasing the probability of leverage linkages.
Finally, our results show a positive relationship between firm centrality and the number of its linkages that are impacted by the COVID-19.

The network modeling of financial connectedness is a useful tool for policy-makers and other authorities in monitoring the financial system.
Moreover, despite being motivated by and applied to a European financial network, the proposed econometric framework is general and can be of interest for studying a wide spectrum of matrix-variate datasets emerging in several fields of data science.

%%%%%%%%%%%%%%%%%%%%%%%%%%%%%%%%%%%%%%%%%%%%%%%%%%%%%%%%%%%%
\bibliographystyle{apalike}
\bibliography{biblio}

\appendix

%{\color{purple}
%\textbf{NOTES:}
%\begin{itemize}
%\item Bayesian Lasso: $\gamma_{1,lk}^2 \sim \mathcal{G}a(a_0,b_0)$
%\item identification: set $\mu_1=0$ and impose $\mu_2 < \ldots < \mu_{M_b}$
%\item identification: impose $\beta_1/(\alpha_1-1) < \ldots < \beta_{M_\sigma}/(\alpha_{M_\sigma}-1)$ oppure re-parametrize IG as location-scale and impose constraint on location
%\item $M_b \geq 3$ and $M_\sigma \geq 3$
%\item add covariate `lecpoas'
%\end{itemize}
%}

\section{Proof of the results in the paper}

\subsection{Model properties}
\begin{proof}[Proof of Eq.~\ref{eq:marginal_likelihood}]
\begin{align*}
P(Y_{lk,t} | \boldsymbol{\mu}_{lk},\boldsymbol{\gamma}_{lk}^2, \boldsymbol{\alpha}_{lk},\boldsymbol{\beta}_{lk}) & = \int \!\!\!\int \! P(Y_{lk,t} | B_{lk,1},\ldots,B_{lk,R}, \boldsymbol{\sigma}_{lk}^2) P(\boldsymbol{\sigma}_{lk}^2 | \boldsymbol{\alpha}_{lk},\boldsymbol{\beta}_{lk},\mathbf{q}_{lk}) \: \mathrm{d}\boldsymbol{\sigma}_{lk}^2 \\
 & \quad \cdot P(B_{lk,1},\ldots,B_{lk,R} | \boldsymbol{\mu}_{lk},\boldsymbol{\gamma}_{lk}^2,\mathbf{p}_{lk}) \: \mathrm{d}B_{lk,1} \ldots \mathrm{d}B_{lk,R} \\
 & = \int \!\!\!\int \! P(Y_{lk,t} | B_{lk,1},\ldots,B_{lk,R}, \boldsymbol{\sigma}_{lk}^2) \sum_{m'=1}^{M_\sigma^n} \tilde{q}_{m',lk} P(\boldsymbol{\sigma}_{lk}^2 | \tilde{\boldsymbol{\alpha}}_{m',lk},\tilde{\boldsymbol{\beta}}_{m',lk}) \: \mathrm{d}\boldsymbol{\sigma}_{lk}^2 \\
 & \quad \cdot \sum_{m=1}^{M_b^{n^2}} \tilde{p}_{m,lk} P(B_{lk,1},\ldots,B_{lk,R} | \tilde{\boldsymbol{\mu}}_{m,lk},\tilde{\boldsymbol{\gamma}}_{m,lk}^2) \: \mathrm{d}B_{lk,1} \ldots \mathrm{d}B_{lk,R} \\
 & = \sum_{m'=1}^{M_\sigma^n} \sum_{m=1}^{M_b^{n^2}} \tilde{p}_{m,lk} \tilde{q}_{m',lk} P(Y_{lk,t} | \tilde{\boldsymbol{\alpha}}_{m',lk},\tilde{\boldsymbol{\beta}}_{m',lk}, \tilde{\boldsymbol{\mu}}_{m,lk},\tilde{\boldsymbol{\gamma}}_{m,lk}^2).
\end{align*}
Since:
\begin{align*}
P(\boldsymbol{\sigma}_{lk}^2 | \boldsymbol{\alpha}_{lk},\boldsymbol{\beta}_{lk},\mathbf{q}_{lk}) & = \prod_{i=1}^n P(\sigma_{i,lk}^2 | \boldsymbol{\alpha}_{lk},\boldsymbol{\beta}_{lk},\mathbf{q}_{lk}) \\
 & = \prod_{i=1}^n \sum_{m=1}^{M_\sigma} q_{m,lk} P(\sigma_{i,lk}^2 | \alpha_{m,lk},\beta_{m,lk}) \\
 & = \sum_{i_1=1}^{M_\sigma} \ldots \sum_{i_n=1}^{M_\sigma}  \prod_{u=1}^n q_{i_u,lk} P(\sigma_{u,lk}^2 | \alpha_{i_u,lk},\beta_{i_u,lk}).
\end{align*}
By relabeling the indices using the inverse lexicographic order:
\begin{equation*}
u = 1 + \sum_{\ell=1}^{n} (i_\ell - 1) M_\sigma^{\ell-1}
\end{equation*}
one obtains
\begin{align*}
P(\boldsymbol{\sigma}_{lk}^2 & | \boldsymbol{\alpha}_{lk},\boldsymbol{\beta}_{lk},\mathbf{q}_{lk}) = \sum_{m'=1}^{M_\sigma^n} \tilde{q}_{m',lk} P(\boldsymbol{\sigma}_{lk}^2 | \tilde{\boldsymbol{\alpha}}_{m',lk},\tilde{\boldsymbol{\beta}}_{m',lk}),
\end{align*}
where 
\begin{align*}
P(\boldsymbol{\sigma}_{lk}^2 | \tilde{\boldsymbol{\alpha}}_{m',lk},\tilde{\boldsymbol{\beta}}_{m',lk}) & = \prod_{i=1}^n P(\sigma_{i,lk}^2 | \alpha_{\pi(i,m'),lk},\beta_{\pi(i,m'),lk}) \\
\tilde{q}_{m',lk} & = \prod_{i=1}^n q_{\pi(i,m'),lk}
\end{align*}
with $\pi$ mapping the pair of indices $(i,m') \mapsto m$, and $m \in [1,M_\sigma]$.
A similar argument applies to $P(B_{lk,1},\ldots,B_{lk,R} | \boldsymbol{\mu}_{lk},\boldsymbol{\gamma}_{lk}^2,\mathbf{p}_{lk})$.
\end{proof}

\subsection{Posterior distributions}
In the following, we provide the derivation of the full conditional distribution using the Gibbs sampler.

The combination of the likelihood of the model in Eq.~\eqref{eq:model_observables} and the mixture priors in Eqs.~\eqref{eq:prior_Bijlkr}-\eqref{eq:prior_sigma2_ilk} yields a high-dimensional integral with no closed-form solution.
To address this issue, we follow the data-augmentation principle and introduce two sets of latent allocation variables in the set of observations, $\mathbf{D}_{lk,r}^b = \{ d_{ij,lk,r}^b \: \text{ for } i,j=1,\ldots,n \}$, for $l,k=1,2$, $r=1,\ldots,R$, and $\mathbf{D}_{lk}^\sigma = \{ d_{i,lk}^\sigma \: \text{ for } i=1,\ldots,n \}$, for $l,k=1,2$.
Denote with $\boldsymbol{\alpha}_{lk} = (\alpha_{1,lk},\ldots,\alpha_{M_\sigma,lk})'$, $\boldsymbol{\beta}_{lk} = (\beta_{1,lk},\ldots,\beta_{M_\sigma,lk})'$, $\boldsymbol{\mu}_{lk} = (\mu_{1,lk},\ldots,\mu_{M_b,lk})'$, $\boldsymbol{\gamma}_{lk}^2 = (\gamma_{1,lk}^2,\ldots,\gamma_{M_b,lk}^2)'$.
Thus, we obtain the data-augmented joint posterior distribution for the parameters of the likelihood, $B_{lk,r}$ and $\boldsymbol{\sigma}_{lk}^2$, the first stage hyper-parameters $\boldsymbol{\alpha}_{lk}$, $\boldsymbol{\beta}_{lk}$, $\boldsymbol{\mu}_{lk}$, and $\boldsymbol{\gamma}_{lk}^2$, the latent allocation variables $\mathbf{D}_{lk,r}^b$, $\mathbf{D}_{lk}^\sigma$, and the mixing probabilities, $\mathbf{p}_{lk}$ and $\mathbf{q}_{lk}$, with $l,k=1,2$, and $r=1,\ldots,R$, as:
\begin{align*}
 & \prod_{l=1}^2 \prod_{k=1}^2 \prod_{m=1}^{M_\sigma} P(\alpha_{m,lk}) P(\beta_{m,lk}) P(q_{m,lk}) \prod_{m=1}^{M_b} P(\mu_{m,lk}) P(\gamma_{m,lk}^2) P(p_{m,lk}) \\
 & \cdot \prod_{l=1}^2 \prod_{k=1}^2 \prod_{i=1}^n P(\sigma_{i,lk}^2 | \boldsymbol{\alpha}_{lk}, \boldsymbol{\beta}_{lk}, d_{i,lk}^\sigma) P(d_{i,lk}^\sigma | \mathbf{q}_{lk}) \prod_{j=1}^n \prod_{r=1}^R P(b_{ij,lk,r} | \boldsymbol{\mu}_{lk}, \boldsymbol{\gamma}_{lk}^2, d_{ij,lk,r}^b) P(d_{ij,lk,r}^b | \mathbf{p}_{lk}) \\
 & \cdot \prod_{l=1}^2 \prod_{k=1}^2 \prod_{t=1}^T P(Y_{lk,t} | \mathbf{f}_t, B_{lk,1},\ldots,B_{lk,R}, \boldsymbol{\sigma}_{lk}^2)
\end{align*}

\paragraph{Full conditional distribution of $b_{ij,lk,r}$}
The full conditional distributions of the coefficients are given as follows.
%Let $\mathbf{b}_{lk,r} = \operatorname{vec}(B_{lk,r})$. Then
%\begin{align*}
%P(\mathbf{b}_{lk,r} |-) \propto \mathcal{N}\Big( \overline{\boldsymbol{\mu}}, \: \overline{\Gamma} \Big)
%\end{align*}
%where, defining $\overline{\mathbf{y}}_{lk,r,t} = \operatorname{vec}(Y_{lk,t}-\sum_{s\neq r} B_{lk,s} f_{s,t})$, 
%\begin{align*}
%\overline{\boldsymbol{\mu}} & = \overline{\Gamma} \Big( \operatorname{diag}(\boldsymbol{\gamma}_{d_{lk,r}}^2)^{-1} \boldsymbol{\mu}_{d_{lk,r}} + \sum_{t} \overline{\mathbf{y}}_{lk,r,t}' \big( \operatorname{diag}(\boldsymbol{\sigma}_{lk}^2) \otimes I_n \big)^{-1} f_{r,t} \Big) \\
%\overline{\Gamma} & = \Big( \operatorname{diag}(\boldsymbol{\gamma}_{d_{lk,r}}^2)^{-1} + \sum_{t} \big( \operatorname{diag}(\boldsymbol{\sigma}_{lk}^2) \otimes I_n \big)^{-1} f_{r,t}^2 \Big)
%\end{align*}
For each entry $(i,j)$, define $\epsilon_{ij,lk,r,t} = y_{ij,lk,t} - \sum_{r'\neq r} b_{ij,lk,r'} f_{r',t}$, thus obtaining
\begin{align*}
P(b_{ij,lk,r} |-) & \propto \exp\Big( -\frac{(b_{ij,lk,r} - \mu_{d_{ij,lk,r}^b})^2}{2\gamma_{d_{ij,lk,r}^b}^2} \Big) \prod_{t=1}^T \exp\Big( \frac{(\epsilon_{ij,lk,r,t} - b_{ij,lk,r} f_{r,t})^2}{2\sigma_{i,lk}^2} \Big) \\
 & \propto \mathcal{N}(\overline{\mu}, \: \overline{\gamma}^2)
\end{align*}
where
\begin{align*}
\overline{\gamma}^2 & = \Big( \frac{1}{\gamma_{d_{ij,lk,r}}^2} + \sum_{t=1}^T \frac{f_{r,t}^2}{\sigma_{i,lk}^2} \Big)^{-1} \qquad
\overline{\mu}  = \overline{\gamma}^2 \Big( \frac{\sum_{t=1}^T f_{r,t} \epsilon_{ij,lk,r,t}}{\sigma_{i,lk}^2} + \frac{\mu_{d_{ij,lk,r}}}{\gamma_{d_{ij,lk,r}}^2} \Big)
\end{align*}

\paragraph{Full conditional distribution of $\sigma^2_{i,lk}$}
The full conditional distributions of the noise variances are given by
\begin{align*}
P(\sigma_{i,lk}^2 |-) & \propto (\sigma_{i,lk}^2)^{-\alpha_{d_{i,lk}^\sigma}-1} \exp\Big( -\frac{\beta_{d_{i,lk}^\sigma}}{\sigma_{i,lk}^2} \Big) \\
 & \quad \cdot \prod_{t=1}^T (\sigma_{i,lk}^2)^{-n/2} \exp\Big( -\frac{1}{2} \operatorname{tr}\big( \operatorname{diag}(\boldsymbol{\sigma}_{lk}^2)^{-1} (Y_{lk} - \sum_{r=1}^R B_{lk,r} f_{r,t})' (Y_{lk} - \sum_{r=1}^R B_{lk,r} f_{r,t}) \big) \Big) \\
 & \propto \mathcal{IG}\Big( \alpha_{\tilde{d}_{i,lk},lk} + \frac{Tn}{2}, \: \beta_{\tilde{d}_{i,lk},lk} + \frac{1}{2} \sum_{t=1}^T E_{ii,lk,t} \Big)
%\frac{1}{2} \sum_{t=1} \sum_{j} \tilde{Y}_{ij,lk,t} \Big)
\end{align*}
where $E_{lk,t} = (Y_{lk,t} - \sum_{r} B_{lk,r} f_{r,t})' (Y_{lk,t} - \sum_{r} B_{lk,r} f_{r,t})$.

\paragraph{Full conditional distribution of $d_{ij,lk,r}^b$ and $d_{i,lk}^\sigma$}
The full conditional distributions of the allocation variables are given by
\begin{align*}
%P(d_{ij,lk,r} = m |-) & \propto \prod_{i,j,r : d_{ij,lk,r = m}} p_{m,lk} \mathcal{N}( b_{ij,lk,r} | \mu_{m,lk}, \gamma_{m,lk}^2 ) \\
P(d_{ij,lk,r}^b = m |-) & \propto p_{m,lk} \mathcal{N}(b_{ij,lk,r} | \mu_{m,lk}, \gamma_{m,lk}^2) \\
%P(\tilde{d}_{i,lk} = m |-) & \propto \prod_{i : d_{i,lk = m}} q_{m,lk} \mathcal{IG}( \sigma_{i,lk}^2 | \alpha_{m,lk}, \beta_{m,lk} ) \\
P(d_{i,lk}^\sigma = m |-) & \propto q_{m,lk} \mathcal{IG}( \sigma_{i,lk}^2 | \alpha_{m,lk}, \beta_{m,lk} )
\end{align*}

\paragraph{Full conditional distribution of $p_{m,lk}$ and $q_{m,lk}$}
The full conditional distribution of the mixing probabilities for each mixture are 
\begin{align*}
P(\mathbf{p}_{lk} |-) & \propto \mathcal{D}ir\Big( \phi_b + \sum_{i,j,r} \I(d_{ij,lk,r}^b=1), \ldots, \phi_b + \sum_{i,j,r} \I(d_{ij,lk,r}^b=M_b) \Big) \\
P(\mathbf{q}_{lk} |-) & \propto \mathcal{D}ir\Big( \phi_\sigma + \sum_{i} \I(d_{i,lk}^\sigma =1), \ldots, \phi_\sigma + \sum_{i} \I(d_{i,lk}^\sigma =M_\sigma) \Big)
\end{align*}

\paragraph{Full conditional distribution of $\mu_{m,lk}$}
For each $m=2,\ldots,M_b$, the posterior distributions of the component-specific means are obtained as
\begin{align*}
P(\mu_{m,lk} |-) & \propto \exp\Big( -\frac{\mu_{m,lk}^2}{2\underline{s}^2} \Big) \prod_{\{ i,j,r : d_{ij,lk,r}^b = m \}} \exp\Big( -\frac{(b_{ij,lk,r} - \mu_{m,lk})^2}{2 \gamma_{m,lk}^2} \Big) \\
 & \propto \mathcal{N}( \overline{\mu}, \: \overline{s}^2 )
\end{align*}
where
\begin{align*}
\overline{\mu} = \overline{s}^2 \sum_{\{ i,j,r : d_{ij,lk,r}^b=m \}} \frac{b_{ij,lk,r}}{\gamma_{m,lk}^2} \qquad
\overline{s}^2 = \Big( \frac{1}{\underline{s}^2} + \sum_{\{ i,j,r : d_{ij,lk,r}^b=m \}} \frac{1}{\gamma_{m,lk}^2} \Big)^{-1}
\end{align*}

\paragraph{Full conditional distribution of $\gamma_{m,lk}^2$}
For $m=1$, since $\mu_{1,lk}=0$, the posterior distributions of the component-specific variances are obtained as
\begin{align*}
P(\gamma_{1,lk}^2 |-) & \propto (\gamma_{1,lk}^2)^{\underline{a}_0-1} \exp\Big( -\frac{\gamma_{1,lk}^2}{\underline{b}_0} \Big) \prod_{\{ i,j,r : d_{ij,lk,r}^b=1 \}} (\gamma_{1,lk}^2)^{-1/2} \exp\Big( -\frac{b_{ij,lk,r}^2}{2 \gamma_{1,lk}^2}\Big) \\
 & \propto \text{GiG}\Big( \underline{a}_0 - \frac{\sum_{i,j,r} \I(d_{ij,lk,r}^b=1)}{2}, \: \frac{2}{\underline{b}_0}, \: \sum_{\{i,j,r : d_{ij,lk,r}^b=1 \}} b_{ij,lk,r}^2 \Big)
\end{align*}
Instead, for each $m=2,\ldots,M_b$, the posterior distributions are
\begin{align*}
P(\gamma_{m,lk}^2 |-) & \propto (\gamma_{m,lk}^2)^{-\underline{a}_1-1} \exp\Big( -\frac{\underline{b}_1}{\gamma_{m,lk}^2}\Big) \prod_{\{ i,j,r : d_{ij,lk,r}^b=m \}} (\gamma_{m,lk}^2)^{-1/2} \exp\Big( -\frac{(b_{ij,lk,r}  - \mu_{m,lk})^2}{2 \gamma_{m,lk}^2} \Big) \\
 & \propto \mathcal{IG}\Big( \underline{a}_1 + \frac{\sum_{i,j,r} \I(d_{ij,lk,r}^b=m)}{2}, \: \underline{b}_1 + \sum_{\{ i,j,r : d_{ij,lk,r}^b=m \}} \frac{(b_{ij,lk,r} - \mu_{m,lk})^2}{2} \Big) \\
 & \propto \text{GiG}\Big( -\underline{a}_1 - \frac{\sum_{i,j,r} \I(d_{ij,lk,r}^b=m)}{2}, \: 0, \: 2\underline{b}_1 + \sum_{\{ i,j,r : d_{ij,lk,r}^b=m \}} (b_{ij,lk,r} - \mu_{m,lk})^2 \Big)
\end{align*}

\paragraph{Full conditional distribution of $\alpha_{m,lk}$}
For $m=1,\ldots,M_\sigma$, the posterior distributions of the component-specific shapes are obtained as
\begin{align*}
P(\alpha_{m,lk} |-) \propto \alpha_{m,lk}^{\underline{a}_2-1} \exp\Big( -\frac{\alpha_{m,lk}}{\underline{b}_2} \Big) \Big( \frac{\beta_{m,lk}^{\alpha_{m,lk}}}{\Gamma(\alpha_{m,lk})} \Big)^{\# \{ i: d_{i,lk}^\sigma=m \}} \Big( \prod_{\{ i: d_{i,lk}^\sigma=m \}} \sigma_{i,lk}^2 \Big)^{-\alpha_{m,lk}}
\end{align*}
we sample from this distribution using an adaptive RWMH with proposal.
%Alternatively, using the conjugate prior:
%\begin{align*}
%P(\alpha_{m,lk} | \beta_{m,lk}) \propto  \frac{\beta_{m,lk}^{\alpha_{m,lk} c_2} a_2^{-\alpha_{m,lk}-1}}{\Gamma(\alpha_{m,lk})^{b_2}}
%\end{align*}
%then one gets
%\begin{align*}
%P(\alpha_{m,lk} |-) \propto \frac{\beta_{m,lk}^{\alpha_{m,lk} (c_2  + \#\{ d_{i,lk}^\sigma=m \})} (a_2 \prod_{i:d_{i,lk}^\sigma=m} \sigma_{_i,lk}^2 )^{-\alpha_{m,lk}-1}}{\Gamma(\alpha_{m,lk})^{b_2 + \#\{ d_{i,lk}^\sigma=m \}}}
%\end{align*}

\paragraph{Full conditional distribution of $\beta_{m,lk}$}
For $m=1,\ldots,M_\sigma$, the posterior distributions of the component-specific scales are given by
\begin{align*}
P(\beta_{m,lk} |-) & \propto \beta_{m,lk}^{\underline{a}_3-1} \exp\Big( -\frac{\beta_{m,lk}}{\underline{b}_3} \Big) \prod_{\{ i:d_{i,lk}^\sigma=m  \}} \beta_{m,lk}^{\alpha_{m,lk}} \exp\Big( -\frac{\beta_{m,lk}}{\sigma_{i,lk}^2} \Big) \\
 & \propto \mathcal{G}a\Big( \underline{a}_3 + \alpha_{m,lk} \cdot \# \{ i: d_{i,lk}^\sigma=m \}, \: \underline{b}_3 + \ \sum_{\{ i : d_{i,lk}^\sigma=m \}} \sigma_{i,lk}^2 \Big)
\end{align*}

\section{Additional results}

\begin{figure}
\centering
\begin{tabular}{ccccc}
 & SX5E & V2X & LECPOAS & NCOVEUR \\
\begin{rotate}{90} \hspace*{27pt} {\small return} \end{rotate} &
\includegraphics[trim= 0mm 0mm 0mm 0mm,clip,height= 3.2cm, width= 3.7cm]{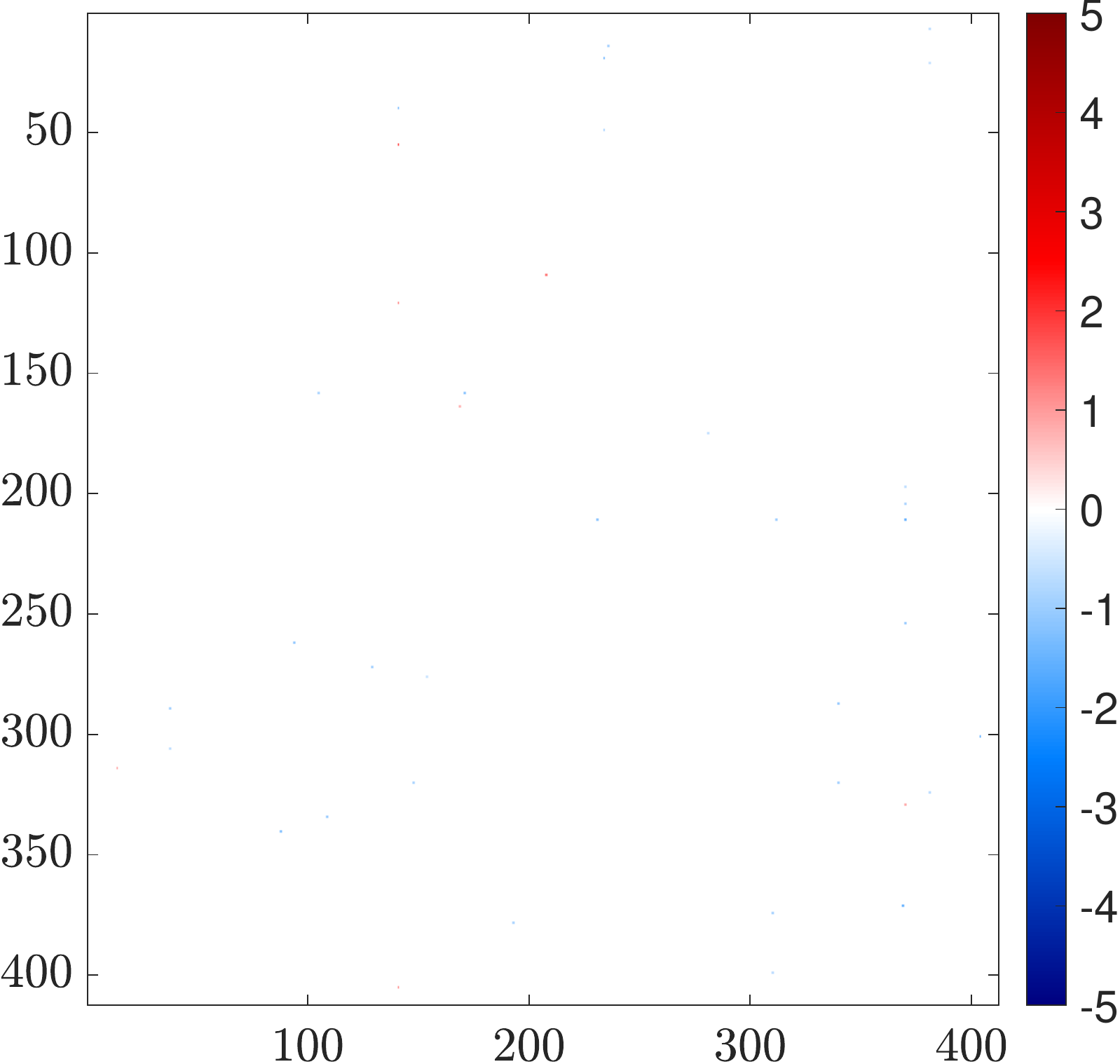} & 
\includegraphics[trim= 0mm 0mm 0mm 0mm,clip,height= 3.2cm, width= 3.7cm]{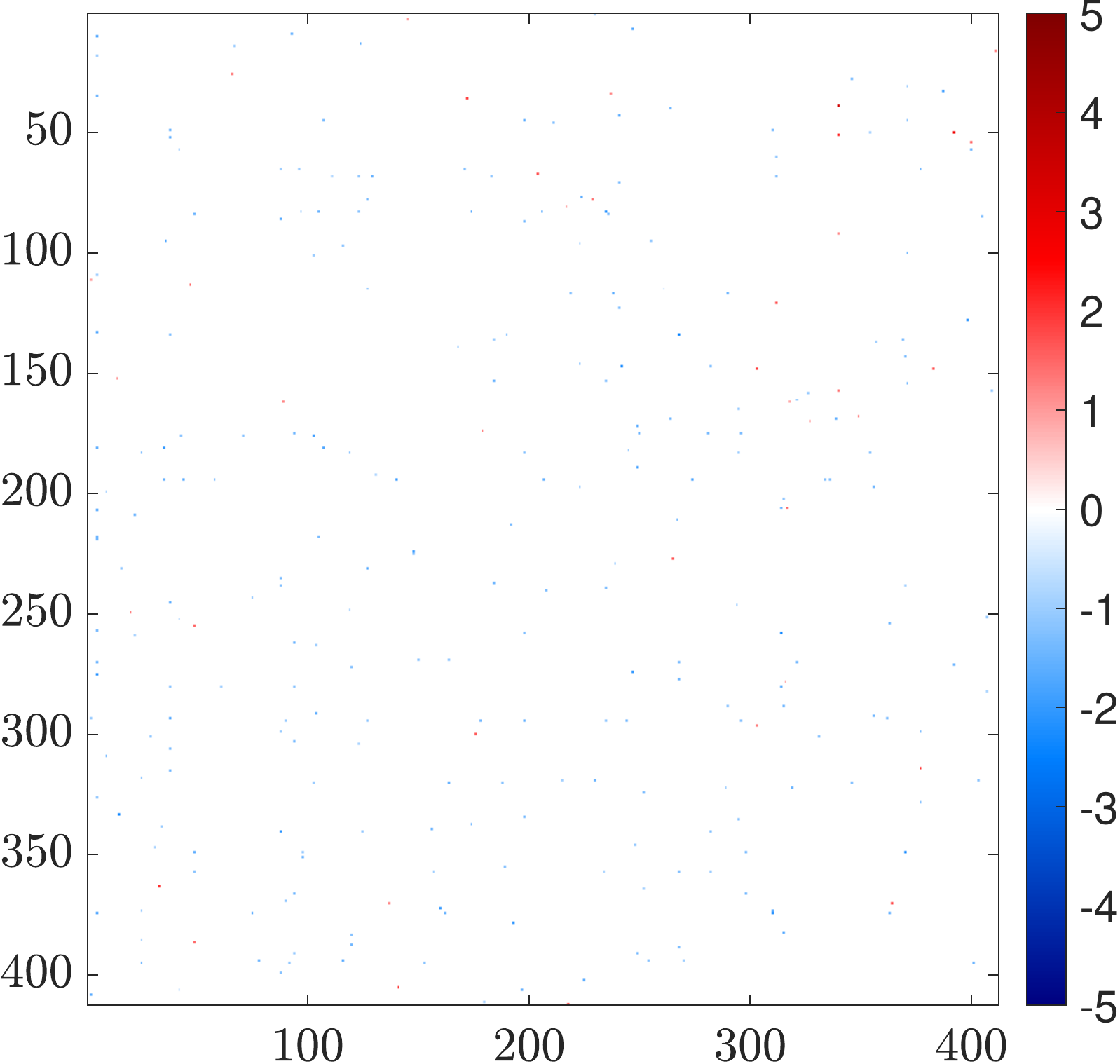} & 
\includegraphics[trim= 0mm 0mm 0mm 0mm,clip,height= 3.2cm, width= 3.7cm]{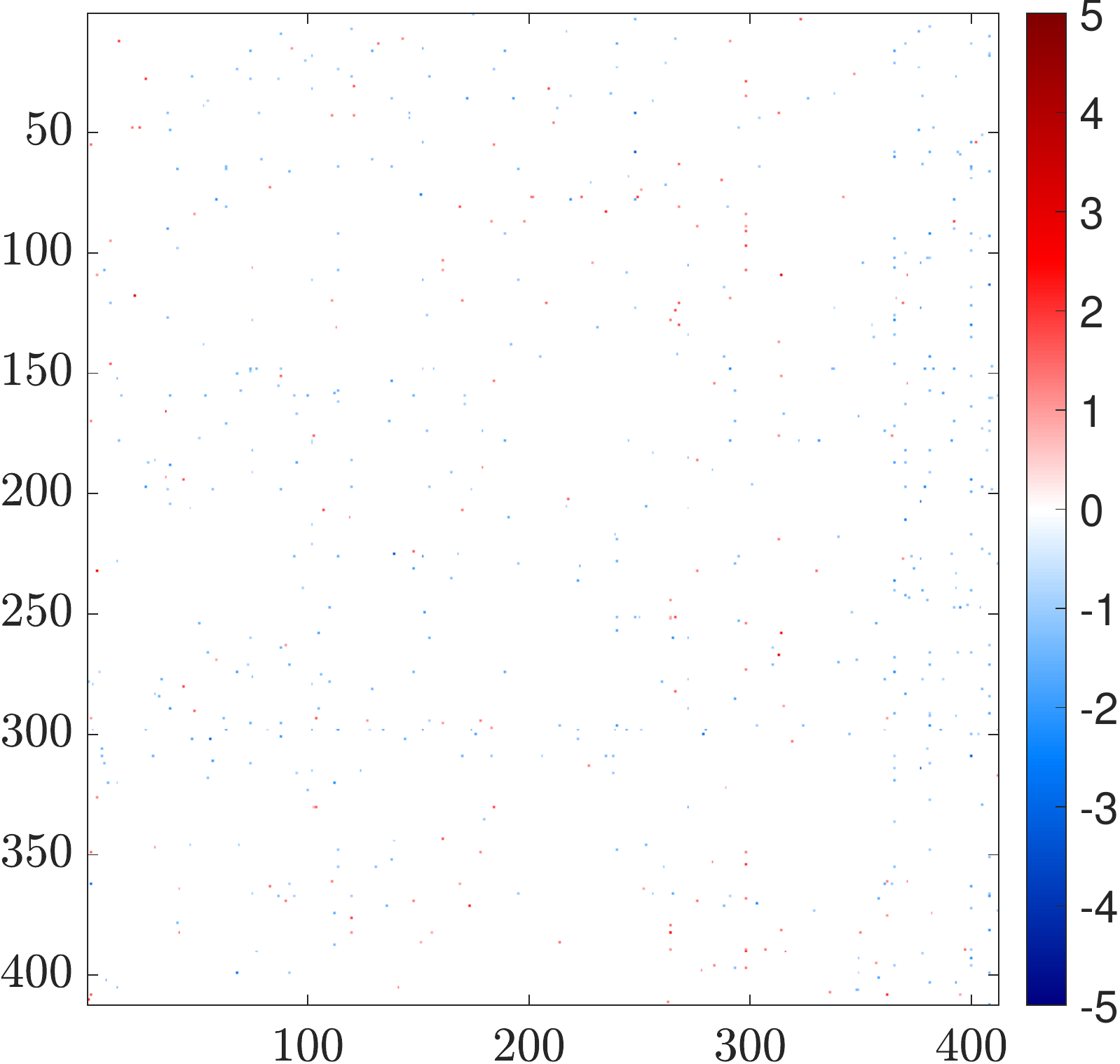} & 
\includegraphics[trim= 0mm 0mm 0mm 0mm,clip,height= 3.2cm, width= 3.7cm]{n412_M2_estimate_significant_11_r5-eps-converted-to.pdf} \\
\begin{rotate}{90} \hspace*{5pt} {\small risk premium} \end{rotate} &
\includegraphics[trim= 0mm 0mm 0mm 0mm,clip,height= 3.2cm, width= 3.7cm]{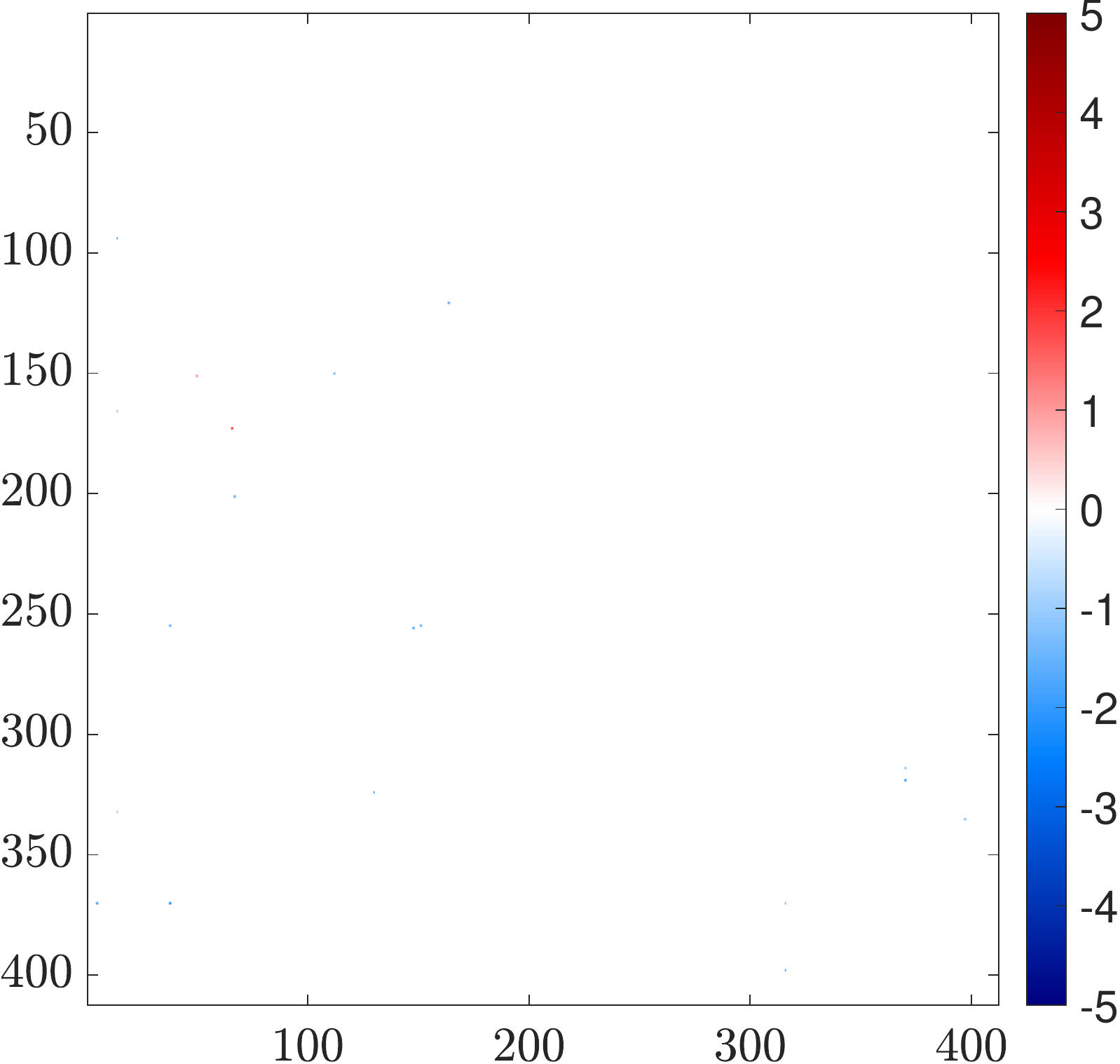} & 
\includegraphics[trim= 0mm 0mm 0mm 0mm,clip,height= 3.2cm, width= 3.7cm]{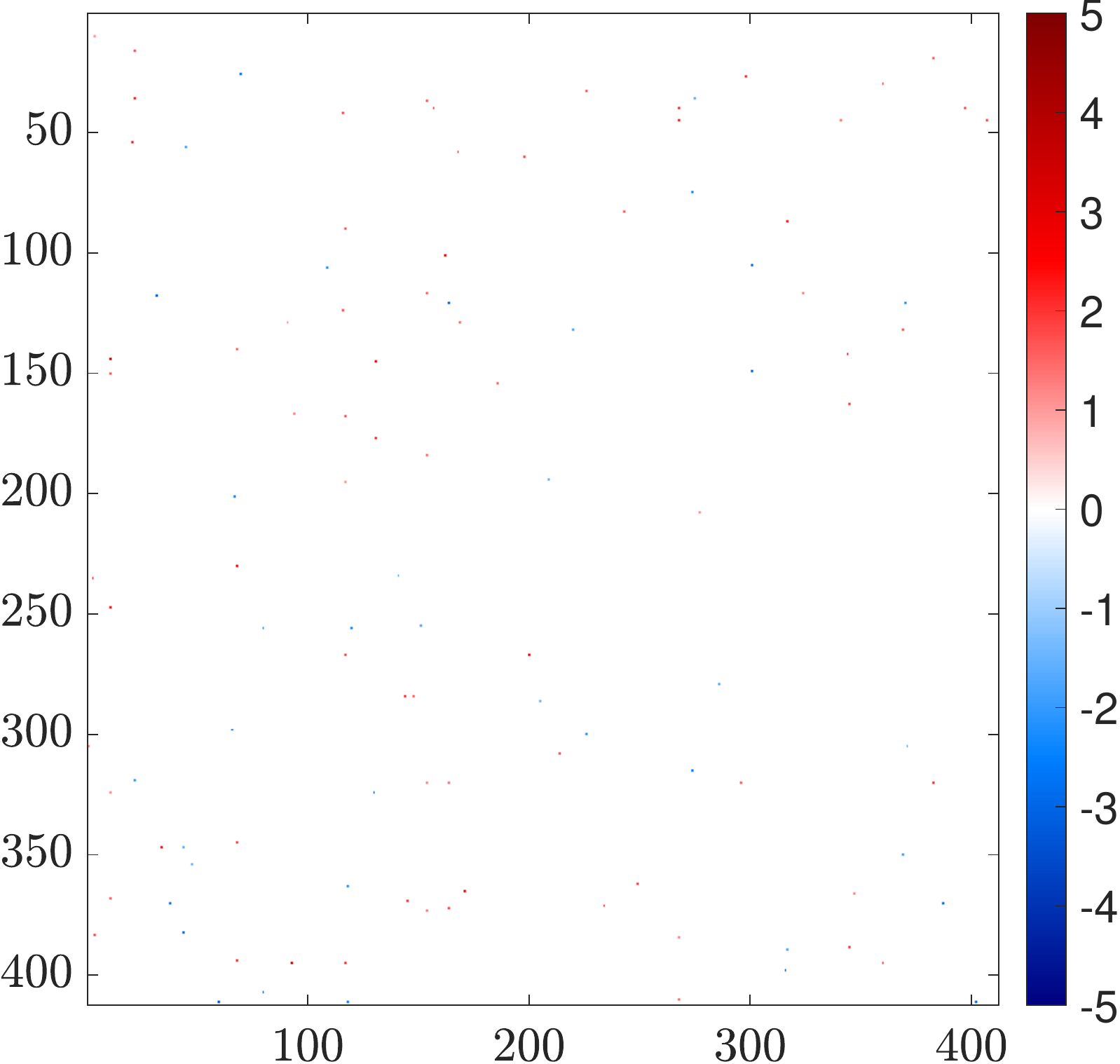} & 
\includegraphics[trim= 0mm 0mm 0mm 0mm,clip,height= 3.2cm, width= 3.7cm]{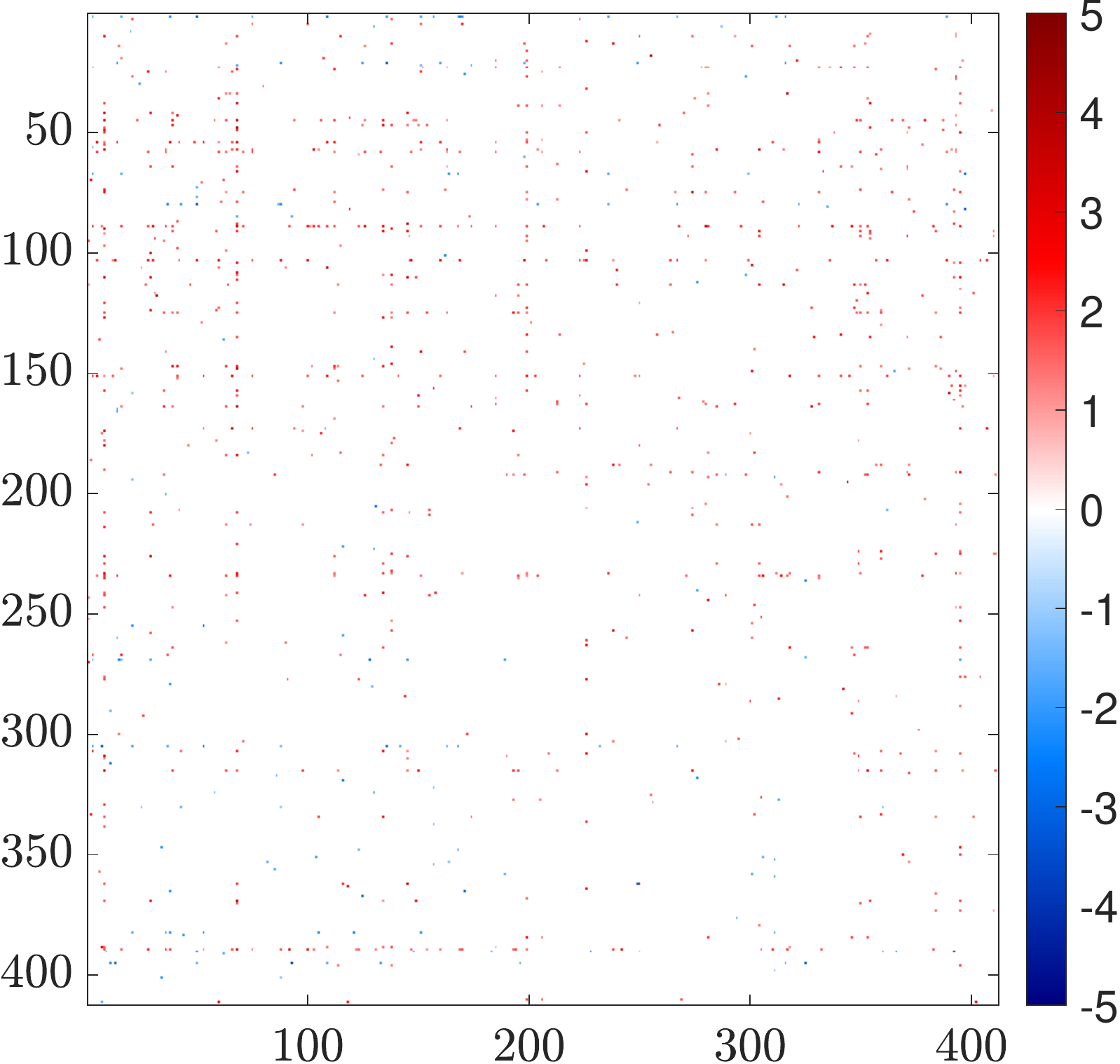} & 
\includegraphics[trim= 0mm 0mm 0mm 0mm,clip,height= 3.2cm, width= 3.7cm]{n412_M2_estimate_significant_12_r5-eps-converted-to.pdf} \\
\begin{rotate}{90} \hspace*{20pt} {\small leverage} \end{rotate} &
\includegraphics[trim= 0mm 0mm 0mm 0mm,clip,height= 3.2cm, width= 3.7cm]{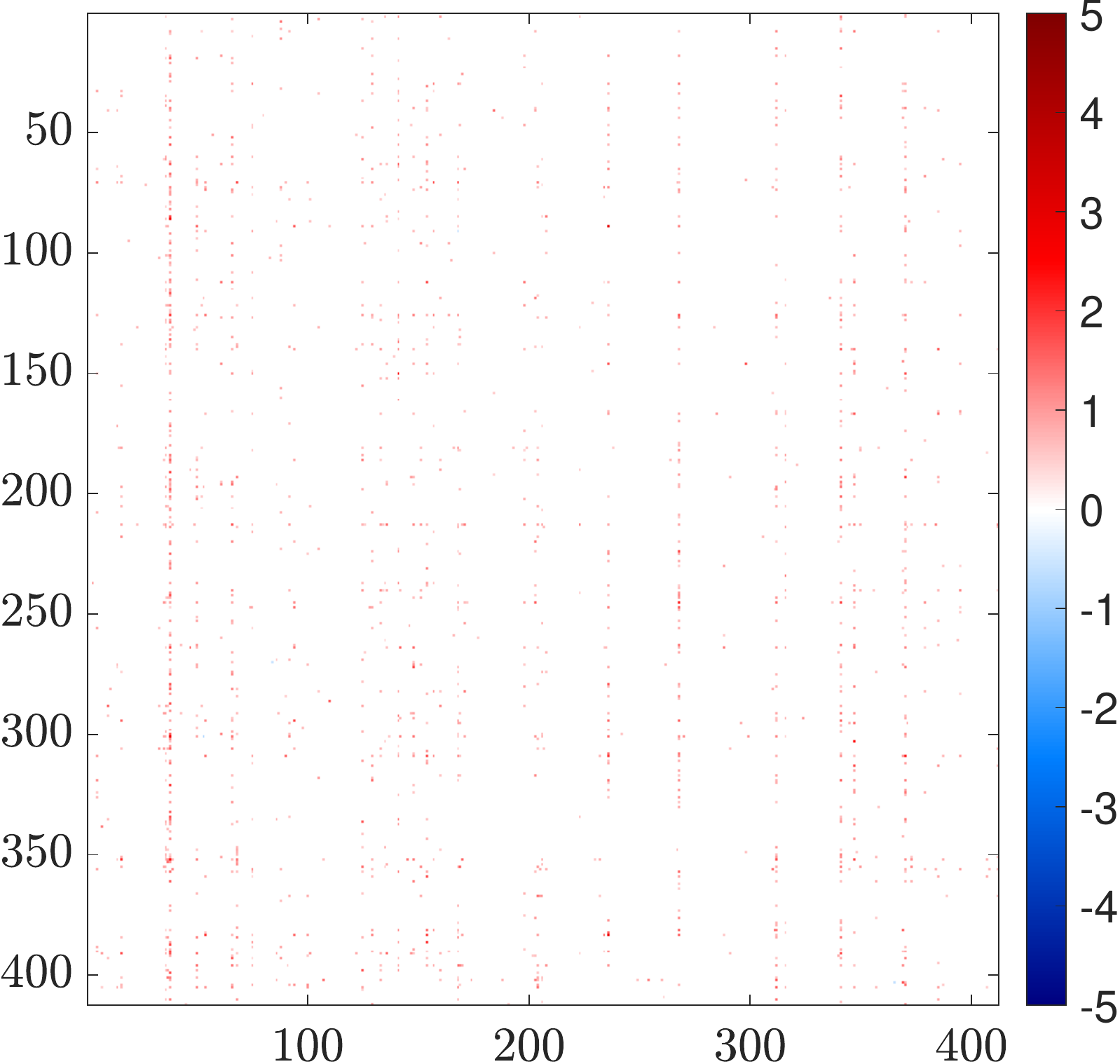} & 
\includegraphics[trim= 0mm 0mm 0mm 0mm,clip,height= 3.2cm, width= 3.7cm]{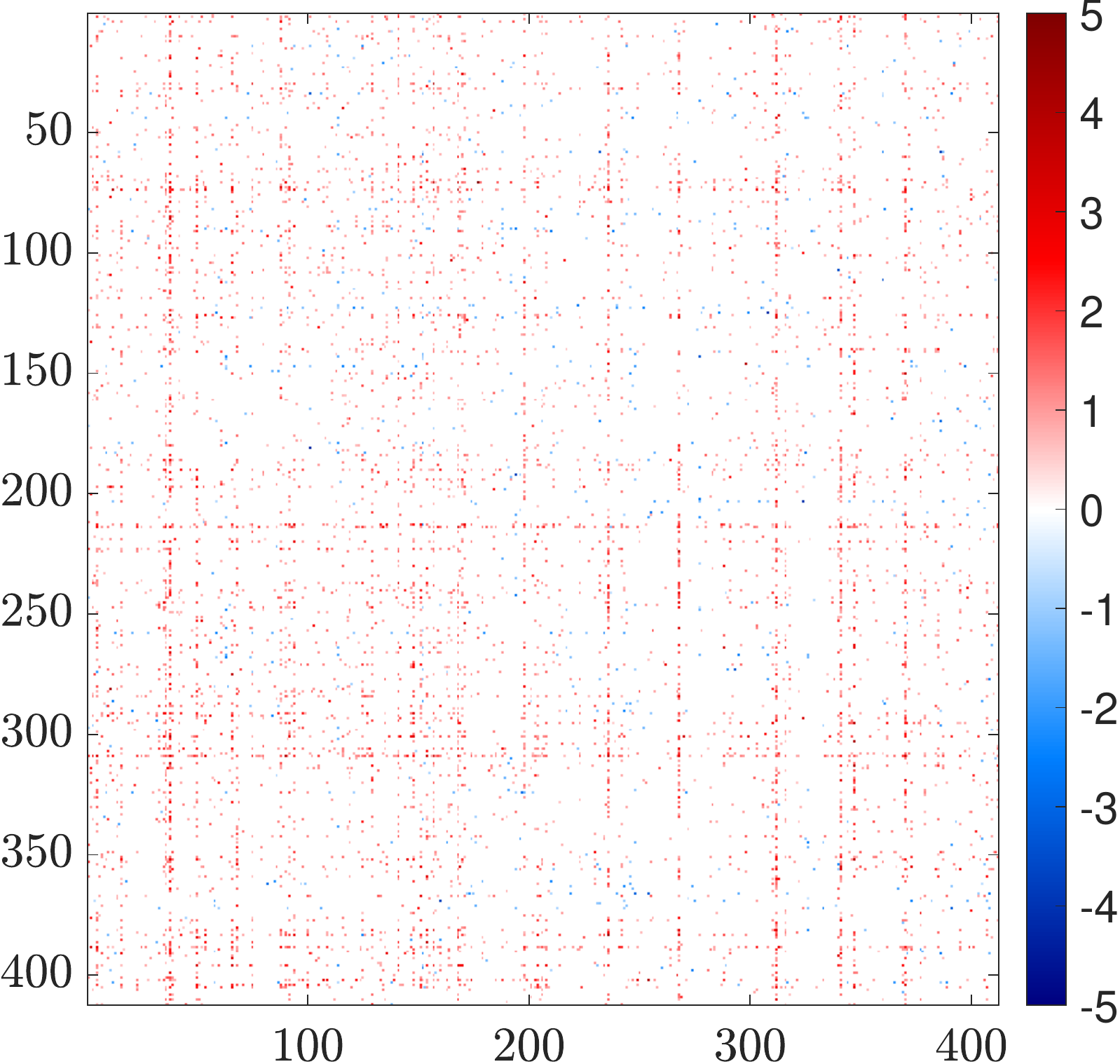} & 
\includegraphics[trim= 0mm 0mm 0mm 0mm,clip,height= 3.2cm, width= 3.7cm]{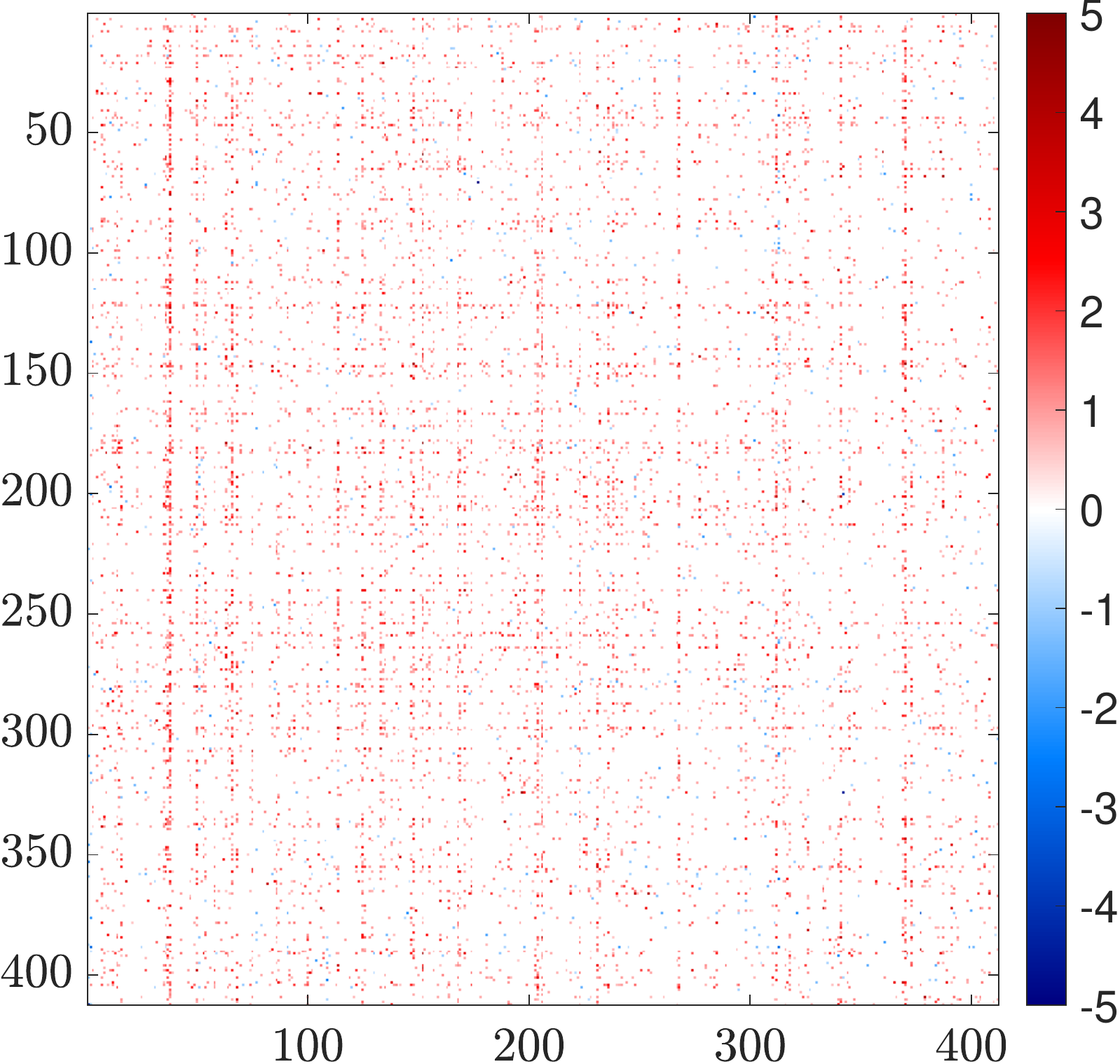} & 
\includegraphics[trim= 0mm 0mm 0mm 0mm,clip,height= 3.2cm, width= 3.7cm]{n412_M2_estimate_significant_21_r5-eps-converted-to.pdf} \\
\begin{rotate}{90} \hspace*{20pt} {\small volatility} \end{rotate} &
\includegraphics[trim= 0mm 0mm 0mm 0mm,clip,height= 3.2cm, width= 3.7cm]{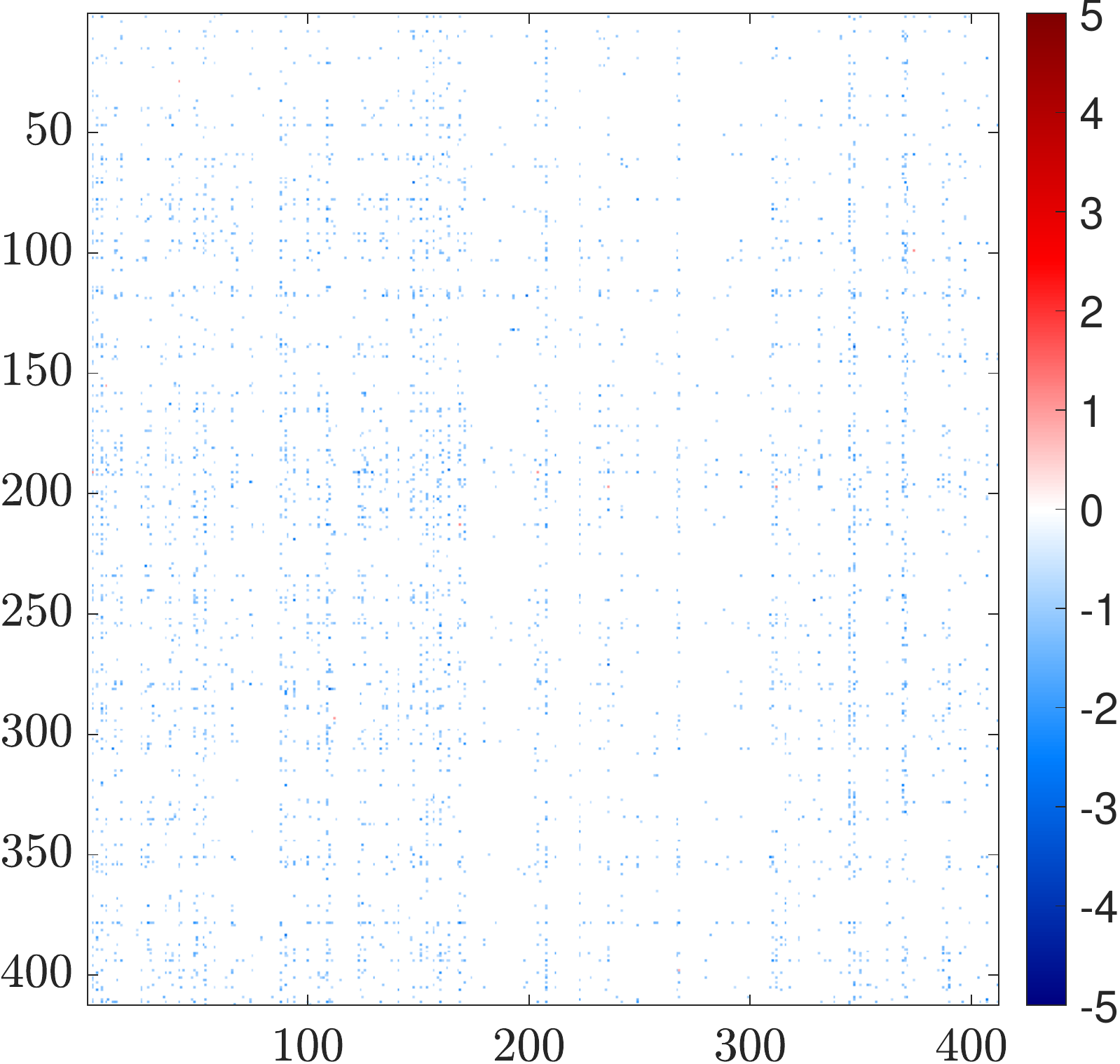} & 
\includegraphics[trim= 0mm 0mm 0mm 0mm,clip,height= 3.2cm, width= 3.7cm]{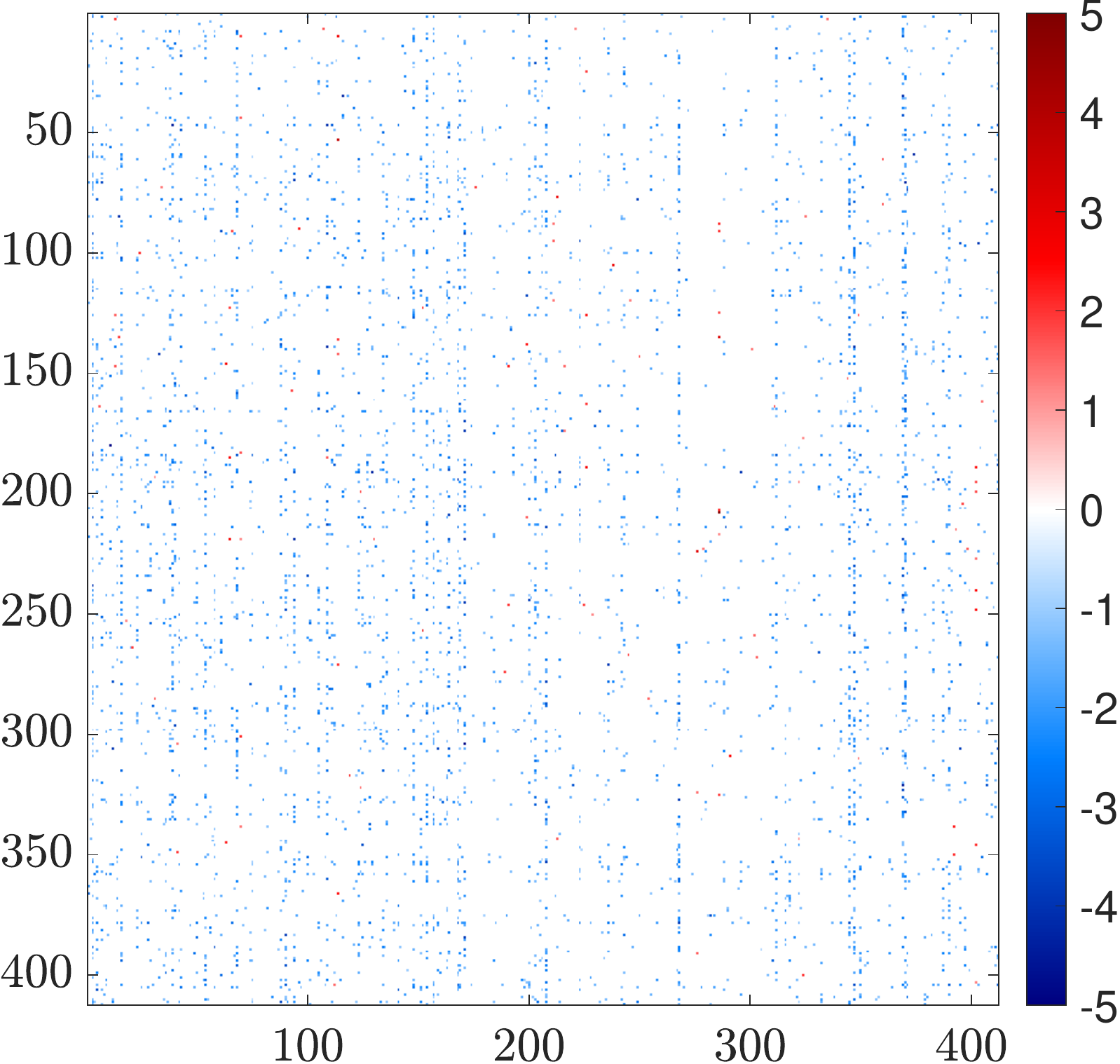} & 
\includegraphics[trim= 0mm 0mm 0mm 0mm,clip,height= 3.2cm, width= 3.7cm]{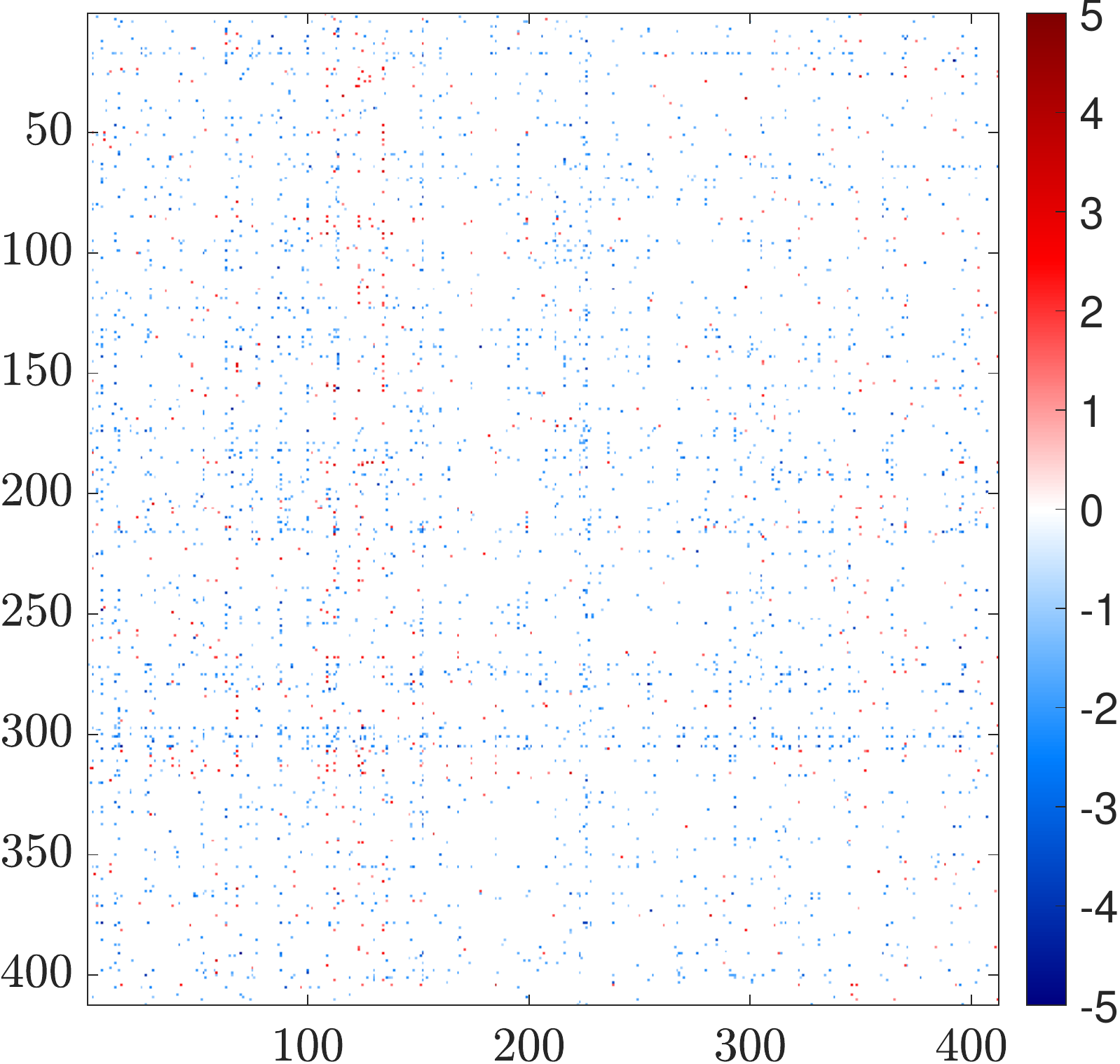} & 
\includegraphics[trim= 0mm 0mm 0mm 0mm,clip,height= 3.2cm, width= 3.7cm]{n412_M2_estimate_significant_22_r5-eps-converted-to.pdf}
\end{tabular}
\caption{Impact of risk factors (column) on financial linkages of the multilayer network.
In each plot, the coefficient in position $(i,j)$ refers to the impact of the risk factor in column on the edge from institution $j$ to institution $i$.
Only non-null coefficients are reported: blue indicates positive impact on edge existence, red indicates negative impact on edge existence.
A coefficient is considered null if its posterior HPDI contains zero.
\label{fig:posterior_mean_apdx}
}
\end{figure}

\end{document}